\documentclass[12pt]{article}

\usepackage{a4wide}
\usepackage{amsmath,amsthm,amssymb,xspace}

\newtheorem{Teorema}{Theorem}[section]
\newtheorem{Teorema *}{Theorem *}
\newtheorem{Teorema **}{Theorem **}
\newtheorem{Lemma}{Lemma}[section]

\newtheorem{Proposition}{Proposition}[section]
\newtheorem{Problema *}{Problem *}

\newtheorem{Definizione}{Definition}[section]
\newtheorem{Definition}{Definition}[section]
\newtheorem{definition}{Definition}[section]
\newtheorem{proposition}{Proposition}[section]
\newtheorem{remark}{Remark}[section]
\newtheorem{Remark}{Remark}[section]

\newcommand{\oneder}[2]{\ensuremath{{\mathop{{%
            \Rightarrow}}\limits^{{#1}}}_{\!\!_{#2}}\!}}
\newcommand{\multider}[1]{\ensuremath{{\mathop{{ %
            \Rightarrow}}\limits^{{#1}}}_{\!_\Re} %
            \raisebox{2pt}{\!\!\!\scriptsize *}}}
\newcommand{\multidernorm}[2]{\ensuremath{{\mathop{{%
          \Rightarrow}}\limits^{{#1}}}%
       \raisebox{2pt}{\!\scriptsize *}}\!_{\!_{#2}}\,}
\newcommand{\prsoneder}[1]{\ensuremath{\oneder{#1}{\Re}}}
\newcommand{\onederseq}[1]{\ensuremath{\oneder{#1}{\Re_{SEQ}}}}
\newcommand{\onederpar}[1]{\ensuremath{\oneder{#1}{\Re_{PAR}}}}
\newcommand{\multiderparaux}[1]{\ensuremath{\multidernorm{#1}{\Re_{PAR,AUX}}}}
\newcommand{\multiderseq}[1]{\ensuremath{\multidernorm{#1}{\Re_{SEQ}}}}
\newcommand{\multiderpar}[1]{\ensuremath{\multidernorm{#1}{\Re_{PAR}}}}

\def\parcomp#1#2{{#1\!\parallel\!#2}}
\def\seqcomp#1#2{{#1.(#2)}}

\def\prsrule#1#2#3{{#1\mathop{{\rightarrow}}\limits^{{#2}}#3}}
\def\prslongrule#1#2#3{{#1\mathop{{\rightarrow}}\limits^{{#2}}#3}}
\def\dollar{\ensuremath{\$}}
\def\dollarrule#1#2{\ensuremath{\prsrule{#1}{\dollar}{#2}}}
\def\cancrule#1#2{\ensuremath{\prsrule{#1}{\#}{#2}}}

\def\prssimpder#1#2#3{{#1\,\prsoneder{#2}\,#3}}
\def\prsder#1#2#3{{\,#1\,\multidernorm{#2}{\Re}#3}}
\def\prsdernorm#1#2#3#4{{\,#1\,\multidernorm{#2}{#3}#4}}
\def\prsderpar#1#2#3{{#1\,\multidernorm{#2}{\Re_{PAR}}#3}}
\def\prsderseq#1#2#3{{#1\,\multidernorm{#2}{\Re_{SEQ}}#3}}
\def\prsderparaux#1#2#3{{#1\,\multidernorm{#2}{\Re_{PAR,AUX}}#3}}

\def\ldsrule#1#2#3{\ensuremath{
   \cfrac[c]{#1}
    {\ #2\ }\text{\footnotesize \ \ensuremath{#3}}}}

\newcommand{\Rule}[1]{\ensuremath{{\mathop{{%
            \rightarrow}}\limits^{{#1}}}}}

\def\ALTL{\text{{\em ALTL}}\xspace}
\def\PRS{\text{{\em PRS}}\xspace}
\def\PRSs{\text{{\em PRS\/}s}\xspace}
\def\BRS{\text{{\em BRS}}\xspace}
\def\BRSs{\text{{\em BRS\/}s}\xspace}

\def\RDHA{\text{{\em RDHA\/}}\xspace}
\def\RDHAs{\text{{\em RDHA\/}s}\xspace}
\def\FSM{\text{{\em FSM\/}}\xspace}
\def\FSMs{\text{{\em FSM\/}s}\xspace}
\def\CHA{\text{{\em CHA\/}}\xspace}
\def\CHAs{\text{{\em CHA\/}s}\xspace}

\def\RSMs{\text{{\em RSM\/}s}\xspace}

\def\npla#1{{\langle #1\rangle}}
\def\size#1{{|#1|}}
\def\runs#1#2{\ensuremath{\text{\emph{runs}}}_{#1}(#2)}
\def\prsruns#1{\ensuremath{\text{\emph{runs}}}{(#1)}}
\def\suffix#1{\ensuremath{\text{\emph{suffix}}}(#1)}
\def\firstact#1{\ensuremath{\text{\emph{firstact}}}(#1)}
\def\buchi{\text{B\"{u}chi}\xspace}

\title{Verification of recursive parallel systems\footnote{This is an english version of preprint number 42 (october 2003) of Dipartimento di Matematica e Applicazioni dell'Universit\`{a} di Napoli}.}
\author{
L.~Bozzelli$^1$ \ \ \ M.~Benerecetti$^2$ \ and \ \  A.~Peron$^2$\\[10pt]
{\small\begin{tabular}{c@{\hspace{1.5cm}}c}
$\!\!^1$Dept. of Mathematics and Applications & $\!\!^2$Dept. of Physical Sciences\\
Universit\`a di Napoli ``Federico II'' & Universit\`a di Napoli ``Federico
II''\\
Napoli, Italy & Napoli, Italy\\
{\tt laura.bozzelli@dma.unina.it} & {\tt \{bene,peron\}@na.infn.it}
\end{tabular}
}}

\date{}
\begin{document}
\maketitle

\begin{abstract}
  In this paper we consider the problem of proving properties of
  infinite behaviour of formalisms suitable to describe (infinite
  state) systems with recursion and parallelism. As a formal setting,
  we consider the framework of Process Rewriting Systems (\PRSs). For
  a meaningfull fragment of \PRSs, allowing to accommodate both
  Pushdown Automata and Petri Nets, we state decidability results for
  a class of properties about infinite derivations (infinite term
  rewritings).  The given results can be exploited for the automatic
  verification of some classes of linear time properties of infinite
  state systems described by \PRSs. In order to exemplify the assessed
  results, we introduce a meaningful automaton based formalism which
  allows to express both recursion and multi--treading.
\end{abstract}

\section{Introduction}
Authomatic verification of systems is nowadays one of the most
investigated topic. A major difficulty to face when considering this
problem comes to tha fact that, reasoning about systems in general may
require to deal with infinite state models. For instance, software
sytems may introduce infinite states both manipulating data ranging
over infinite domains and having unbounded control structures such as
recursive procedure calls and/or dynamic creation of concurrent
processes (e.g. multi--treading). Many different formalisms have been
proposed for the description of infinite state systems. Among the most
popular are the well known formalisms of Context Free Process,
Pushdown Processes, Petri Nets, and Process Algebras. The first two
are models of sequential computation, whereas Petri Nets and Process
Algebra explicitely take into account concurrency.  The model checking
problem for these infinite state formalisms have been studied in the
literature.  As far as context free processes and Pushdown Automata
are concerned, decidability of the modal $\mu$--calculus, the most
powerful of the modal and temporal logics used for verification, has
been established (e.g. see \cite{BS94,hun94}).  As far as models of
concurrency, certain linear time properties are already undecidable
for small classes of Petri Nets (see \cite{esp97}, for a systematic
picture of decidability issues about model checking infinite states
concurrent models for both linear and branching time logics).

In order to exemplify our decidability results for formalisms
involving recursion and parallelisms, in this paper we shall consider
automata--based formalisms enriched with hierarchy (procedure call)
concurrency and communication among concurrent components, which have
gained popularity in recent years (e.g. see \emph{Statecharts\/}
\cite{h87}, \emph{ROOM\/} \cite{sgw94}, and \emph{UML\/}
\cite{bjr97}). Pure formalisms underlying the above mentioned
specification languages have been recently proposed and studied.
In \cite{asy99} \emph{Communicating Hierarchical Automata} (\CHAs)
have been introduced which extend \emph{Finite State Machines} (\FSMs)
with hierarchy and concurrency (though remaining in the finite--state
case). In \cite{alur01}, the finite state formalism \CHAs has been
strengthened, precisely to capture the expressive power of Pushdown
Automata (\emph{Recursive State Machine}). Efficient algorithms for
model checking and reachability analysis have been studied for
Pushdown Automata (e.g. see \cite{bouajjani97,ehrs00}), for \CHAs
\cite{ay98} and for Recursive State Machines \cite{alur01}.  In
\cite{lanotte03}, an extension of both \CHAs and \RSMs, called
\emph{Dynamical Hierarchical Machines}, is presented, which allow to
model code mobility via communication and dynamic activation of
state-transition machines. This formalism turns out to be Turing
equivalent, since unbounded hierarchy (i.e. recursive call) together
with the unrestricted ability of dynamically activating parallel
components at any hierarchical level allows for easily simulating a
double Pushdown Automaton. In this paper we shall consider a
restricted version of Dynamic Hierarchical Machines, where recursion
and dynamic activation of parallel components (multi--treads) is
allowed, while communication among parallel components is restricted in
such a way to prevent Turing equivalence.

Verification of formalisms which accommodate both parallelism and
recursion is a challenging problem. To formally study this kind of
systems, recently the formal framework of Process Rewrite Systems
(\PRSs) has been introduced \cite{mayr98}.  This framework, which
based on term rewriting, subsumes many common inifinite states models
such us Pushdown Systems, Petri Nets, Process Algebra, etc.  As we
shall see, also restricted Dynamic Hierarchical Automata can be easily
encoded into \PRSs. The decidability results already known in the
litterature for the general framework of \PRSs concern reachability,
i.e. properties of finite sequences of term rewriting (derivations).

In this paper we extend the known decidability results for a
relevant syntactic fragment of \PRSs to properties of infinite
derivations, thus allowing for automatic verification of some
classes of linear time properties.  Since this result is obtained
within the general formalism of the considered fragment of \PRSs,
it applies not only to the specific context of restricted Dynamic
Hierarchical Automata, but, more significantly, also to any
specification formalism which can be accommodated within this
fragment.  The fragment we consider is that of \PRSs in normal
form, where every rewrite rule either only deal with procedure
calls  (this kind of rules allows to capture Pushdown Processes),
or only deal with dynamic activation of processes and
synchronization (this kind of rules allows to capture Petri Nets).
A \PRS in normal form is extended with a notion of acceptance
\emph{a la} B\"uchi. A subset of rewrite rules is labelled as
'accepting' and an infinite derivation is accepting if there is an
accepting rewriting rule which is applied infinitely often in that
derivation. We prove that it is effectively decidable the problem
whether, for a given set of rewriting rules, there is an infinite
accepting derivation. We prove also that it is decidable whether
there is an infinite derivation devoid of any application of an
accepting rewrite rule, and whether there is an infinite
derivation involving a positive (finite) number of applications of
accepting rewriting rules.

The rest of the paper is structured as follows.  In Section 2, we
introduce the formalism of Dynamic Hierarchical Automata.  In
Section 3, we recall the framework of Process Rewriting Systems,
we summarize some decidability results for reachability problems
in the context of \PRSs, and show how DHAs can be embedded in PRS.
In Section 4, it is shown how decidability results about infinite
derivations can be used to check properties about infinite
executions of infinite state systems modelled by \PRSs.  In
Section 5, we prove decidability of the three problems about
infinite derivations in $\PRSs$ in normal form, mentioned above.


\section{Dynamic Hierarchical Automata}

In \cite{asy99} \emph{Communicating Hierarchical Automata}
(\CHAs) have been introduced which extend \emph{Finite State Machines}
(\FSMs) with
\emph{hierarchy}, \emph{concurrency\/} and \emph{communication\/}
among concurrent components.
A \CHA is a collection of Finite State Machines
(\FSMs). Hierarchy is achieved by injecting \FSMs into states of
other \FSMs. Whenever a \FSM state $s$ is entered, if such a
state contains a \FSM $M$, then $M$ starts running. The state $s$
can be left when $M$ reaches a final state. From this perspective,
entering the state $s$ can be viewed as a procedure call, with $M$
acting the part of the procedure. Finiteness of states of \CHAs
is guaranteed by syntactically forbidding recursive injection of
\FSMs into states. Concurrency is achieved by composing \FSMs in
parallel and by letting them run contemporaneously. Concurrent
machines communicate by synchronizing on transitions with the same
input label. The form of communication is a form of global
synchronization: if a parallel component performs a transition
labelled by an input symbol $a$, all of the other components
having $a$ in their input alphabet must perform a transition
labelled by $a$. In \cite{alur01}, the finite state formalism
\CHAs has been strengthened, from the expressive power viewpoint,
in such a way that the expressive power of Pushdown Automata is
precisely reached. Such a formalism is called \emph{Recursive
State Machine} ($RSM$, for short). The additional expressive power
is obtained by allowing recursive injection of \FSMs into states
(i.e. by admitting a form of recursive 'procedure call').
Actually, with respect to \CHAs, $RSMs$ does not allow explicit
representation of parallelism.

In \cite{lanotte03} an extension of both \CHAs and $RSMs$ is presented
called \emph{Dynamical Hierarchical Machines} ($DHMs$). $DHMs$
allow the explicit representation of hierarchy, pa\-rallelism and
communication. Moreover, in order to model aspects of code
mobility the form of communication between parallel components
allows sending and receiving \FSMs (communication in \CHAs takes
the form of pure communication). A \FSM received from a parallel
component can be dynamically activated in different ways: either
in parallel with the receiving component or in parallel with a
component at a lower/upper level (with respect to the receiving
component). Hierarchy and dynamical activation allows to easily
simulate a Pushdown Automaton. Moreover, parallelism and the
unrestricted ability of dynamically activating parallel components
at any hierarchical level allows to easily simulate a double
Pushdown Automaton thus reaching the expressive power of Turing
Machines.

In this section we introduce an extension of $RSMs$, called
\emph{Restricted Dynamical Hierarchical Automata} (\RDHAs for
short), which allows recursive injection of \FSMs into states (as
in $RSMs$) and dynamical parallel activations of \FSMs (as in
$DHMs$). As in \CHAs and $RSMs$, a \RDHA is a collection of
\FSMs, and \FSMs can be (recursively) injected into states. A
transition of a \FSM is decorated by a pair of symbols, the
former being an input symbol (belonging to an alphabet
$\Upsilon$), the latter representing an action. The possible
actions are: $NIL$, representing the null action; $HALT$,
representing the termination action; a channel symbol in $\Gamma$,
representing a synchronization request on a channel name;
$NEW(i,p)$, representing the dynamic activation of the $i$-th
\FSM in its initial state $p$. A transition is triggered by the
input symbol and, when performed, produces the corresponding
action.

In order to obtain a formalism which is less expressive than
$DHMs$, parallelism communication and dynamic activation are
presented in a restricted form. Actually, there is no explicit
syntactical construct for parallelism in \RDHAs. Parallelism is
the result of dynamic activation of sequential machines.
During its evolution, a
\FSM can dynamically activate in parallel with itself a (possibly
unbound) number of \FSMs. When a \FSM $A$ activates an another
\FSM $A'$, $A$ and $A'$ are put in parallel at the same
hierarchical level (the father of $A'$ is the father of $A$).
\newline
Synchronization has the form of handshaking between two parallel
components. A transition in a \FSM labelled by a synchronization
request on a channel $\alpha$ can be performed only if there is a
\FSM, activated in parallel with it, able to perform a transition
with a synchronization request on the same channel name.
Synchronization between parallel components of different
hierarchical level (i.e. interlevel communication) is not allowed.
Moreover, synchronization is allowed only if the involved
parallel components are not waiting for return of a procedure call
(i.e. if they are leaves in the hierarchy of activations).
\newline
A \FSM $A'$ injected into a box $b$ of a \FSM $A$ is
disactivated either when it is in a final state associated with a
transition departing from $b$ (a kind of procedure termination
with value return) or when it performs the $HALT$ action
(procedure termination without value return).

\begin{Definizione}
Let $\Upsilon$ and $\Gamma$ be finite alphabets for input symbols
and channels, respectively. A {\em Restricted Dynamic Hierarchical
Automaton} $\mathrm{(}$\RDHA for short$\mathrm{)}$ over
$\Upsilon$ is a collection of {\em sequential machines\/}
$A_{1},\ldots,A_{n}$, with
\[
A_{i}= \npla{Q_{i}\cup{}B_{i},Y_{i},Q_{i}^{0},Q_{i}^{T},\delta_{i}}, \mbox{ where}
\]
\begin{itemize}
\item $Q_{i}$ is the finite set of {\em nodes\/};
\item $B_{i}$ is
the finite set of {\em boxes\/} $\mathrm{(}$we assume $B_i\cap Q_i
=\emptyset$$\mathrm{)}$;
\item $Q_{i}^{0}\subseteq{}Q_{i}$ is the
set of {\em initial nodes\/};
\item $Q_{i}^{T}\subseteq{}Q_{i}$ is
the set of {\em exit nodes\/};
\item $Y_{i}:B_{i}\rightarrowtail{}\{1,\ldots,n\}$ is the {\em hierarchy
function\/} associating boxes with sequential machines
$A_{1},\ldots,A_{n}$;
\item $\delta_{i} \subseteq ((Q_i\setminus
Q^T_i) \cup (B_i \times \bigcup^n_{j=1}Q_j)) \times \Upsilon
\times Sy \times (Q_i \cup (B_i \times \bigcup^n_{j=1}Q_j) )$ is
the {\em transition relation\/}, with
\[
Sy = \{NIL,HALT\} \cup \Gamma \cup \{NEW(i,p) \mid  i=1,\ldots, n
\textrm{ and } p \in Q^0_i \};
\]
for any $\npla{u,a,\xi,v}
\in \delta_i$, the following constraints are fulfilled:
\begin{itemize}
\item if $\xi \not\in \{NIL,HALT\}$, then $u,v \in Q_i$; \item if
$\xi = HALT$, then $v \in Q^T_i$; \item if $u \not\in Q_i$, then
$v \in Q_i$; \item if $v \not\in Q_i$, then $u \in Q_i$; \item if
$u$ has the form $\npla{b,q}$, then $q\in Q^T_j$, with $j=Y_i(b)$;
\item if $v$ has the form $\npla{b,q}$, then $q\in Q^0_j$, with
$j=Y_i(b)$;
\end{itemize}
\end{itemize}
As a further constraint we require that $Q_i \cap Q_j = \emptyset$ and
$B_i \cap B_j = \emptyset$ for all $1 \leq i < j \leq n$.
\end{Definizione}

When an activation of a \FSM $A_i$ in a state $u$ performs a transition
$\npla{u,a,NIL,\npla{b,q}}$, it enters the box $b$ and
activates an instance of the \FSM $A_{Y_i(b)}$ (a procedure call) in its initial
node $q$ and waits for the termination of $A_{Y_i(b)}$.
The \FSM $A_{Y_i(b)}$ terminates when it reaches an exit node
$q_t$ and can be deallocated if either $q_t$ is reached by
a transition of the form $\npla{u',a',HALT,q_t}$ (termination
without value return), or if $A_i$ has a transition
of the form $\npla{\npla{b,q_t},a',NIL,u'}$ (termination
with value return).
\newline
When an activation of a \FSM $A_i$ in a state $u$ performs a transition
$\npla{u,a,NEW(j,q),v}$, it enters the node $v$ and
activates in parallel with itself an instance of $A_j$ in its initial node $q$
(a dynamic activation). The two activations of $A_i$
and $A_j$ can run asynchronously in parallel.
\newline
An activation of a \FSM $A_i$ in a state $u$ can perform a
transition $t_1=\npla{u,a,\gamma,v}$, with $\gamma$ a channel name
in $\Gamma$, only if there is a parallel activation of a \FSM
$A_j$ in a state $u'$ which can perform a transition of the form
$t_2=\npla{u',a,\gamma,v'}$ (synchronization on the same channel
name and the same input symbol). Both $A_i$ and $A_j$ have to perform transitions $t_1$ and
$t_2$, simultaneously.
\newline
Notice that, as a consequence of constraints imposed on the
transition relations, actions of synchronization and dynamic
creation can be performed only starting from nodes (i.e. leaves in
the hierarchy of \FSM activations).

In order to give a formal semantics of \RDHAs we have to
introduce the notion of configuration. A configuration is a tree
which describes the collection of instances of \FSMs
instantaneously activated, together with the hierarchy of
activations (caller - called relationship). In particular, a node
of the configuration tree is either a box (non leaf node of the
tree) or a node (leaf node of the tree) of the  considered \RDHA,
representing the current state of a \FSM activation instance. A
configuration tree is described by an algebra of terms composed of
node and box symbols by means of a binary operation of {\em
sequential} composition (i.e. procedure call) denoted by
($\_.(\_)$)  and  a binary operation of {\em parallel} composition
denoted by $\parcomp {\_}{\_}$. (We recall that a $NEW$ action
results in a parallel composition.)

\begin{Definizione}
Let ${\cal A} = \{A_{1},\ldots,A_{n}\}$, be a \RDHA with $A_{i}\npla{Q_{i}\cup{}B_{i},Y_{i},Q_{i}^{0},Q_{i}^{T},\delta_{i}}$. The
set of {\em configuration terms\/} of ${\cal A}$, written
$Conf({\cal A})$, is inductively defined as follows:
\begin{itemize}
\item $\varepsilon\in Conf({\cal A})$ $\mathrm{(}$the {\em empty
configuration\/}$\mathrm{)}$;
\item $\bigcup^{n}_{i=1}Q_i
\subseteq Conf({\cal A})$;
\item $\seqcomp{b}{t} \in Conf({\cal
A})$, for $b \in \bigcup^{n}_{i=1}B_i$ and $t \in Conf({\cal A})$;
\item $\parcomp{t_1}{t_2} \in Conf({\cal A})$, for $t_1,t_2 \in
Conf({\cal A})$.
\end{itemize}
\end{Definizione}

In the following we shall consider equal configuration terms of
$Conf({\cal A})$ up to commutativity and associativity of parallel
composition. Moreover, the configuration term $\varepsilon$ will
be treated as the identity element for parallel and sequential
composition. More precisely, for a \RDHA ${\cal A}$,
configuration terms of ${\cal A}$ are considered equal up to a notion of
equivalence $\approx_{{\cal A}}$ defined as the least equivalence
fulfilling the following requirements:
\begin{itemize}
\item $\parcomp{t_1}{t_2}\approx_{{\cal A}} \parcomp{t_2}{t_1}$, for all
$t_1,t_2 \in Conf({\cal A})$;
\item $\parcomp{t_1}{(\parcomp{t_2}{t_3} )}\approx_{{\cal A}} \parcomp{(\parcomp{t_1}{t_2})}{t_3}$,
for all $t_1,t_2, t_3 \in Conf({\cal A})$;
\item $\parcomp{t}{\varepsilon} \approx_{{\cal A}}t$, for any $t\in
Conf({\cal A})$;
\item $b.(\varepsilon)\approx_{{\cal A}}b$;
\item $b.(t_{1})\approx_{{\cal A}}b.(t_{2})$, for all $t_1,t_2 \in Conf({\cal A})$
such that $t_{1}\approx_{\Phi}t_{2}$.
\end{itemize}

In the following we shall define the semantics of \RDHAs in terms
of Labelled Transition Systems ($LTSs$) defined in the well known
style of Structured Operational Semantics (see \cite{P81}). The
LTS for a \RDHA ${\cal A}$ is the triple $\npla{Conf_{{\cal
A}},\Upsilon,\prsrule{}{a}{}}$, where the  states are the
configuration terms, the set of labels is $\Upsilon$ and the
transition relation $\prsrule{}{a}{} \subseteq Conf_{{\cal A}}
\times \Upsilon \times Conf_{{\cal A}}$ is defined by the
following set of rules. (In the following  $\delta$ stands for
$\bigcup^{n}_{i=1} \delta_i$.)

\begin{eqnarray}
\ldsrule{}{\prsrule{q}{a}{q'}}{\npla{q,a,NIL,q'}
\in \delta}
\\
\ldsrule{}{\prsrule{q}{a}{b.(q')}}{\npla{q,a,NIL,\npla{b,q'}}
\in \delta}
\\
\ldsrule{}{\prsrule{b.(q)}{a}{q'}}{\npla{\npla{b,q},a,NIL,q'}
\in \delta}
\\
\ldsrule{}{\prsrule{q}{a}{\varepsilon}}{\npla{q,a,HALT,q'}
\in \delta}
\\
\ldsrule{}{\prsrule{q}{a}{\parcomp{q'}{p}}}{\npla{q,a,NEW(j,p),q'}
\in \delta}
\\
\ldsrule{ }
{\prsrule{ \parcomp{q_1}{q_2} }{a}{ \parcomp{q'_1}{q'_2} }}
{\npla{q_1,a,\alpha,q'_1}, \npla{q_2,a,\alpha,q'_2} \in \delta
}
\\
\ldsrule{ \prsrule{t'}{a}{t''} }
{\prsrule{ \parcomp{t}{t'} }{a}{ \parcomp{t}{t''} }}
{a\in \Upsilon}
\\
\ldsrule{ \prsrule{t'}{a}{t''} }
{\prsrule{b.(t')}{a}{b.(t'')}}
{a\in \Upsilon}
\end{eqnarray}

Axiom 1 considers the case of a basic transition inside a \FSM.
Axiom 2 considers the case of a procedure call: the \FSM
$A_{Y_i(b)}$, with $b\in B_{i}$, is activated in its initial state
$q'$. Axiom 3 (resp. 4) considers the case of a procedure
termination with (resp. without) value return. Axiom 5 considers
the case of a transition with dynamic activation; notice that the
newly activated \FSM is put in parallel with the activating one.
Rule 6 deals with
synchronization: two requests of synchronization on a common
channel name are synchronized. Rule 7 allows a parallel component,
which does not perform synchronization requests, to freely
(asynchronously) evolve. Rule 8 allows a machine, which does not perform
synchronization requests, to freely evolve in the context of
procedural call.




\section{Process Rewriting Systems and \RDHAs}

In this section we recall the framework of \emph{Process Rewriting
Systems} (\PRSs). In this setting we rephrase the LTS semantics of
\RDHAs given in the previous section. We conclude the section by
summarizing some decidability results, known in the literature,
for the problem of model checking of systems described by \PRSs.

\subsection{Process Rewrite Systems}

In this section we recall the notion of Process Rewrite System, as
introduced in~\cite{mayr98}. The idea is that a process (and
its current state) is described by a term. The behaviour of a process is
given by rewriting the corresponding term by means of a finite set
of rewriting rules.

\begin{definition}[Process Term] Let $Var$ be a finite set of process
  variables. The set $T$ of {\em process terms} over $Var$ is
  inductively defined as follows:
\begin{itemize}
\item $Var\subseteq T$;
\item $\varepsilon\in{}T$;
\item $\parcomp{t_{1}}{t_{2}}\in{}T$, for all $t_{1},t_{2}\in{}T$;
\item  $\seqcomp{X}{t}\in{}T$, for all $X\in{}Var$ and $t\in{}T$,
\end{itemize}
where $\varepsilon$ denotes the empty term, ``\,$\parallel$'' denotes
parallel composition, and ``$.()$'' denotes sequential
composition\footnote{\cite{mayr98} also allows terms of the form
  $t_{1}.(t_{2})$, where $t_{1}$ is a parallel composition of
  variables. In the current context this generalization is not
  relevant.}.
\end{definition}

We denote with $T_{SEQ}$ the subset of terms in $T$ devoid of any
occurrence of parallel composition operator,
and with $T_{PAR}$ the subset of terms
in $T$ devoid of any occurrence of the sequential composition operator.
Notice that we have $T_{PAR}\cap T_{SEQ} = Var\cup\{\varepsilon\}$.

In the rest of the paper we only consider process terms modulo
commutativity and associativity of ``$\parallel$'', moreover
$\varepsilon$ will act as the identity for both parallel and
sequential composition.
Therefore, we introduce the relation $\approx_{T}$, which is the
smallest equivalence relation on $T$ such that for all $t_1,t_2,
t_3\in{}T$ and $X\in{}Var$:
\begin{itemize}
\item $\parcomp{t_{1}}{t_{2}}\approx_{T}\parcomp{t_{2}}{t_{1}}$,
  $\parcomp{t_{1}}{(\parcomp{t_{2}}{t_{3}})} \approx_{T}
  \parcomp{(\parcomp{t_{1}}{t_{2}})}{t_3}$, and
    $\parcomp{t_1}{\varepsilon}\approx_{T}t_1$;
\item $\seqcomp{X}{\varepsilon}\approx_{T}X$, and if $t_{1}\approx_{T}t_{2}$,
  then $\seqcomp{X}{t_{1}}\approx_{T}\seqcomp{X}{t_{2}}$.
\end{itemize}

In the paper, we always confuse terms and their equivalence classes
(w.r.t.  $\approx_{T}$). In particular, $t_1 = t_2$ (resp., $t_1 \not=
t_2$) will be used to mean that $t_1$ is equivalent (resp., not
equivalent) to $t_2$.
\begin{Definition}[Process Rewrite System]
  A {\em Process Rewrite System} $\mathrm{(}$or \PRS, or {\em Rewrite System}$\mathrm{)}$ over the
  alphabet $\Sigma$ and the set of process variables $Var$ is a finite
  set of rewrite rules $\Re\subseteq{}T\times\Sigma\times{}T$ of the
  form $\prsrule{t}{a}{t'}$, where $t$ $\mathrm{(}$$\neq\varepsilon$$\mathrm{)}$ and $t'$
  are terms in $T$, and $a\in\Sigma$.
\end{Definition}

The semantics of a \PRS $\Re$ is given by a Labelled Transition
System $\npla{T,\Sigma,\prslongrule{}{}{}}$, where the set of
states is the set of terms $T$ of $\Re$, the set of actions is the
alphabet $\Sigma$ of $\Re$, and the transition relation
$\prslongrule{}{}{} \subseteq T \times \Sigma \times T$ is the
smallest relation satisfying the following inference rules:

\ldsrule{}{\prslongrule{t}{a}{t'}}{(\prsrule{t}{a}{t'})\in\Re}\hspace*{1cm}
\ldsrule{\prslongrule{t_1}{a}{t_1'}}
     {\prslongrule{\parcomp{t_1}{t}}{a}{\parcomp{t_1'}{t}}}{\forall{}t\in{}T}
     \hspace*{1cm}
\ldsrule{\prslongrule{t_1}{a}{t_1'}}
     {\prslongrule{\seqcomp{X}{t_1}}{a}{\seqcomp{X}{t_1'}}}{\forall{}X\in{}Var}
\newline\newline

 For a \PRS $\Re$ with set of terms $T$ and LTS
$\npla{T,\Sigma,\prslongrule{}{}{}}$, a \emph{path in} $\Re$ from
$t\in{}T$ is a path in $\npla{T,\Sigma,\prslongrule{}{}{}}$ from
$t$, i.e. a (finite or infinite) sequence of LTS edges
$\prsrule{t_{0}}{a_{0}}{t_{1}} \prsrule{}{a_{1}}{t_{2}}
\prsrule{}{a_{2}}{}$ such that $t_0=t$ and
$\prsrule{t_{j}}{a_{j}}{t_{j+1}} \in \prsrule{}{}{}$ for any $j$.
A \emph{run in $\Re$} from $t$ is a maximal path from $t$, i.e. a
path from $t$ which is either infinite or has the form
$\prsrule{t_{0}}{a_{0}}{t_{1}} \prsrule{}{a_{1}}{\ldots}
\prsrule{}{a_{n-1}}{t_n}$ and there is no edge
$\prsrule{t_n}{a_n}{t'} \in \prsrule{}{}{}$, for any $a_n\in
\Sigma$ and $t' \in T$. We write $runs_{\Re}(t)$ (resp.,
$runs_{\Re,\infty}(t)$) to refer to the set of runs (resp.,
infinite runs) in $\Re$ from $t$, and $runs(\Re)$  to refer to the
set of all the runs  in $\Re$.

\bigskip

The LTS semantics induces, for a rule $r\in \Re$, the following
notion of \emph{one--step derivation} by $r$. The \emph{one--step
derivation} by $r$ relation, \prsoneder{r}, is the least relation
such that:
\begin{itemize}
\item $t$ \prsoneder{r} $t'$, for $r=\prsrule{t}{a}{t'}$ \item
$t_{1}$$\parallel$$t$ \prsoneder{r} $t_{2}$$\parallel$$t$, if
$t_{1}$ \prsoneder{r} $t_{2}$ and $t\in{}T$ \item $X.(t_{1})$
\prsoneder{r} $X.(t_{2})$, if $t_{1}$ \prsoneder{r} $t_{2}$ and
$X\in{}Var$
\end{itemize}

A \emph{finite derivation} in $\Re$ from a term $t$ to a term $t'$
(through a finite sequence $\sigma = r_{1}r_{2}\ldots{}r_{n}$ of
rules in $\Re$), is a sequence $d$ of one--step derivations $t_0$
\prsoneder{r_1} $t_1$ \prsoneder{r_2} $t_2\ldots$
\prsoneder{r_{n-1}} $t_{n-1}$ \prsoneder{r_n} $t_n$, with $t_0=t$,
$t_n=t'$ and $t_i$ \prsoneder{r_{i+1}} $t_{i+1}$ for all
$i=0,\ldots,n-1$. The derivation $d$ is a \emph{$n$--step
derivation} (or a \emph{derivation of length $n$}), and for
succinctness is denoted by $t$ \multider{\sigma} $t'$. Moreover,
we say that $t'$ is \emph{reachable} in $\Re$ from  term $t$
(through derivation $d$). If $\sigma$ is empty, we say that $d$ is
a {\em null derivation\/}.
\newline
A \emph{infinite derivation} in $\Re$ from a term $t$ (through an
infinite sequence $\sigma = r_{1}r_{2}\ldots $ of rules in $\Re$),
is an infinite sequence of one step derivations $t_0$
\prsoneder{r_1} $t_1$ \prsoneder{r_2} $t_2\ldots$ such that
$t_0=t$ and $t_i$ \prsoneder{r_{i+1}} $t_{i+1}$  for all $i \geq
0$. For succinctness such derivation is denoted by $t$
\multider{\sigma}.
\newline
Notice that there is a strict correspondence between the notion of
derivation from a term $t$ and that of path from the term $t$. In
fact, we have that there is a path $\prsrule{t_{0}}{a_{0}}{t_{1}}
\prsrule{}{a_{1}}{t_{2}\ldots}$  from $t_0$ in $\Re$ iff there
exists a derivation $t_0$ \prsoneder{r_1} $t_1$ \prsoneder{r_2}
$t_2\ldots$ from $t_0$ in $\Re$, with $a_i = label(r_i)$, for any
$i$ (where for a rule $r \in \Re$ with $r = \prsrule {t}{a}{t'}$,
$label(r)$ denotes the label $a$ of $r$).

\bigskip

We rephrase now the semantics
of \RDHAs in the setting of \PRSs.
Let  ${\cal A}= \{A_{1},\ldots,A_{n}\}$, be a \RDHA with
$A_{i}= \npla{Q_{i}\cup{}B_{i},Y_{i},Q_{i}^{0},Q_{i}^{T},\delta_{i}}$.
The \PRS for ${\cal A}$, written $PRS_{\cal A}$, is given as follows:
\begin{enumerate}
\item $Var = \{X_q : q\in \bigcup^n_{i=1}Q_i\} \cup
\{X_b : b\in \bigcup^n_{i=1}B_i\}$ is the set of variables indexed
over the set of nodes and boxes;
\item the alphabet $\Sigma$ equals the alphabet $\Upsilon$ of ${\cal A}$
\item the set of rules $\Re$ is the union of the following sets:
\begin{enumerate}
\item $\{\prsrule{X_q}{a}{X_{q'}} : \npla{q,a,NIL,q'}\in
\bigcup^n_{i=1}\delta_i\}$
\item $\{\prsrule{X_q}{a}{\epsilon} : \npla{q,a,HALT,q'}\in
\bigcup^n_{i=1}\delta_i\}$
\item $\{\prsrule{X_q}{a}{X_b.(X_p)} : \npla{q,a,NIL,\npla{b,p}}\in
\bigcup^n_{i=1}\delta_i\}$
\item $\{\prsrule{X_b.(X_p)}{a}{X_q} : \npla{\npla{b,p},a,NIL,q}\in
\bigcup^n_{i=1}\delta_i\}$
\item $\{\prsrule{ \parcomp{X_{q_1}}{X_{q_2}} }{a}{\parcomp{X_{q'_1}}{X_{q'_2}} } :
\npla{q_1,a,\gamma,q'_1}, \npla{q_2,a,\gamma,q'_2}, \in
\bigcup^n_{i=1}\delta_i\}$
\end{enumerate}
\end{enumerate}

It is easy to show that  the given translation of a \RDHA ${\cal
A}$ into $PRS_{\cal A}$, is correct in the sense that the LTS for
a ${\cal A}$ and the LTS for $PRS_{\cal A}$ are isomorphic.

The embedding of \RDHAs into
\PRSs suggests an immediate interpretation of
\PRS format rules.
Rules involving sequential
composition allow one to model
procedure call and termination: in particular
a rule of the form $\prsrule{X}{a}{Y.(t)}$ allows to
model procedure call, and a rule of the
form $\prsrule{Y.(t)}{a}{Z}$ allows to
model procedure termination (possibly with value return if
$t \neq \varepsilon$). Rules involving parallel composition
allow to model dynamic process activation and synchronization
among parallel process: the former can be expressed by
rules having the form $\prsrule{t_1}{a}{\parcomp{t_1}{t_2}}$,
whereas the latter by rules having the form
$\prsrule{\parcomp{t_1}{t_2}}{a}{\parcomp{t'_1}{t'_2}}$.

In the following, we shall consider PRS in a
syntactical restricted form called \emph{normal form}.

\begin{definition}[Normal Form] A \PRS $\Re$ is said to be in {\em
    normal form} if every rule $r\in\Re$ has one of the following forms:
\begin{description}
\item[PAR rules:] Any rule devoid of sequential composition;
\item[SEQ rules:] $\prsrule{X}{a}{Y.(Z)}$, $\prsrule{X.(Y)}{a}{Z}$ or
  $\prsrule{X}{a}{Y}$, or $\prsrule{X}{a}{\varepsilon}.$
\end{description}
\noindent with $X,Y,Z\in{}Var$.
A \PRS where all the rules are SEQ rules is called \emph{sequential} \PRS.
Similarly, a \PRS where all the rules are PAR rules is called
\emph{parallel} \PRS.
\end{definition}

With reference to our embedding of $RDHA$ into $PRS$, notice that
the $PRS_{\cal A}$ for a $RDHA$  ${\cal A}$ is in normal form and consists of
both sequential and parallel rules.

The sequential and parallel fragments of $PRS$ are significant: in
\cite{mayr98} it is shown that sequential \PRS\/s are semantically
equivalent (via \emph{bisimulation equivalence}) to Pushdown
Automata (PDA), while parallel \PRS\/s are semantically equivalent
to Petri Nets (PN). Moreover, from the fact that Pushdown systems
and Petri Nets  are not comparable (see \cite{M96,BCS96}) it
follows that \PRSs in normal form are strictly more expressive
than both their sequential and parallel fragment.

\subsection{Decidability results for \PRSs}

In this section we will summarize decidability results on \PRSs
which are known in the literature and which will be exploited in
further sections of the paper.

\paragraph{Verification of \ALTL (Action--based LTL)}\

Given a finite set $\Sigma$ of atomic propositions, the set of
formulae $\varphi$ of \ALTL over $\Sigma$ is defined as follows:
\begin{displaymath}
\textrm{$\varphi::= true$ $|$ $\neg\varphi$ $|$
$\varphi_{1}\wedge\varphi_{2}$ $|$
  $\npla{a}\varphi$
  $|$ $\varphi_{1}U\varphi_{2}$ $|$ $G\varphi$ $|$ $F\varphi$}
\end{displaymath}

\noindent where $a\in\Sigma$

In order to give semantics to \ALTL formulae on a \PRS $\Re$, we
need some additional notation. Given a path $\pi=t_{0}$
$\mathop{\rightarrow}\limits^{a_{0}}$ $t_{1}$
$\mathop{\rightarrow}\limits^{a_{1}}$ $t_{2}$
$\mathop{\rightarrow}\limits^{a_{2}}\ldots$ in $\Re$, $\pi^{i}$
denotes the suffix of $\pi$ starting from the $i$--th term in the
sequence, i.e. the path $t_{i}$
$\mathop{\rightarrow}\limits^{a_{i}}$ $t_{i+1}$
$\mathop{\rightarrow}\limits^{a_{i+1}}\ldots$. The set of all the
suffixes of $\pi$ is denoted by $\suffix{\pi}$ (notice that if
$\pi$ is a run in $\Re$, then $\pi^{i}$ is also a run in $\Re$,
for each $i$.)  If the path $\pi=t_{0}$
$\mathop{\rightarrow}\limits^{a_{0}}$ $t_{1}$
$\mathop{\rightarrow}\limits^{a_{1}}\ldots$ is \emph{non--trivial}
(i.e., the sequence contains at least two terms) $\firstact{\pi}$
denotes $a_{0}$, otherwise we set $\firstact{\pi}$ to an element
non in $\Sigma$.

\ALTL formulae over a \PRS $\Re$ are interpreted in terms of the
set of \PRS runs satisfying the given \ALTL formula.  The
\emph{denotation of a formula $\varphi$} relative to $\Re$, in
symbols $[[\varphi]]_{\Re}$, is  defined inductively as follows:
\begin{itemize}
\item $[[true]]_{\Re}= \prsruns{\Re}$ \item
$[[\neg\varphi]]_{\Re}=\prsruns{\Re}\setminus[[\varphi]]_{\Re}$
\item $[[\varphi_{1}\wedge\varphi_{2}]]_{\Re}=
  [[\varphi_{1}]]_{\Re}\,\cap\,[[\varphi_{2}]]_{\Re}$
\item $[[\npla{a}\varphi]]_{\Re}=\{\pi\in{}\prsruns{\Re}\ \mid\
  \firstact{\pi}=a \text{ and } \pi^{1}\in{}[[\varphi]]_{\Re}\}$
\item $[[\varphi_{1}U\varphi_{2}]]_{\Re} = \{\begin{array}[t]{l}
    \pi\in{}\prsruns{\Re}\ \mid\ \text{for some } i\geq 0,
    \pi^{i} \text{ is defined and }
 \pi^{i}\in{}[[\varphi_{2}]]_{\Re}, \text{ and }\\ %
    \text{for all } j<i, \pi^{j}\in{}[[\varphi_{1}]]_{\Re}\
    \}\end{array}$
\item $[[G\varphi]]_{\Re}=\{\pi\in{}\prsruns{\Re}\ \mid\ \suffix{\pi}
  \subseteq [[\varphi]]_{\Re}\}$
\item $[[F\varphi]]_{\Re}=\{\pi\in{}\prsruns{\Re}\ \mid\
  \suffix{\pi}\cap[[\varphi]]_{\Re}\not=\emptyset\}$

\end{itemize}

For any term $t\in{}T$ and \ALTL formula $\varphi$, we say that
$t$ satisfies $\varphi$ (resp., satisfies $\varphi$ restricted to
infinite runs) (w.r.t $\Re$), in symbols $t \models_{\Re}\varphi$
(resp., $t \models_{\Re,\infty}\varphi$), if
$\runs{\Re}{t}\subseteq[[\varphi]]_{\Re}$ (resp.,
$\runs{\Re,\infty}{t}\subseteq[[\varphi]]_{\Re}$).

The model-checking problem (resp., model--checking problem
restricted to infinite runs) for \ALTL and \PRSs is the problem of
deciding if, given a \PRS $\Re$, a \ALTL formula $\varphi$ and a
term $t$ of $\Re$, $t\models_{\Re}\varphi$ (resp.,
$t\models_{\Re,\infty}\varphi$). The following are well--known
results:

\begin{proposition}[see~\cite{mayr98}]\label{prop:parallel-altl}
  The model--checking problem for \ALTL and  parallel \PRSs, possibly
  restricted to infinite runs, is decidable.
\end{proposition}


\begin{proposition}[see~\cite{alur01,bouajjani97,mayr98}]\label{prop:sequential-altl}
The model--checking problem for \ALTL and  sequential \PRSs,
possibly restricted to infinite runs, is decidable.
\end{proposition}


\paragraph{Verification of the reachable property}\

A \emph{state property} of a \PRS $\Re$ over the alphabet $\Sigma$, is
a formula of the propositional language over the set of atomic
propositions of the form $EN(a)$ for each $a\in\Sigma$, defined as
follows:

\begin{displaymath}
\textrm{$\varphi::= EN(a)$ $|$ $\neg\varphi$ $|$
$\varphi_{1}\wedge\varphi_{2}$ $|$
$\varphi_{1}\vee\varphi_{2}$}
\end{displaymath}

\noindent where $a\in\Sigma$

The intuitive meaning of the atomic proposition $EN(a)$ is that action
$a$ is currently enabled.  The semantics of a formula in this language
is given in terms of the set of process terms satisfying the formula.
Therefore, the \emph{denotation} $[[\varphi]]_{\Re}$ of the state
formula $\varphi$ is defined as follows:

%
\begin{itemize}
\item 
  $[[EN(a)]]_{\Re}=\{t\in{}T \mid 
  r: \prsrule{t}{a}{t'}\in\Re, \text{ for some } t'\in T\}$
\item $[[\varphi_{1}\wedge\varphi_{2}]]_{\Re} =
  [[\varphi_{1}]]_{\Re}\cap[[\varphi_{2}]]_{\Re}$
\item $[[\varphi_{1}\vee\varphi_{2}]]_{\Re} =
  [[\varphi_{1}]]_{\Re}\cup[[\varphi_{2}]]_{\Re}$
\item $[[\neg\varphi]]_{\Re}=T\setminus[[\varphi]]_{\Re}$
\end{itemize}

For any term $t\in{}T$ and state formula $\varphi$, we say that $t$
satisfies $\varphi$ (w.r.t $\Re$), in symbols $t
\models_{\Re}\varphi$, if $t\in[[\varphi]]_{\Re}$.

Given a state formula $\varphi$ and a process term $t$, the
\emph{reachable state property problem in $\Re$} w.r.t $t$ and
$\varphi$ is the problem of deciding whether there exists a process
term $t'$ reachable from $t$ in $\Re$, with $t'\models\varphi$.

\begin{proposition}[see~\cite{mayr98}]\label{prop:reachprop}
  The reachable state property problem for \PRS is decidable.
\end{proposition}




\section{Verification of properties about infinite runs in \PRS.}\label{sec:mc}



Our goal is to show decidability of some problems about infinite
derivations of \PRSs in normal form, and show how decidability of
these problems can be used to check interesting properties of
infinite state systems modelled by \PRSs in normal form. For this
reason we introduce the notion  of \buchi Rewrite System (\BRS).
Intuitively, a \BRS is a \PRS where we can distinguish between
non--accepting  rules and accepting  rules.
\begin{Definition}[B\"{u}chi Rewrite System]
  A \emph{B\"{u}chi Rewrite System} $\mathrm{(BRS)}$ over
  a finite set of process variables $Var$ and an alphabet $\Sigma$ is
  a pair $\npla{\Re,\Re_{F}}$, where $\Re$ is a \PRS over $Var$
  and $\Sigma$, and
  $\Re_{F}\subseteq\Re$ is the set of accepting rules.
\end{Definition}

A B\"{u}chi Rewrite System $\npla{\Re,\Re_{F}}$ is called a \BRS
\emph{in normal form} (resp., \emph{sequential} \BRS,
\emph{parallel} \BRS), if the underlying \PRS $\Re$ is a \PRS in
normal form (resp., parallel \PRS, sequential \PRS).
\newline
\begin{Definition}[Acceptance in B\"{u}chi Rewrite Systems]\label{Def:OldAcceptance}
  Let us consider a \BRS $M=\npla{\Re,\Re_{F}}$. An \emph{infinite}
  derivation $t$ \multider{\sigma} in $\Re$ from $t$ is said to
  be \emph{accepting} $\mathrm{(}$in $M\mathrm{)}$ if
  $\sigma$ contains infinite occurrences of accepting
  rules.\newline
  A \emph{finite}
  derivation $t$ \multider{\sigma} $t'$ in $\Re$ from $t$ is said to
  be \emph{accepting} $\mathrm{(}$in $M\mathrm{)}$ if
  $\sigma$ contains some occurrence of accepting rule.
\end{Definition}

\noindent{}The main result of this paper is the following:

\noindent \emph{Given a \BRS $\npla{\Re,\Re_F}$ in normal form and
a process
  variable $X$ it is decidable whether:
\begin{description}
\item[Problem 1:] there exists an infinite accepting derivation from
  $X$;
\item[Problem 2:] there exists an infinite derivation from $X$, not
  containing occurrences of accepting rules;
\item[Problem 3:] there exists an infinite derivation from $X$,
  containing a finite non--null number of occurrences of accepting
  rules.
  \end{description}
}

Before proving this result in Section~\ref{sec:infinite}, we show
how a solution to these problems can be effectively employed to
perform model checking of some linear time properties of infinite
runs (from process variables) in \PRSs in normal form. In
particular we consider the following small \ALTL fragment
\setcounter{equation}{0}
\begin{equation}\label{Eq:OldFragmentALTL}
  \varphi ::= F\,\psi\ |\ GF\,\psi \ |\ \neg \varphi
\end{equation}
where $\psi$ denotes a \ALTL \emph{propositional}
formula\footnote{The set of  \ALTL propositional formulae $\psi$
over the set
$\Sigma$ of atomic propositions (or actions) is so defined:\\
\hspace*{4cm}$\psi ::= <$$a$$>true$$ \ | \psi \wedge \psi \ |\
\neg \psi$ (where $a\in{}\Sigma$)}. For succinctness we denote a
\ALTL propositional formula of the form $<$$a$$>true$ (with
$a\in\Sigma$) simply by $a$.
\newline

The fragment  allows us to express some useful properties on
infinite runs. Examples are simple \emph{safety properties} such
as $G\,\bigvee_{i=1}^{n}a_i$ (resp., $G\,\bigwedge_{i=1}^{n}\neg
a_i$), meaning that the system only executes (resp,.  never
executes) actions from the set $\{a_1,...,a_n\}$; \emph{guarantee
properties} such as $F\,\bigvee_{i=1}^{n}a_i$ (resp.,
$F\,\bigwedge_{i=1}^{n}\neg a_i$), meaning that the system
eventually executes [resp., does not only execute] actions from
the set $\{a_1,...,a_n\}$; \emph{response
  properties} such as $GF\bigvee_{i=1}^{n}a_i$ (resp.,
$GF\bigwedge_{i=1}^{n}\neg a_i$), meaning that the system
infinitely often executes actions from the set (resp., outside the
set) $\{a_1,...,a_n\}$; and \emph{persistence properties} such as
$GF\bigvee_{i=1}^{n}a_i$ (resp., $GF\bigwedge_{i=1}^{n}\neg a_i$),
meaning that the system executes almost always (resp., finitely
often) some actions in the set $\{a_1,...,a_n\}$.\newline

To prove the decidability of the model--checking problem
restricted to infinite runs for this fragment of \ALTL we need
some
definitions.\\
Given a propositional formula $\psi$ over $\Sigma$ we denote by
$[[\psi]]_{\Sigma}$ the subset of $\Sigma$ inductively defined as
follows
\begin{itemize}
    \item $\forall{a}\in\Sigma$  $[[a]]_{\Sigma}=\{a\}$
\item $[[\neg\psi]]_{\Sigma}=\Sigma\setminus[[\psi]]_{\Sigma}$
\item $[[\psi_{1}\wedge\psi_{2}]]_{\Sigma}=
  [[\psi_{1}]]_{\Sigma}\,\cap\,[[\psi_{2}]]_{\Sigma}$
\end{itemize}

Evidently, given a \PRS $\Re$ over $\Sigma$, a \ALTL propositional
formula $\psi$ and an infinite run $\pi$ of $\Re$ we have that
$\pi\in[[\psi]]_{\Re}$ iff $firstact(\pi)\in[[\psi]]_{\Sigma}$.\\
Given a rule $r=t$\Rule{a}$t'\in\Re$ we say that $r$
\emph{satisfies} the propositional formula $\psi$ if
$a\in[[\psi]]_{\Sigma}$. We denote by $AC(\psi)$  the set of the
rules in $\Re$ that satisfy $\psi$. \newline
 The following is a  model--checking procedure for
the fragment of \ALTL defined above with input a \PRS $\Re$ in
normal form, a temporal formula $\varphi$, and a process variable
$X$. Let us denote by  $\psi$  the propositional formula
associated to $\varphi$.

\begin{itemize}
\item Build the \BRS $\npla{\Re,\Re_F}$, where $\Re_F = AC(\psi)$;
\item Then if $\varphi$ is of the form:
\begin{description}
\item[$F\,\psi$: ]
  $X\models_{\Re,\infty}F\,\psi$ if, and only if there does not exists an
  infinite derivation  in $\npla{\Re,\Re_F}$ starting from $X$
  \emph{not} containing occurrences of accepting rules. This amounts to
  solving Problem 2.
\item[$\neg\,F\,\psi$: ]
  $X\models_{\Re,\infty}\neg\,F\,\psi$ if, and only if there does not exists
  an infinite derivation  in $\npla{\Re,\Re_F}$ starting from $X$
  containing occurrences of accepting rules. This amounts to solving
  a combination of Problem 1 and Problem 3.
\item[$GF,\psi$: ]
  $X \models_{\Re,\infty} GF\,\psi$ if, and only if there does not
  exists an infinite derivation  in $\npla{\Re,\Re_F}$ starting
  from $X$ containing a finite number of occurrences of
  accepting rules. This amounts to solving a combination of Problem 2
  and Problem 3.
\item[$\neg\,GF\,\psi$: ]
  $X \models_{\Re,\infty} \neg\,GF\,\psi$ if, and only if there does
  not exists an infinite derivation  starting from $X$ containing
  an infinite number of occurrences of accepting rules. This amounts to
  solving Problem 1.
\end{description}
\end{itemize}

 So, we obtain the following result.

\begin{Teorema}\label{Theorem:ResultOnALTL}
\noindent{}The model--checking problem for  \PRSs in normal form
and the fragment \ALTL
$\mathrm{(}$\ref{Eq:OldFragmentALTL}$\mathrm{)}$ restricted to
infinite runs from process variables is decidable.
\end{Teorema}


\section{Decidability results on infinite derivations}\label{sec:decidability}

In this section we prove the main results of the paper, namely the
decidability of the problems about infinite derivations stated in
Section~\ref{sec:mc}. Therefore, in Subsection~\ref{sec:finite} we
report some preliminary results on the decidability of some
properties about  derivations of parallel and sequential \BRSs
which are necessary to carry out the proof of the main result,
which is given in Subsection~\ref{sec:infinite}.

\subsection{Decidability results on derivations of parallel and sequential \BRSs}\label{sec:finite}

In this section we establish simple decidability results on
derivations of parallel and sequential \PRSs. These results are
the basis for the decidability proof of the problems 1-3.

\begin{proposition}\label{prop:decfinite}
  Given a parallel \BRS $\npla{\Re',\Re_{F}'}$ over $Var$ and the
  alphabet $\Sigma$, and two variables $X,Y\in Var$, it is decidable
  whether:
\begin{enumerate}
\item 
  there exists a derivation in $\Re'$ of the form
  $\prsdernorm{X}{\sigma}{\Re'}{\parcomp{t}{Y}}$ for some term $t$.
\item 
  there exists an accepting $\mathrm{(}$resp., non
  accepting$\mathrm{)}$ finite derivation in $\Re'$ of the form
  $\prsdernorm{X}{\sigma}{\Re'}{\parcomp{t}{Y}}$, for some term $t$,
  with $|\sigma|>0$.
\item 
  there exists an accepting $\mathrm{(}$resp.,
  non--accepting$\mathrm{)}$ finite derivation in $\Re'$ of the form
  $\prsdernorm{X}{\sigma}{\Re'}{\varepsilon}$.
\item 
  there exists an accepting $\mathrm{(}$resp.,
  non--accepting$\mathrm{)}$ finite derivation in $\Re'$ of the form
  $\prsdernorm{X}{\sigma}{\Re'}{Y}$.
\end{enumerate}
\end{proposition}

\begin{proof}
  Parallel \PRSs are semantically equivalent to Petri
  Nets~\cite{vanleeuwen98}. The first problem is, therefore,
  reducible to the \emph{partial reachability problem} for Petri Nets,
  which has been proved to be decidable in~\cite{mayr84,lipt76}.

  To prove decidability of the remaining problems, we exploit
  decidability of the model-checking problem for full Action--based
  LTL in parallel \PRSs (see Proposition~\ref{prop:parallel-altl}).

  Let us consider the second problem.  To show decidability of this
  problem, we start from $\Re'$ and build a new parallel \PRS $\Re''$
  over the alphabet $\overline{\Sigma} = \{f,nf,Y\}$, in the following
  way. We substitute every accepting (resp., non--accepting) rule in
  $\Re'$ of the form $\prsrule{t}{a}{t'}$, with the rule
  $\prsrule{t}{f}{t'}$ (resp., $\prsrule{t}{\emph{nf}}{t'}$).
  Finally, we add the rule $r = \prsrule{Y}{Y}{Y}$.

  The reason to add the rule $\prsrule{Y}{Y}{Y}$ is to allow us to
  express reachability of variable $Y$ as a ALTL formula.   Similarly, the
  addition of the rules of the form $\prsrule{t}{f}{t'}$
  [$\prsrule{t}{\emph{nf}}{t'}$] allows us to express in ALTL the
  application of accepting [non--accepting] rules along a run.
  The second problem is, therefore, reducible to the problem of checking
  whether there exists a run $\pi\in{}\runs{\Re''}{X}$
  satisfying the following LTL formula:
\begin{displaymath}
\varphi:=F\,\npla{f}F\,\npla{Y}true\quad\quad %
\text{[resp., }
\varphi:=\npla{\emph{nf\,}}\,\bigl(\,\npla{\emph{nf\,}}true\,
U\,\npla{Y}true\,\bigr)  \text{ ]}
\end{displaymath}

The formula $F\,\npla{f}F\,\npla{Y}true$ intuitively means that, at
least one accepting rule is eventually applied, and, after that, the
rule labelled $Y$ is eventually applied (in other words, $Y$ is
reachable after some accepting rule application).  On the other hand,
the formula $\npla{\emph{nf\,}}\,\bigl(\,\npla{\emph{nf\,}}true\,
U\,\npla{Y}true\,\bigr)$ means that, after a non-accepting rule (here
we look for derivations in $\Re'$ with length strictly greater than
$0$), $Y$ is reached by applying only non--accepting rules.
In terms of ALTL model--checking, the second problem corresponds to
checking whether, for all $\pi\in{}\runs{\Re''}{X}$,
$\pi\notin[[\varphi]]_{\Re''}$, or, in other words, to checking
whether $X\models_{\Re''}\neg\varphi$. If the result of this check is
true, the second problem has a negative answer, otherwise, the answer
is positive.

Let us now consider the third problem. Similarly to the problem above,
starting from $\Re'$, we build a new \PRS $\Re''$, this time on the
new alphabet $\overline{\Sigma}=\{f,nf\}\bigcup{}Var$, as follows.  We
substitute every accepting (resp., non--accepting) rule in $\Re'$ of
the form $\prsrule{t}{a}{t'}$ with the rule $\prsrule{t}{f}{t'}$
(resp., $\prsrule{t}{\emph{nf}\,}{t'}$).  Finally, for all $Y\in{}Var$
we add the rule $\prsrule{Y}{Y}{Y}$.  Notice that, by construction,
a term $t$ has no successor in $\Re''$ if and only if $t=\varepsilon$.
Let now $\varphi_{1}$ be the following LTL formula:
\begin{displaymath}
      \varphi_{1} = \bigvee_{Y\in Var}\Bigl(\npla{Y}true\Bigr) \vee
      \npla{\emph{nf}\,}true \vee \npla{f}true
\end{displaymath}
The negation of $\varphi_{1}$ (namely, $\neg\varphi_{1}$) means
that no rule can be applied, in other words the system has
terminated.  It is now easy to see that the problem is reducible
to the following LTL model-checking problem in $\Re''$
\begin{displaymath}
X\models_{\Re''}\neg\,F\,\bigl(\npla{f}(F\,\neg\varphi_{1})\bigr)
       \quad\quad \text{[resp., } X \models_{\Re''}
       \neg\,\bigl(\npla{\emph{nf}\,}true\,U\,\neg\,\varphi_{1}\bigr) \text{ ]}
\end{displaymath}

\noindent whose intuitive meaning is that it can never be the case
that from $X$ the system can eventually reach termination after some
application of accepting rules [resp., the system cannot reach
termination by applying only non--accepting rule].

Finally, let us consider the fourth problem. Starting from $\Re'$, we
build a new \PRS $\Re''$ over the alphabet
$\overline{\Sigma}=\{f,nf,\varepsilon\}\bigcup{}Var$, as follows.  We
substitute every accepting (resp., non--accepting) rule in $\Re'$ of
the form $t$$\mathop{{ \rightarrow}}\limits^{{a}}$$t'$ with the rule
$t$$\mathop{{ \rightarrow}}\limits^{{f}}$$t'$ (resp., $t$$\mathop{{
    \rightarrow}}\limits^{{nf}}$$t'$). For every ${}Z\in{}Var$, we
add the rule $Z\mathop{\rightarrow}\limits^{Z}Z$. Finally, we
add the rule
$Y\mathop{\rightarrow}\limits^{\varepsilon}\varepsilon$. Again, by
construction, in $\Re''$ a term $t$ has no successor if, and only
if, $t=\varepsilon$.
Let now $\varphi_{2}$ be the following LTL formula,
\begin{displaymath}
       \varphi_{2} =   \bigvee_{Y\in Var}\Bigl(\npla{Y}true\Bigr)
   \vee\npla{f}true\vee\npla{\emph{nf}\,}true\vee\npla{\varepsilon}true
\end{displaymath}
It is easy to see that the problem is now reducible to the following
LTL model-checking problem in $\Re''$
\begin{eqnarray}
       X\models_{\Re''}\neg\,\Bigl( F(\npla{f}true)\wedge\nonumber 
        \bigl((\npla{\emph{nf}\,}true\vee\npla{f}true)\,U\,
        (\npla{Y}\npla{\varepsilon}\,\neg\,\varphi_{2}) \bigr)\Bigr)
\end{eqnarray}
%
\begin{eqnarray}
      \text{[resp., } X\models_{\Re''}\neg\bigl( \npla{\emph{nf}\,}true\,U\,
      (\npla{Y}\npla{\varepsilon}\neg\,\varphi_{2})\bigr) \text{ ]} \nonumber
\end{eqnarray}

\noindent meaning that it not the case that some accepting rule is
eventually applied ($F(\npla{f}true)$), while only rules in $\Re'$
(either accepting or non--accepting rules) are applied
($(\npla{\emph{nf}\,}true\vee\npla{f}true)$) until $Y$ is
eventually reached and followed by immediate termination
($(\npla{Y}\npla{\varepsilon}\,\neg\,\varphi_{2})$) [resp., it is
not the case that $Y$ is eventually reached and followed by
immediate termination by applying only non--accepting rules].
\end{proof}

\noindent{}Let us now define an additional notion of reachability in a
sequential \PRS, and show that it is decidable whether two terms are
reachable according to this notion. As we shall see in the next
section, this decidability result will be needed to prove decidability
of the problems on infinite derivations we are interested in.

\begin{definition}\label{def:reach-seq} Given a sequential \PRS $\Re$ over $Var$, and
  variables $X,Y\in{}Var$, we say that $Y$ is \emph{reachable from $X$
    in $\Re$}, if there exists a term $t\in{}T\setminus
  \{\varepsilon\}$ of the form
  $X_{1}.(X_{2}.(\ldots{}X_{n}.(Y)\ldots))$ $\mathrm{(}$with $n$
  possibly equals to zero$\mathrm{)}$ such that
  $\prsdernorm{X}{}{\Re}{t}$.
\end{definition}

\begin{proposition}\label{prop:rech-seq} Let
  $\npla{\Re_{SEQ},\Re_{SEQ,F}}$ be a sequential \BRS over $Var$ and
  the alphabet $\Sigma$. Given any two process variables $X$ and $Y$
  in $Var$, it is decidable whether:
\begin{enumerate}
\item $Y$ is reachable from $X$ in $\Re_{SEQ}$. \item $Y$ is
reachable from $X$ in $\Re_{SEQ}$ through a non accepting
  derivation.
\end{enumerate}
\end{proposition}

\begin{proof}
  The proof relies on the decidability of the model-checking problem
  for LTL and sequential \PRSs (see
  Proposition~\ref{prop:sequential-altl}).

  First, we construct a new sequential \PRS $\Re'_{SEQ}$ over the
  alphabet $\overline{\Sigma}=\{f,nf,Y\}$, in the following way.  We
  replace every accepting (resp., not accepting) rule in $\Re_{SEQ}$
  of the form $t$$\mathop{{ \rightarrow}}\limits^{{a}}$$t'$ with the
  rule $t$$\mathop{{ \rightarrow}}\limits^{{f}}$$t'$ (resp.,
  $t$$\mathop{{ \rightarrow}}\limits^{{nf}}$$t'$). Finally, we add the
  rule $r=Y\mathop{\rightarrow}\limits^{Y}Y$.

  Now, the first problem can be restated as the problem of deciding,
  given two variables $X$ and $Y$, whether the following property is
  satisfied:
\begin{description}
\item[A.] There exists a derivation in $\Re_{SEQ}$ of the form $X$
  \multiderseq{} $t$ for some term $t\in{}T_{SEQ}\setminus
  \{\varepsilon\}$ with $t=X_{1}.(X_{2}.(\ldots{}X_{n}.(Y)\ldots))$.
\end{description}
Satisfaction of Property A can be expressed by the following LTL
satisfaction problem: Property A is satisfied if, and only if,
there exists a run $\pi\in{}\runs{\Re'_{SEQ}}{X}$
satisfying the following LTL formula:
\begin{displaymath}
  \varphi:= F(\npla{Y}true)
\end{displaymath}
Therefore, Property A is not satisfied if, and only if, for all
$\pi\in{}\runs{\Re'_{SEQ}}{X}$,
$\pi\notin[[\varphi]]_{\Re'_{SEQ}}$, that is if, and only if,
$X\models_{\Re'_{SEQ}}\neg\varphi$.

Finally, consider the second problem. This problem can be restated
as the problem of deciding, given two variables $X$ and $Y$,
whether the following property is satisfied:
\begin{description}
\item[B.] There exists a finite non--accepting derivation in
  $\Re_{SEQ}$ of the form $X$ \multiderseq{} $t$, for some term
  $t\in{}T_{SEQ}\setminus \{\varepsilon\}$, with $t=X_{1}.(X_{2}.(\ldots{}X_{n}.(Y)\ldots))$.
\end{description}

As it was the case for Property A, the satisfaction of Property B
is reducible to the following LTL satisfaction problem in
$\Re'_{SEQ}$:
\begin{displaymath}
  X\models_{\Re'_{SEQ}}\neg\,
  \bigl((\npla{\emph{nf}\,}true)\,U\,(\npla{Y}true)\bigr)
\end{displaymath}
\end{proof}

\begin{proposition}\label{prop:reducCond}
    Let us consider a sequential $\mathrm{(}$resp., parallel$\mathrm{)}$ \BRS
  $\npla{\Re',\Re'_{F}}$ over $Var$ and the
  alphabet $\Sigma$.
  Given $X\in{}Var$, it is decidable whether  the following
  condition is satisfied:
    \begin{itemize}
    \item there exists in $\Re'$ an infinite accepting derivation
      $\mathrm{(}$resp., an infinite derivation devoid of accepting rules,
      an infinite derivation containing a finite non--null number of accepting rule
      occurrences$\mathrm{)}$ from $X$.
    \end{itemize}
\end{proposition}

\begin{proof}
  The proof relies on decidability of the model-checking problem for
  LTL and sequential \PRSs (resp. parallel \PRSs) restricted to
  infinite runs (see Propositions~\ref{prop:sequential-altl} and
  \ref{prop:parallel-altl}).

  We first construct a new sequential (resp., parallel) \PRS $\Re''$
  over the alphabet $\overline{\Sigma}=\{f,nf\}$ as follows.  We
  replace every accepting (resp., non--accepting) rule in $\Re'$ of
  the form $t$$\mathop{{ \rightarrow}}\limits^{{a}}$$t'$ with the rule
  $t$$\mathop{{ \rightarrow}}\limits^{{f}}$$t'$ (resp.,
  $t$$\mathop{{\rightarrow}}\limits^{\emph{nf}\,}$$t'$).

  Let us first consider the problem of deciding whether
\begin{description}
\item[A.] There exists an accepting infinite derivation in $\Re'$
 from $X$.
\end{description}

\noindent The negation of Property A can be expressed by the following ALTL
formula

\begin{displaymath}
        \varphi:= \neg\,G\,F\,(\npla{f}true)
\end{displaymath}

\noindent Therefore, Property A is not satisfied if, and only if,
$X\models_{\Re',\infty}\varphi$.\\

Now, let us consider the problem of deciding whether
\begin{description}
\item[B.] There exists an infinite derivation in $\Re'$ from $X$
  devoid of accepting rules.
\end{description}
The negation of Property B can be expressed by the following
formula LTL
\begin{displaymath}
        \varphi:= \neg\,G\,(\npla{\emph{nf}\,}true)
\end{displaymath}

\noindent Property B is, therefore, not satisfied if, and only if,
$X\models_{\Re',\infty}\varphi$.\\

Finally, let us consider the problem of deciding whether

\begin{description}
\item[C.] There exists an infinite derivation in $\Re'$  from $X$
containing a finite non--null number of
  accepting rule occurrences.
\end{description}

\noindent The negation of Property C can be expressed by the following
formula LTL
\begin{displaymath}
        \varphi:= \neg\,F\,\bigl(\npla{f}\,(G\,\npla{\emph{nf}\,}true)\bigr)
\end{displaymath}
\noindent Again, Property C is not satisfied if, and only if,
$X\models_{\Re',\infty}\varphi$.
\end{proof}

\begin{Teorema}\label{theo:cond}
Let us consider a sequential \BRS
  $\npla{\Re_{SEQ},\Re_{SEQ,F}}$ and a parallel \BRS $\npla{\Re_{PAR},\Re_{PAR,F}}$
  over $Var$ and the
  alphabet $\Sigma$.
  Given $X\in{}Var$, it is decidable whether one of the following
  conditions is satisfied:
    \begin{itemize}
    \item there exists a variable $Y\in{}Var$ reachable $\mathrm{(}$resp., reachable
      through a non--accepting derivation, reachable$\mathrm{)}$ from $X$ in
      $\Re_{SEQ}$, and there exists in $\Re_{PAR}$ an infinite
      accepting derivation $\mathrm{(}$resp,. an infinite derivation devoid of
      accepting rule occurrences, an infinite derivation containing a finite non--null number of
      accepting rule occurrences$\mathrm{)}$ from $Y$.
    \item there exists in $\Re_{SEQ}$ an infinite accepting derivation
      $\mathrm{(}$resp., an infinite derivation devoid of accepting rule occurrences,
      an infinite derivation containing a finite non--null number of accepting rule
      occurrences$\mathrm{)}$ from $X$.
    \end{itemize}
\end{Teorema}

\begin{proof}
  The result follows directly from Propositions~\ref{prop:rech-seq}
  and~\ref{prop:reducCond}
\end{proof}

\subsection{Decidability of properties about infinite derivations}
\label{sec:infinite}

To prove decidability of Problems 1--3 stated in the previous section,
we show that each of those problems can be reduced to (a
combination of) two similar, but simpler, problems: the first is a
decidability problem on infinite derivations restricted to
parallel \BRSs; the second is a  decidability problem on infinite
derivations restricted to sequential \BRSs. Since each of those
restricted problems is  decidable (see theorem \ref{theo:cond}),
decidability of Problems 1--3 is entailed.

In particular, we show that, given a \BRS $\npla{\Re,\Re_{F}}$ in
normal form over $Var$ and the alphabet $\Sigma$, it is possible
to effectively construct two \BRSs, a parallel \BRS
$\npla{\Re_{PAR},\Re_{PAR,F}}$ and a sequential \BRS
$\npla{\Re_{SEQ},\Re_{SEQ,F}}$, in such a way that:
\begin{enumerate}
\item Problem 1 is reducible to the problem of deciding, given a
  process variable $X$, if one of the following conditions is
  satisfied:
    \begin{itemize}
    \item There exists a variable $Y\in{}Var$ reachable from $X$ in
      $\Re_{SEQ}$, and there exists an infinite accepting derivation
      in $\Re_{PAR}$ from $Y$.
    \item There exists an infinite accepting derivation in $\Re_{SEQ}$
      from $X$.
    \end{itemize}
  \item Problem 2 is reducible to the problem of deciding, given a
    process variable $X$, if one of the following conditions is
    satisfied:
    \begin{itemize}
    \item There exists a variable $Y\in{}Var$ reachable from $X$ in
      $\Re_{SEQ}$ through a non--accepting derivation, and there exists
      an infinite derivation in $\Re_{PAR}$ from $Y$ not containing
      accepting rule occurrences.
    \item There exists an infinite derivation in $\Re_{SEQ}$ from $X$
      not containing accepting rule occurrences.
    \end{itemize}
  \item Problem 3 is reducible to the problem of deciding, given a
    process variable $X$, if one of the following conditions is
    satisfied:
    \begin{itemize}
    \item There exists a variable $Y\in{}Var$ reachable from $X$ in
      $\Re_{SEQ}$, and there exists an infinite derivation in
      $\Re_{PAR}$ from $Y$ containing a finite non--null number
      of accepting rule occurrences.
    \item There exists an infinite derivation in $\Re_{SEQ}$ from $X$
      containing a finite non--null number of accepting rule occurrence.
    \end{itemize}
\end{enumerate}

\setcounter{equation}{0} 
%
In the following, $\Re^{P}$ (resp., $\Re^{P}_{F}$) denotes the set
$\Re$ (resp., the set $\Re_{F}$) restricted to PAR rules.

Before illustrating the main idea underlying our approach, we need few
additional definitions and notation, which allows us to look more in
detail at the structure of derivations.  The following definition
introduces the notion of level of application of a rule in a
derivation:

\begin{definition}
  Let $\prssimpder{t}{r}{t'}$ be a single--step derivation in $\Re$.
  We say that $r$ is \emph{applied at level 0} in
  $\prssimpder{t}{r}{t'}$, if $t = \parcomp{\overline{t}}{s}$,
  $t'=\parcomp{\overline{t}}{s'}$ $\mathrm{(}$for some $\overline{t}, s, s'\in
  T$$\mathbb{)}$, and $r = \prsrule{s}{a}{s'}$, for some $a\in\Sigma$.

  We say that $r$ is \emph{applied at level $k > 0$ in}
  $\prssimpder{t}{r}{t'}$, if $t = \parcomp{\overline{t}}{X.(s)}$, $t'
  = \parcomp{\overline{t}}{X.(s')}$ $\mathrm{(}$for some
  $\overline{t}, s,s'\in T$$\mathbb{)}$, $\prssimpder{s}{r}{s'}$, and
  $r$ is applied at level $k-1$ in $\prssimpder{s}{r}{s'}$.
\end{definition}

The definition above extends in the obvious way to $n$--step
derivations and to infinite derivations. The next definition
introduces the notion of subderivation starting from a term.

\begin{Definition}[Subderivation]
  Let $\overline{t}$ \multider{\lambda}
  $t$$\parallel$$X.(s)$ \multider{\sigma} be a finite or infinite
  derivation in $\Re$ starting from $\overline{t}$. The
  \emph{subderivation $d'=s$ \multider{\mu} of
    $d=t$$\parallel$$X.(s)$ \multider{\sigma} from $s$} is inductively
  defined as follows:

\begin{enumerate}
\item if $d$ is the null derivation or $s = \varepsilon$, then
$d'$ is
  the null derivation;
\item if $\sigma = r\sigma'$, and $d$ is of the form
\begin{displaymath}
\textrm{
     $t$$\parallel$$X.(Z)$ \prsoneder{r} $t$$\parallel$$Y$
    \multider{\sigma'}  $\quad\mathrm{(}$with $r =X.(Z)$\Rule{a}$Y$$\mathrm{)}$}
\end{displaymath}
\noindent  then $d'$ is the null derivation. \item if $\sigma =
r\sigma'$, and $d$ is of the form
\begin{displaymath}
\textrm{
     $t$$\parallel$$X.(s)$ \prsoneder{r} $t$$\parallel$$X.(s')$
    \multider{\sigma'}   $\quad\mathrm{(}$with $s$ \prsoneder{r} $s'$$\mathrm{)}$}
\end{displaymath}
\noindent then $d' = s$ \prsoneder{r} $s'$ \multider{\mu'}, with
$s'$ \multider{\mu'} the subderivation of $t$$\parallel$$X.(s')$
\multider{\sigma'} from $s'$;
 \item if
$\sigma = r\sigma'$, and $d$ is of the form
\begin{displaymath}
\textrm{
     $t$$\parallel$$X.(s)$ \prsoneder{r} $t'$$\parallel$$X.(s)$
    \multider{\sigma'}   $\quad\mathrm{(}$with $t$ \prsoneder{r} $t'$$\mathrm{)}$}
\end{displaymath}
\noindent then $d'$ is the subderivation of $t'$$\parallel$$X.(s)$
\multider{\sigma'} from $s$;
\end{enumerate}
Moreover, we say that $d'$ is a subderivation of $\overline{t}$
\multider{\lambda}
  $t$$\parallel$$X.(s)$ \multider{\sigma}.
\end{Definition}

Clearly, in the definition above $\mu$ is a subsequence of
$\sigma$. Moreover, if $k$ is the level of application of a rule
occurrence of $\mu$ in the derivation $d$ then, $k>0$, and this
occurrence is applied in the subderivation $d'=s$ \multider{\mu}
at level $k-1$.\newline
 Moreover, we say that a subderivation
$\prsder{s}{\rho}{}$ of $\prsder{X}{\sigma}{}$ from $s$ is a
\emph{maximal subderivation} in $\prsder{X}{\sigma}{}$, if there
is no subderivation $\prsder{\overline{s}}{\overline{\rho}}{}$ of
$\prsder{X}{\sigma}{}$ from $\overline{s}$, with $\rho$ a proper
subsequence of $\overline{\rho}$.

Given a sequence $\sigma=r_{1}r_{2} \ldots{}r_{n} \ldots{}$ of
rules in $\Re$, and a subsequence $\sigma'=r_{k_{1}}r_{k_{2}}
\ldots{} r_{k_{m}} \ldots{}$ of $\sigma$, $\sigma\setminus
\sigma{}'$ denotes the sequence obtained by removing from $\sigma$
all and only the occurrences of rules in $\sigma'$ (namely, those
$r_{i}$ for which it exists a $j=1,\ldots,|\sigma'|$, with
$k_{j}=i$).

 \begin{definition}
   The class $\Xi_{PAR}$ $\mathrm{(}$resp.
   $\Pi_{PAR}$$\mathrm{)}$ is
   the class of derivations $\prsder{t}{\sigma}{}$ in $\Re$ \emph{\textbf{not}}
   satisfying the following property:
 \begin{itemize}
 \item the derivation $\prsder{t}{\sigma}{}$ can be written in the
   form $\prsder{t}{\sigma}{\prsder{\parcomp{t'}{X.(s)}}{\rho}{}}$
   such that the subderivation of
   $\prsder{\parcomp{t'}{X.(s)}}{\rho}{}$ from $s$ is an infinite
   derivation $\mathrm{(}$resp., an accepting infinite
   derivation$\mathrm{)}$.
 \end{itemize}
\end{definition}




Let us sketch the main ideas at the basis of our technique. To fix
the ideas, let us consider the problem 1. Moreover, let us focus
first on the class of derivations $\Pi_{PAR}$, showing how it is
possible to mimic accepting infinite derivations in $\Re$ from a
variable, belonging to this class, by using only PAR rules
belonging to an extension of the parallel \BRS
$\npla{\Re^{P},\Re^{P}_{F}}$ denoted by
$\npla{\Re_{PAR},\Re_{PAR,F}}$. More precisely, we show that
\begin{description}
    \item[A.]  if $p$ \multider{\sigma} (with
$p\in{}T_{PAR}$) is an infinite accepting derivation in
$\Pi_{PAR}$ then, there exists an infinite accepting derivation in
$\npla{\Re_{PAR},\Re_{PAR,F}}$ from $p$, and vice versa.
\end{description}

With reference to Problems 2 (resp., 3), within
$\npla{\Re_{PAR},\Re_{PAR,F}}$ it will also be possible to
simulate an infinite derivation in $\Re$ from $p\in{}T_{PAR}$
belonging to $\Xi_{PAR}$, and not containing accepting rule
occurrences (resp., containing a finite non--null number of
accepting rule occurrences), through an infinite derivation in
$\Re_{PAR}$ from $p$, not containing accepting rule occurrences
(resp., containing a finite non--null number of accepting rule
occurrences), and vice versa.

Suppose now that the accepting infinite derivation $p$
\multider{\sigma} belongs to $\Pi_{PAR}$. Then, all its possible
subderivations contain all, and only, the rule occurrences in
$\sigma$ applied at a level $k$ greater than $0$ in $p$
\multider{\sigma}. If $\sigma$ contains only PAR rule occurrences
the statement A is evident since $\npla{\Re_{PAR},\Re_{PAR,F}}$ is
an extension of $\npla{\Re^{P},\Re^{P}_{F}}$. Otherwise,
$\prsder{p}{\sigma}{}$ can be written in the form:
 \begin{equation}\label{eq:derivation}
 \prsder{p}{\lambda}{\prssimpder{\parcomp{t}{Z'}}{r}
   {\prsder{\parcomp{t}{Y.(Z)}}{\omega}{}}}
 \end{equation}
 where $r = \prsrule{Z'}{a}{Y.(Z)}$,  $\lambda$ contains only occurrences of rules
 in $\Re_{P}$, and $t\in{}T_{PAR}$. Let $\prsder{Z}{\rho}{}$ be  the subderivation of
 $\prsder{\parcomp{t}{Y.(Z)}}{\omega}{}$ from $Z$.

Since $p$ \multider{\sigma} is in $\Pi_{PAR}$,
$\prsder{Z}{\rho}{}$ does not contain infinite occurrences of
accepting rules. Thus,
 only one of the following three cases may occur:
\begin{description}
\item[A] $\prsder{Z}{\rho}{}$ leads to the term $\varepsilon$, and
  $\prsder{p}{\sigma}{}$ is of the form
  \begin{equation}\label{eq:form1}
  \prsder{p}{\lambda}
     {\prssimpder{\parcomp{t}{Z'}}{r}
          {\prsder{\parcomp{t}{Y.(Z)}}{\omega_{1}}
             {\prsder{\parcomp{\overline{t}}{Y}}{\omega_{2}}{}}}}
  \end{equation}
  where $\rho$ is a subsequence of $\omega_1$ and $t$
  \multider{\omega_{1}/\rho} $\overline{t}$.  The infinite derivation above is
  accepting if, and only if, the following infinite derivation,
  obtained by anticipating (by interleaving) the application of the
  rules in $\rho$ before the application of the rules in $\xi=
  \omega_1\setminus\rho$, is accepting
  \begin{equation}\label{eq:form1-2}
  \prsder{p}{\lambda}
     {\prssimpder{\parcomp{t}{Z'}}{r}
          {\prsder{\parcomp{t}{Y.(Z)}}{\rho}
             {\prsder{\parcomp{t}{Y}}{\xi}
             {\prsder{\parcomp{\overline{t}}{Y}}{\omega_{2}}{}}}}}
  \end{equation}
  where $\prsder{t}{\xi}{\overline{t}}$. 

  The idea is to collapse the derivation
  $\prssimpder{Z'}{r}{\prsder{Y.(Z)}{\rho}{Y}}$ into a
  single accepting PAR rule of the form $\dollarrule{Z'}{Y}$, if
  $r\rho$ is accepting, or into a non--accepting PAR rule of the form
  $\cancrule{Z'}{Y}$, if $r\rho$ is non--accepting.

  Notice that in the step from (\ref{eq:form1}) to (\ref{eq:form1-2}),
  we exploit the fact that the properties on infinite derivations we
  are interested in are insensitive to permutation
  of rule applications within a derivation.

  Now, we can apply recursively the same reasoning to the infinite
  accepting derivation in $\Re$ from $t$$\parallel$$Y\in{}T_{PAR}$
  \begin{equation}
    \textrm{$t$$\parallel$$Y$ \multider{\xi} $\overline{t}$$\parallel$$Y$
      \multider{\omega_{2}}}
  \end{equation}
  which belongs to $\Pi_{PAR}$.
\item[B] The subderivation $\prsder{Z}{\rho}{}$ leads to a variable
  $W$ and $\prsder{p}{\sigma}{}$ can be written as:
  \begin{equation}\label{eq:form2}
  \prsder{p}{\lambda}
     {\prssimpder{\parcomp{t}{Z'}}{r}
          {\prsder{\parcomp{t}{Y.(Z)}}{\omega_{1}}
             {\prssimpder{\parcomp{\overline{t}}{Y.(W)}}{r'}
                 {\prsder{\parcomp{\overline{t}}{W'}}{\omega_{2}}{}}}}}
  \end{equation}

  where $r' = \prsrule{Y.(W)}{b}{W'}$ (with $W'\in{}Var$),
   $\rho$ is a subsequence of $\omega_1$ and $t$ \multider{\omega_{1}/\rho} $\overline{t}$.

  The derivation above is accepting if and only if the following
  derivation
  is accepting
  \begin{equation}\label{eq:form2-1}
  \prsder{p}{\lambda}
     {\prssimpder{\parcomp{t}{Z'}}{r}
          {\prsder{\parcomp{t}{Y.(Z)}}{\rho}
             {\prsder{\parcomp{t}{Y.(W)}}{r'}
               {\prsder{\parcomp{t}{W'}}{\xi}
                 {\prsder{\parcomp{\overline{t}}{W'}}{\omega_{2}}{} }}}}}
  \end{equation}

with $\xi=\omega_{1}/\rho$.

  In this case we shall collapse the derivation
  $\prssimpder{Z'}{r}{\prsder{Y.(Z)}{\rho}{\prssimpder{Y.(W)}{r'}{W'}}}$
  into a single accepting PAR rule of the form $\dollarrule{Z'}{W'}$,
  if $r\rho r'$ is accepting, or into a non--accepting PAR rule of the
  form $\cancrule{Z'}{W'}$, otherwise.

  Now, we can apply recursively the same reasoning to the infinite
  accepting derivation in $\Re$ from $t$$\parallel$$W'\in{}T_{PAR}$
  \begin{equation}
    \textrm{$t$$\parallel$$W'$ \multider{\xi} $\overline{t}$$\parallel$$W'$
      \multider{\omega_{2}}}
  \end{equation}
  which belongs to $\Pi_{PAR}$.
\item[C] In this case $\prsder{Z}{\rho}{}$ does not influence the
  applicability of rules in $\omega \setminus \rho$ in the derivation $
  {\prsder{\parcomp{t}{Y.(Z)}}{\omega}{}}$
  (i.e. the rule applications occurring in $\rho$ can be arbitrarily
  interleaved with any rule application in $\omega/\rho$).  In other
  terms, we have $\prsder{t}{\omega\setminus\rho}{}$ that is still an
  infinite accepting derivation in $\Pi_{PAR}$. On the other hand, if
  $\prsder{Z}{\rho}{}$ contains some occurrence of accepting rule or
  $r$ is an accepting rule, we cannot abstract this information away.
  In fact, it might be the case that the infinite number of accepting
  rules occurring in $\prsder{p}{\sigma}{}$ is due to an infinite
  number of occurrences of subderivations like $\prsder{Z}{\rho}{}$.
  For this reason, we keep track of a possible occurrence of accepting
  rule in the sequence $r\rho$ by adding a new variable $Z_{ACC}$ and
  an accepting rule of the form $\prsrule{Z'}{}{Z_{ACC}}$.

 \end{description}

 In other words, we are going to build a parallel \BRS
 $\npla{\Re_{PAR},\Re_{PAR,F}}$ where all the maximal finite accepting
 [resp., non--accepting] subderivations and the maximal infinite
 subderivations containing a finite number of accepting rule
 occurrences are abstracted away by PAR rules not occurring in $\Re$,
 according to the intuitions given above.

 The extended \BRS $\npla{\Re_{PAR},\Re_{PAR,F}}$ is constructed in
 two steps.  In the first step, $\npla{\Re^{P},\Re^{P}_{F}}$ is
 extended with PAR rules of the form $\prsrule{X}{a}{Y}$, where
 $X,Y\in{}Var$, and $a\in\{\$,\#\}$, in such a way that it is possible
 to keep track of subderivations of the forms \textbf{A} and
 \textbf{B}. In the second step, rules of the form
 $\prsrule{X}{}{Z_{ACC}}$ or of the form $\prsrule{X}{}{Z_{NOT\_ACC}}$
 (with $X\in{}Var$) are added, so as to be able to keep track of
 derivations of the form \textbf{C}.\\
 The first extension of $\npla{\Re^{P},\Re^{P}_{F}}$, denoted by $\npla{\overline{\Re}_{PAR},
   \overline{\Re}_{PAR,F}}$, is defined as follows.

\begin{definition}\label{def:parallel1}
  The \BRS $\npla{\overline{\Re}_{PAR}, \overline{\Re}_{PAR,F}}$ is
  the least parallel \BRS over $Var$ and the alphabet $\Sigma' = \Sigma
  \bigcup \{\$,\#\}$ satisfying the following properties:

\begin{enumerate}
\item $\overline{\Re}_{PAR}\supseteq \Re^{P}$, and
  $\overline{\Re}_{PAR,F}\supseteq\Re^{P}_{F}$;
\item $\cancrule{X}{Y} \in{} \overline{\Re}_{PAR}$, if there is a rule
  $\prsrule{X}{a}{Y.(Z)}\in{}\Re\setminus\Re_{F}$, and
  $\prsdernorm{Z}{\sigma}{\overline{\Re}_{PAR}}{\varepsilon}$ with
  $\sigma$ non--accepting in $\overline{\Re}_{PAR}$;
\item $\dollarrule{X}{Y} \in{} \overline{\Re}_{PAR,F}$, if there is a
  rule $r = \prsrule{X}{a}{Y.(Z)}\in{}\Re$, and
  $\prsdernorm{Z}{\sigma}{\overline{\Re}_{PAR}}{\varepsilon}$ and,
  either $\sigma$ is accepting in $\overline{\Re}_{PAR}$, or
  $r\in\Re_F$;
\item $\cancrule{X}{W'}\in\overline{\Re}_{PAR}$, if there are rules
  $\prsrule{X}{a}{Y.(Z)}\in\Re\setminus\Re_F$ and
  $\prsrule{Y.(W)}{b}{W'}\in\Re\setminus\Re_F$, and
  $\prsdernorm{Z}{\sigma}{\overline{\Re}_{PAR}}{W}$, with
  $\sigma$ non--accepting;
\item $\dollarrule{X}{W'}\in\overline{\Re}_{PAR,F}$, if there are
  rules $r = \prsrule{X}{a}{Y.(Z)}\in\Re$ and $r' =
  \prsrule{Y.(W)}{b}{W'}\in\Re$, and
  $\prsdernorm{Z}{\sigma}{\overline{\Re}_{PAR}}{W}$ and, either
  $\sigma$ is accepting in $\overline{\Re}_{PAR}$, or $r\in\Re_F$, or
  $r'\in\Re_F$.
\end{enumerate}
\end{definition}

\begin{Lemma}\label{lemma:algo}
  The parallel \BRS $\npla{\overline{\Re}_{PAR},
    \overline{\Re}_{PAR,F}}$ can be effectively constructed.
\end{Lemma}

\begin{proof}
  Figure~\ref{fig:algo} reports 
  the procedure {\sc Build--parallel--BRS($\npla{\Re,\Re_F}$)}, which, starting
  from  $\npla{\Re,\Re_F}$, builds a parallel \BRS
  $\npla{\Re_{PAR,AUX},\Re_{PAR,AUX,F}}$.  The algorithm {\sc
    Build--parallel--BRS($\npla{\Re,\Re_F})$} employs three auxiliary sets of
  rules $\overline{\Re}$, $\overline{\Re}_{F}$ and $RuleSEQ$, and a
  flag.

  From Proposition~\ref{prop:decfinite} follows that the conditions
  checked in each of the \textbf{if} statements in lines 9, 16, 22 and
  29 are decidable, therefore, the procedure is  effective.

\begin{figure}[htbp]
\ \\\noindent \textbf{Algorithm}
  B{\small{}UILD}--{\small{}PARALLEL}--BRS($\npla{\Re,\Re_F}$)\vspace{4pt}
\\\footnotesize
1 $\quad\Re_{PAR,AUX}:=\{r\in\Re \mid r \emph{ is a PAR rule}\}$;\\
2 $\quad\Re_{PAR,AUX,F}:=\{r\in\Re_{F} \mid r \emph{ is a PAR rule}\}$;\\
3 $\quad\overline{\Re}:=\overline{\Re}_{F}:=\emptyset$;\\
4 $\quad\emph{RuleSEQ}:=\{\prsrule{X}{a}{Y.(Z)}\in\Re\}$;\\
5 $\quad$\textbf{repeat}\\
6 $\quad\quad\quad{}$\emph{flag:=false;}\\
7 $\quad\quad\quad{}$\textbf{while} \emph{RuleSEQ}$\neq\emptyset$
   \textbf{do}\\
8 $\quad\quad\quad\quad\quad{}$\emph{extract a rule}
       $\prsrule{X}{a}{Y.(Z)}$ \emph{from RuleSEQ};\\
9 $\quad\quad\quad\quad\quad{}$\textbf{if}
           $\prsderparaux{Z}{\sigma}{\varepsilon}, \textbf{ and }
           \textbf{(}
           \emph{$\sigma$ is accepting }
           \textbf{ or } \prsrule{X}{a}{Y.(Z)}\in{}\Re_{F}\textbf{)
           }$ \textbf{then}\\
10 $\quad\quad\quad\quad\quad\quad\quad{}$\textbf{if}
$\dollarrule{X}{Y}\notin{}\overline{\Re}$ \textbf{then}\\
11
$\quad\quad\quad\quad\quad\quad\quad\quad\quad{}$$\Re_{PAR,AUX}:=\Re_{PAR,AUX}\
\bigcup\ \{\dollarrule{X}{Y}\}$;\\
12 $\quad\quad\quad\quad\quad\quad\quad\quad\quad{}$
      $\overline{\Re} :=           \overline{\Re}\ \bigcup\ \{\dollarrule{X}{Y}\}$; \\
13 $\quad\quad\quad\quad\quad\quad\quad\quad\quad{}$
          $\Re_{PAR,AUX,F} := \Re_{PAR,AUX,F}\ \bigcup\
           \{\dollarrule{X}{Y}\}$; \\
14 $\quad\quad\quad\quad\quad\quad\quad\quad\quad{}$
   $\overline{\Re}_{F} :=   \overline{\Re}_{F}\ \bigcup\ \{\dollarrule{X}{Y}\}$; \\
15 $\quad\quad\quad\quad\quad\quad\quad\quad\quad{}$
\emph{flag:=true};\\
16 $\quad\quad\quad\quad\quad{}$\textbf{if}
           $\prsrule{X}{a}{Y.(Z)}\notin{}\Re_{F} \textbf{ and }
           \prsderparaux{Z}{\sigma}{\varepsilon} \textbf{ and }
           \sigma \emph{ is not accepting }$
           \textbf{then}\\
17 $\quad\quad\quad\quad\quad\quad\quad{}$\textbf{if}
$\cancrule{X}{Y}\notin{}\overline{\Re}$ \textbf{then}\\
18
$\quad\quad\quad\quad\quad\quad\quad\quad\quad{}$$\Re_{PAR,AUX}:=\Re_{PAR,AUX}\
\bigcup\ \{\cancrule{X}{Y}\}$;\\
19 $\quad\quad\quad\quad\quad\quad\quad\quad\quad{}$
$\overline{\Re}:=\overline{\Re}\ \bigcup\ \{\cancrule{X}{Y}\}$;
           \\
20
$\quad\quad\quad\quad\quad\quad\quad\quad\quad{}$\emph{flag:=true;}\\
21 $\quad\quad\quad\quad\quad{}$\textbf{for each}
    $\prsrule{Y.(W)}{b}{W'}\in\Re$ \textbf{do}\\
22 $\quad\quad\quad\quad\quad\quad\quad{}$\textbf{if} $\begin{array}[t]{l} %
              \prsderparaux{Z}{\sigma}{W} \textbf{ and }
              \textbf{(}\sigma
              \emph{ is accepting } \textbf{ or }
              \prsrule{X}{a}{Y.(Z)}\in{}\Re_{F} \textbf{ or } \\
              \hspace*{3.7cm}\prsrule{Y.(W)}{b}{W'}\in{}\Re_{F}\textbf{)}
              \ \ \ \ \ \textbf{ then } \end{array}$ \\
23 $\quad\quad\quad\quad\quad\quad\quad\quad\quad{}$\textbf{if}
$\dollarrule{X}{W'}\notin{}\overline{\Re}$ \textbf{then}\\
24
$\quad\quad\quad\quad\quad\quad\quad\quad\quad\quad\quad{}$$\Re_{PAR,AUX}:=\Re_{PAR,AUX}\
\bigcup\ \{\dollarrule{X}{W'}\}$;\\
25 $\quad\quad\quad\quad\quad\quad\quad\quad\quad\quad\quad{}$
      $\overline{\Re} :=           \overline{\Re}\ \bigcup\ \{\dollarrule{X}{W'}\}$; \\
26 $\quad\quad\quad\quad\quad\quad\quad\quad\quad\quad\quad{}$
          $\Re_{PAR,AUX,F} := \Re_{PAR,AUX,F}\ \bigcup\
           \{\dollarrule{X}{W'}\}$; \\
27 $\quad\quad\quad\quad\quad\quad\quad\quad\quad\quad\quad{}$
   $\overline{\Re}_{F} :=   \overline{\Re}_{F}\ \bigcup\ \{\dollarrule{X}{W'}\}$; \\
28 $\quad\quad\quad\quad\quad\quad\quad\quad\quad\quad\quad{}$
     \emph{flag:=true};\\
29 $\quad\quad\quad\quad\quad\quad\quad{}$\textbf{if}
    $\begin{array}[t]{l}
         \prsrule{X}{a}{Y.(Z)}\notin{}\Re_{F} \textbf{ and }
         \prsrule{Y.(W)}{b}{W'}\notin{}\Re_{F} \textbf{ and } \\
         \prsderparaux{Z}{\sigma}{W}\ \textbf{and}\
         \sigma
             \emph{ is not accepting}\ \ \ \ \ \ \textbf{ then } \end{array}$\\
30 $\quad\quad\quad\quad\quad\quad\quad\quad\quad{}$\textbf{if}
    $\cancrule{X}{W'}\notin{}\overline{\Re}$ \textbf{then}\\
31
$\quad\quad\quad\quad\quad\quad\quad\quad\quad\quad\quad{}$$\Re_{PAR,AUX}:=\Re_{PAR,AUX}\
\bigcup\ \{\cancrule{X}{W'}\}$;\\
32 $\quad\quad\quad\quad\quad\quad\quad\quad\quad\quad\quad{}$
$\overline{\Re}:=\overline{\Re}\ \bigcup\ \{\cancrule{X}{W'}\}$;
           \\
33
$\quad\quad\quad\quad\quad\quad\quad\quad\quad\quad\quad{}$\emph{flag:=true;}\\
34 $\quad\quad\quad\quad\quad{}$\textbf{done} {$\ \
     \displaystyle\triangleright$~} for\\
35 $\quad\quad\quad{}$\textbf{done}  {$\ \
\displaystyle\triangleright$~} while\\
36 $\quad\quad\quad{}$\emph{RuleSEQ}$:=\{\prsrule{X}{a}{Y.(Z)}\in\Re\};$\\
37 $\quad{}$\textbf{until} $flag=false$\\
 \caption{Algorithm to turn a \BRS
in normal form into a parallel \BRS.}\label{fig:algo}
\end{figure}

Let us show that the algorithm terminates. To see this, it
suffices to prove that the number of iterations of the
\textbf{repeat} loop is finite. Recall that the termination
condition of this loop is \emph{flag} = \emph{false}. At the
beginning of every iteration the \emph{flag} is set to
\emph{false}.  Moreover \emph{flag} is reset to \emph{true} either
when a new rule of the form $\dollarrule{X}{Y}$ is added to
$\overline{\Re}$ (lines 10--15, or lines 23--28), or when a new
rule of the form $\cancrule{X}{Y}$ is added to $\overline{\Re}$
(lines 17--20, or lines 30--33). Since the set of rules of the
form $\dollarrule{X}{Y}$ or $\cancrule{X}{Y}$ (with $X$,$Y$ in
$Var$) is finite (being $Var$ finite), termination immediately
follows.

We now prove that $\npla{\Re_{PAR,AUX},\Re_{PAR,AUX,F}}$ is a
parallel \BRS satisfying the properties of
Definition~\ref{def:parallel1}.
The following properties are clearly satisfied:
\begin{description}
\item[a)] After initialization (lines 1--4)
  $\Re_{PAR,AUX}=\overline{\Re}\ \cup\ \{r\in\Re \mid r \text{ is a PAR
    rule}\}$ (where for simplicity we consider the lines 11--12,
  18--19, 24--25, 31--32 a single atomic instruction).

  Moreover, $\Re_{PAR,AUX,F}$ $=\overline{\Re}_{F}\ \cup\
  \{r\in\Re_{F} \mid r \text{ is a PAR rule}\}$ (where for simplicity
  we consider the lines 13--14, 26--27 a single atomic instruction).
\item[b)] $\overline{\Re}_{F}\subseteq\overline{\Re}$.
\end{description}

With $\overline{\Re}_{PAR}$ (resp., $\overline{\Re}_{PAR}$) we denote
the set $\Re_{PAR,AUX}$ (resp. $\Re_{PAR,AUX,F}$) at termination of
the algorithm. We show that
$\npla{\overline{\Re}_{PAR},\overline{\Re}_{PAR,F}}$ satisfies
Properties 1-5 of Definition~\ref{def:parallel1}. Property 1 is
clearly satisfied as a consequence of lines 1 and 2 of the algorithm.

Let us prove Property 2.  Let $\prsrule{X}{a}{Y.(Z)}\in\Re\setminus
\Re_{F}$ and
$\prsdernorm{Z}{\sigma}{\overline{\Re}_{PAR}}{\varepsilon}$, with
$\sigma$ non accepting (i.e., not containing occurrences of rules in
$\overline{\Re}_{PAR,F}$).  We have to show that
$\cancrule{X}{Y}\in\overline{\Re}_{PAR}$.  In particular, we show that
$\cancrule{X}{Y}\in\overline{\Re}$.  Let us consider the last
iteration of the \textbf{repeat} loop.
Since any update of the sets
$\Re_{PAR,AUX}\overline{},\overline{\Re},\Re_{PAR,AUX,F},$
$\overline{\Re}_{F}$ (the \emph{flag} is set \emph{true}) involves
a new iteration of this loop, it follows that at this step
$\Re_{PAR,AUX}=\overline{\Re}_{PAR}$,
$\Re_{PAR,AUX,F}=\overline{\Re}_{PAR,F}$, and they will not be
updated anymore.  Now the rule $\prsrule{X}{a}{Y.(Z)}$ is examined
during an iteration of the inner \textbf{while} loop.  During this
iteration, since $\Re_{PAR,AUX}=\overline{\Re}_{PAR}$ and
$\Re_{PAR,AUX,F}=\overline{\Re}_{PAR,F}$, 
the condition of the \textbf{if} statement in line 16 must be
satisfied.  On the other hand, since $\Re_{PAR,AUX}$ and
$\overline{\Re}$ cannot be update anymore,
the condition of the \textbf{if} statement in line 17 cannot be
satisfied.  Therefore, $\cancrule{X}{Y}\in\overline{\Re}$, and
Property 2 is proved. Following a similar reasoning, we can easily
prove that also Properties 3--5 are satisfied.

Finally, it is easy to see that $\npla{\overline{\Re}_{PAR},
  \overline{\Re}_{PAR,F}}$ is the least parallel \BRS over $Var$
and $\Sigma'$ satisfying Properties 1-5 of
Definition~\ref{def:parallel1}.
\end{proof}

Now, let us consider the parallel \BRS
$\npla{\overline{\Re}_{PAR},\overline{\Re}_{PAR,F}}$ computed by
the algorithm of Lemma 5.1. As anticipated, in order to simulate
subderivations of the form \textbf{C}, we need to add additional
PAR rules in $\npla{\overline{\Re}_{PAR},\overline{\Re}_{PAR,F}}$.
We need of the following decidability result.

\begin{proposition}\label{prop:finiteacc}
  Given a \BRS $\npla{\Re,\Re_{F}}$ in normal form, and a variable
  $X\in{}Var$, it is decidable whether there exists a finite accepting
  derivation in $\Re$ from $X$.
\end{proposition}

\begin{proof}
  We show that the problem is reducible to the reachable property
  problem for \PRSs, which is decidable (see
  Proposition~\ref{prop:reachprop}).

  Starting from $\Re$, we build a new \PRS $\Re'$ in the following
  way.  Consider the alphabet $\overline{\Sigma} = \{f,\emph{nf}\}$,
  containing only two symbols.  Substitute every accepting (resp., not
  accepting) rule in $\Re$ of the form $\prsrule{t}{a}{t'}$ with the
  rule $\prsrule{t}{f}{t'}$ (resp., $\prsrule{t}{\emph{nf}}{t'}$).

  Clearly, there exists a finite accepting derivation in $\Re$ from
  $X$ if and only if there exists a term $t$ reachable from $X$ in
  $\Re'$ satisfying the state property $EN(f)$. This concludes the
  proof.
\end{proof}

Now, we consider the set of variables $\hat{Var} = Var\bigcup\{Z_{ACC},
Z_{NOT\_ACC}\}$.  Moreover, with $T$ [resp., $T_{PAR}$,
$T_{SEQ}$] we refer to the set of process terms [resp., the set
of terms in which no sequential composition occurs, the set of
terms in which no  parallel composition occurs] built over $\hat{Var}$.



The following definition provides an extension of
$\npla{\overline{\Re}_{PAR},\overline{\Re}_{PAR,F}}$ suitable to
our purposes.

\begin{definition}[Rewrite System $\Re_{PAR}$]
  The \BRS $\npla{\Re_{PAR},\Re_{PAR,F}}$ is the parallel \BRS defined
  from $\npla{\Re,\Re_{F}}$ and
  $\npla{\overline{\Re}_{PAR},\overline{\Re}_{PAR,F}}$ as follows:
\begin{itemize}
\item $\Re_{PAR}=
  \begin{array}[t]{l}%
    \overline{\Re}_{PAR}\ \cup \\[4pt]%
    \{\prsrule{X}{}{Z_{ACC}} \mid
    \begin{array}[t]{l}%
      \exists{} r=\prsrule{X}{a}{Y.(Z)}\in\Re \text{ such that either }
      r\in\Re_{F} \text{ or there} \\[4pt]
      \text{exists a finite accepting derivation in }
      \Re \text{ from } Z\}\  \cup \end{array}\\
    \{X\rightarrow{}Z_{NOT\_ACC} \mid \exists{}  r =
    \prsrule{X}{a}{Y.(Z)}\in\Re\setminus\Re_{F}\}\end{array}$
\item $\Re_{PAR,F}=\overline{\Re}_{PAR,F}\ \cup\
  \{\prsrule{X}{}{Z_{ACC}}\in\Re_{PAR}\}$
\end{itemize}
\end{definition}

\begin{Lemma}
 $\npla{\Re_{PAR},\Re_{PAR,F}}$  can be
 built effectively.
 \end{Lemma}
\begin{proof} It follows directly from Proposition~\ref{prop:finiteacc}.
\end{proof}

\setcounter{equation}{0}
 \begin{remark}\label{remark:2}
   Notice that $\Re_{PAR}\setminus \overline{\Re}_{PAR}$ contains
   rules of the form $\prsrule{X}{}{Z_{ACC}}$ or of the form
   $\prsrule{X}{}{Z_{NOT\_ACC}}$, and every rule in
   $\Re_{PAR}$ does not contain in the left-hand side any occurrence of
   $Z_{ACC}$ and $Z_{NOT\_ACC}$. Therefore, it immediately follows that
   for all $t\in T$:
\begin{equation}
\prsderpar{t}{\sigma}{\varepsilon}
\ \ \Longleftrightarrow \ \ \prsdernorm{t}{\sigma}{\overline{\Re}_{PAR}}{\varepsilon}
\end{equation}

\noindent and for all $X,Y\in{}Var$
\begin{equation}
\prsderpar{X}{\sigma}{Y}\ \ \Longleftrightarrow\ \
\prsdernorm{Y}{\sigma}{\overline{\Re}_{PAR}}{Y}
\end{equation}

\noindent From (1)--(2), it follows immediately that
$\npla{\Re_{PAR},\Re_{PAR,F}}$ still satisfies properties 2-5 of
Definition~\ref{def:parallel1}.

%
\end{remark}




Now, let us go back to Problem 1 and consider an infinite accepting
derivation of the form $X$ \multider{\sigma}, with $X\in{}Var$. If $X$
\multider{\sigma} belongs to $\Pi_{PAR}$, as we have seen, it is
possible to mimic that derivation with an accepting infinite
derivation in $\npla{\Re_{PAR},\Re_{PAR,F}}$, and vice versa.

Let us now assume that $X$ \multider{\sigma} does not belong to the
class $\Pi_{PAR}$. In this case, the derivation $X$ \multider{\sigma}
can be written in the form $X$ \multider{} $t$$\parallel$$Y.(Z)$
\multider{\rho}, with $Z\in{}Var$, and where the subderivation of
$t$$\parallel$$Y.(Z)$ \multider{\rho} from $Z$ is an infinite
accepting derivation in $\Re$ from $Z$. To mimic this kind of
derivations, we build, starting from the \BRSs
$\npla{\Re_{PAR},\Re_{PAR,F}}$ and $\npla{\Re,\Re_{F}}$, a sequential
\BRS $\npla{\Re_{SEQ},\Re_{SEQ,F}}$ according to the following
definition:

\begin{definition}[Rewrite System $\Re_{SEQ}$]
  The \BRS $\npla{\Re_{SEQ},\Re_{SEQ,F}}$ is the sequential \BRS
  defined from $\npla{\Re,\Re_F}$ over the alphabet
  $\Sigma'=\Sigma\cup\{\$,\#\}$ as follows:
\begin{itemize}
\item $\Re_{SEQ}=\begin{array}[t]{l}%
    \{\prsrule{X}{a}{Y.(Z)}\in\Re\}\ \cup \\
    \{\cancrule{X}{Y} \mid \begin{array}[t]{l}
         X,Y\in{}Var \text{ and there exists a non--accepting
         derivation }  \\[4pt]
    \prsderpar{X}{\sigma}{\parcomp{p}{Y}} \text{ in } \Re_{PAR} \text{
         for some } p\in{}T_{PAR}
    \text{ with } \size{\sigma}\geq1\}\ \cup \end{array}\\[4pt]
  \{\dollarrule{X}{Y} \mid \begin{array}[t]{l}
    X,Y\in{}Var \text{ and there exists an accepting derivation }
     \\[4pt]
    \prsderpar{X}{\sigma}{\parcomp{p}{Y}} \text{ in } \Re_{PAR} \text{ for some } p\in{}T_{PAR} \}
     \end{array}
  \end{array}$

\item $\Re_{SEQ,F}=\{\prsrule{X}{a}{Y.(Z)}\in\Re_{F}\}\ \cup\
  \{\dollarrule{X}{Y}\in\Re_{SEQ}\}$
\end{itemize}
\end{definition}

\begin{Lemma}
$\npla{\Re_{SEQ},\Re_{SEQ,F}}$ can be built effectively.
\end{Lemma}

\begin{proof}
  Follows directly from the definition of
  $\npla{\Re_{SEQ},\Re_{SEQ,F}}$ and Proposition~\ref{prop:decfinite}.
\end{proof}

Soundness and completeness of the procedure described above is stated
by the following theorem, whose proof is reported in the appendix.

\begin{Teorema}[Soundness and Completeness]\label{theo:sound2}
  Given $X\in{}Var$, there exists an infinite accepting derivation in
  $\Re$ from $X$ $\mathrm{(}$resp., an infinite derivation devoid of accepting
  rule occurrences, an infinite derivation with a finite non--null number of
  accepting rule occurrences$\mathrm{)}$ if, and only if, one of the following conditions is
  satisfied:
\begin{enumerate}
\item there exists a variable $Y\in{}Var$ reachable
$\mathrm{(}$resp., reachable
      through a non--accepting derivation, reachable$\mathrm{)}$ from $X$ in
      $\Re_{SEQ}$, and there exists in $\Re_{PAR}$ an infinite
      accepting derivation $\mathrm{(}$resp,. an infinite derivation devoid of
      accepting rule occurrences, an infinite derivation containing a finite
non--null number of
      accepting rule occurrences$\mathrm{)}$ from $Y$.
\item there exists in $\Re_{SEQ}$ an infinite accepting derivation
      $\mathrm{(}$resp., an infinite derivation devoid of accepting rule occurrences,
      an infinite derivation containing a finite non--null number of accepting rule
      occurrences$\mathrm{)}$ from $X$.
\end{enumerate}
\end{Teorema}





This  result, together with Theorem~\ref{theo:cond},
allow us to conclude that Problems 1-3, stated at the beginning of this
section, are decidable.


\bibliographystyle{plain}


\appendix
\newpage
\appendix
\begin{LARGE}
\textbf{APPENDIX}
\end{LARGE}

\section{Definitions and simple properties}


In this section we give some definitions and deduce simple properties
that will be used in Appendices~B and~C for the proof of
Theorem~\ref{theo:sound2}.

In the following $\hat{Var}$ denotes the set of variables
$Var\cup\{Z_{ACC},Z_{NOT\_ACC}\}$, $T$ denotes the set of terms over
$\hat{Var}$, and $T_{PAR}$ (resp., $T_{SEQ}$) the set of terms in $T$
not containing sequential (resp., parallel) composition.

\begin{Definition}
The set of {\em subterms\/} of a term $t\in{}T$, denoted by $SubTerms(t)$, 
is defined inductively as follows:
\begin{itemize}
\item $SubTerms(\varepsilon)=\{\varepsilon\}$;
\item $SubTerms(X)=\{X\}$, for all  $X\in{}\hat{Var}$;  
\item $SubTerms(X.(t))=SubTerms(t)\cup\{X.(t)\}$,
for all $X.(t)\in{}T$ with  $t\neq\varepsilon$;
\item $SubTerms($$t_{1}$$\parallel$$t_{2})=\bigcup_{(t_{1}',t_{2}')\in{}S}
(SubTerms(t_{1}')\cup{} SubTerms(t_{2}'))$ $\cup$ $\{t_{1}$$\parallel$$t_{2}\}$,
\newline
with $S=\{(t_{1}',t_{2}')\in{}T\times{}T \mid t_{1}',t_{2}'\neq\varepsilon$
     and $t_{1}$$\parallel$$t_{2}={}t_{1}'$$\parallel$$t_{2}'\}$ and
$t_{1},t_{2}\in{}T\setminus \{\varepsilon\}$.
\end{itemize}
\end{Definition}

\begin{Definition} The set of terms obtained from a term $t\in T$
{\em substituting\/}
an occurrence of a subterm $st$ of $t$ with a term $t'\in T$, denoted
by $t[st\rightarrow{}t']$, is defined inductively as follows:
\begin{itemize}
    \item $t[t\rightarrow{}t']=\{t'\}$;
    \item $X.(t)[st\rightarrow{}t']=
     \{X.(s) \mid s\in{}t[st\rightarrow{}t']\}$, for all terms
$X.(t)\in{}T$ with $t\neq\varepsilon$ and
     $st\in{}SubTerms(X.(t))\setminus \{X.(t)\}$;

    \item $t_{1}$$\parallel$$t_{2}[st\rightarrow{}t']=$
$\{\parcomp{t''}{t'_2} \mid (t'_1,t'_2) \in T \times T, t'_1,t'_2 \neq
\varepsilon$, $\parcomp{t'_1}{t'_2} = \parcomp{t_1}{t_2}$,
$st \in SubTerms(t'_1)$, $t''\in{}t_{1}'[st\rightarrow{}t']\}$, for
all $t_1,t_2\in T\setminus\{\varepsilon\}$ and $st \in
SubTerms(\parcomp{t_1}{t_2})\setminus \{\parcomp{t_1}{t_2}\}$.
\end{itemize}
\end{Definition}

\begin{Definition}
  For a term $t\in{}T$, the set of terms $SEQ(t)$ is the subset of
  $T_{SEQ}\setminus\{\varepsilon\}$ defined inductively as follows:
\begin{itemize}
    \item $SEQ(\varepsilon)=\emptyset$;
    \item $SEQ(X)=\{X\}$, for all $X\in{}\hat{Var}$;
    \item $SEQ(X.(t))=\{X.(t') \mid t'\in{}SEQ(t)\}$, for all
      $X\in{}\hat{Var}$ and $t\in{}T\setminus \{\varepsilon\}$;
\item $SEQ(t_{1}$$\parallel$$t_{2})=SEQ(t_{1})\cup{}SEQ(t_{2})$.
    \end{itemize}
\end{Definition}

\begin{Definition}
  Let $\sigma_{1}$ and $\sigma_{2}$ be finite sequences of rules in
  $\Re$ $\mathrm{(}$the empty sequence is denoted by $\epsilon\mathrm{)}$.  The
  \emph{interleaving of $\sigma_{1}$ and $\sigma_{2}$}, in symbols
  $Interleaving(\sigma_{1},\sigma_{2})$, is the set of rule sequences
  in $\Re$ inductively defined as follows:
\begin{itemize}
\item $Interleaving(\epsilon,\sigma)=
  Interleaving(\sigma,\epsilon)=\{\sigma\}$;
\item $Interleaving(r_{1}\sigma_{1},r_{2}\sigma_{2})=
      \begin{array}[t]{l}\{r_{1}\sigma\mid\sigma\in{}
     Interleaving(\sigma_{1},r_{2}\sigma_{2})\}\ \bigcup\ \\ 
    \{r_{2}\sigma\mid\sigma\in{}
     Interleaving(r_{1}\sigma_{1},\sigma_{2})\}\end{array}$ 

where $r_{1}$ and $r_{2}$ are rules in $\Re$.
\end{itemize}

\end{Definition}
For a term $t\in{}T_{SEQ}\setminus \{\varepsilon\}$ having the form
$t=X_{1}.(X_{2}.(\ldots{}X_{n}.(Y)\ldots))$, with $n\geq0$,
we denote the variable $Y$ by $last(t)$. Given two terms
 $t,t'\in{}T_{SEQ}\setminus\{\varepsilon\}$, with
 $t=X_{1}.(X_{2}.(\ldots{}X_{n}.(Y)\ldots))$ and
 $t'=X_{1}'.(X_{2}'.(\ldots{}X_{k}'.(Y')\ldots))$, we denote with
 $t\circ{}t'$ the term 
 $X_{1}.(X_{2}.(\ldots{}X_{n}.(X_{1}'.(X_{2}'$
 $.(\ldots{}X_{k}'.(Y')\ldots)))\ldots))$. Notice that $t\circ{}t'$  is
the only term  in $t[Y\rightarrow{}t']$ and that the operation $\circ$ on 
terms in $T_{SEQ} \setminus \{\varepsilon\}$ is associative.

\begin{Remark}
\label{rem1}
For terms $t,t'\in{}T$, with $t'\neq\varepsilon$  and $st\in{}SubTerms(t)$,
it holds that if $s\in{}t[st\rightarrow{}t']$, then 
$t'\in{}SubTerms(s)$.
\end{Remark}


\begin{Proposition}
\label{Prop:Subterms1} The following properties hold:
\begin{description}
    \item{P1.} If $t$ \multider{\sigma} $t'$ and $t\in{}SubTerms(s)$,
for some $s\in{}T$, then it holds $s$ \multider{\sigma} $s'$, 
for all $s'\in{}s[t\rightarrow{}t']$;
\item{P2.} If $t$ \multider{\sigma} is an infinite derivation in $\Re$ and
$t\in{}SubTerms(s)$, for some $s\in{}T$, then it holds $s$ \multider{\sigma}.
\end{description}
\end{Proposition}

\begin{proof}
In the proof we exploit the following property (which can be easily checked):
\begin{description}
    \item[A.] If $t$ \prsoneder{r} $t'$ and $t\in{}SubTerms(s)$, for some
$s\in{}T$, then  it holds $s$ \prsoneder{r} $s'$ for all $s'\in{}s[t\rightarrow{}t']$
\end{description}
Let us prove the property P1 reasoning by induction on the length of
$\sigma$. \\
\textbf{Base Step}: $|\sigma|=0$. In this case Property~P1 is obvious.\\
\textbf{Induction Step}: $|\sigma|>0$. The derivation $t$
\multider{\sigma} $t'$ can be written in the form
\begin{displaymath}
\textrm{$t$ \multider{\sigma'} $\overline{t}$ \prsoneder{r} $t'$
with $|\sigma'|=|\sigma|-1.$}
\end{displaymath}
Let $s\in{}T$ be a term with $t\in{}SubTerms(s)$. By inductive
hypothesis, it holds that $s$ \multider{\sigma'} $\overline{s}$, for
all $\overline{s}\in{}s[t\rightarrow\overline{t}]$. Since
$\overline{t}$ \prsoneder{r} $t'$ and
$\overline{t}\in{}SubTerms(\overline{s})$ (see Remark~\ref{rem1}),
from Property~\textbf{A} we deduce that, for all
$s'\in\overline{s}[\overline{t}\rightarrow{}t']$, it holds
$\overline{s}$ \prsoneder{r} $s'$.  Moreover, one can easily prove
that, for all $s'\in{}s[t\rightarrow{}t']$, there exists a
$\overline{s}\in{}s[t\rightarrow\overline{t}]$ such that
$s'\in{}\overline{s}[\overline{t}\rightarrow{}t']$. This immediately
proves the thesis.  \newline Now, let us prove Property~P2. The
infinite derivation $t$ \multider{\sigma} can be written in the form
\begin{displaymath}
\textrm{$t$ \prsoneder{r_{0}} $t_{1}$ \prsoneder{r_{1}} $t_{2}$
\prsoneder{r_{2}}$\ldots$}
\end{displaymath}
Now, let $s\in{}T$ be a term with $t\in{}SubTerms(s)$.  From
Property~\textbf{A}, it follows that $\prssimpder{s}{r_{0}}{s_{1}}$,
for all $s_{1}\in{}s[t\rightarrow{}t_{1}]$.  Moreover, from
Remark~\ref{rem1}, we deduce that $t_{1}\in{}SubTerms(s_{1})$.
Therefore, by repeating the reasoning above, it is possible to define
a sequence of terms, $(s_{n})_{n\in{}N}$, such that
\begin{displaymath}
\textrm{$s$ \prsoneder{r_{0}} $s_{1}$ and $s_{n}$ \prsoneder{r_{n}}
$s_{n+1}$, for all $n>0$,}
\end{displaymath}
thus proving the thesis.
\end{proof}


\begin{Proposition}\label{Prop:Subterms2}
If
 $t,t'\in{}T_{SEQ}\setminus \{\varepsilon\}$ are terms with
$last(t)$ \multider{\rho}
 $t'$, then it holds that
 \begin{description}
    \item{P1.} $t$ \multider{\rho} $t\circ{}t'$;
    \item{P2.} $t''\circ{}t$ \multider{\rho} $t''\circ{}t\circ{}t'$
 for all $t''\in{}T_{SEQ}\setminus\{\varepsilon\}$.
 \end{description}
\end{Proposition}

\begin{proof}
  Let us prove Property~P1. We know that $last(t)$ \multider{\rho}
  $t'$ and $last(t)\in{}SubTerms(t)$. By observing that
  $t[last(t)\rightarrow{}t']=
  \{t\circ{}t'\}$, from Property~\ref{Prop:Subterms1} we obtain the thesis.\\
  Let us prove Property~P2. By Property~P1, we have that $t$
  \multider{\rho} $t\circ{}t'$.  Moreover, for all
  $t''\in{}T_{SEQ}\setminus\{\varepsilon\} $ we have that
  $t\in{}SubTerms(t''\circ{}t)$ and
  $t''\circ{}t[t\rightarrow{}t\circ{}t']= \{t''\circ{}t\circ{}t'\}$.
  From Property~\ref{Prop:Subterms1} we have the thesis.
\end{proof}


\section{Proof  of the sufficient condition of Theorem \ref{theo:sound2}}

In order to prove the \emph{if} direction of Theorem~\ref{theo:sound2}
we need the following
Lemmata~\ref{Lemma:From-Signed-RPAR-To-R}--\ref{Lemma:From-RSEQ-To-R}.

\begin{Lemma}\label{Lemma:From-Signed-RPAR-To-R}
  If $r=t$\Rule{c}$t'\in{}\overline{\Re}_{PAR}\setminus\Re$, then
  there exists a finite derivation in $\Re$ of the form $t$
  \multider{\sigma} $t'$, with $|\sigma|>0$. Moreover, $\sigma$ is
  accepting $\mathrm{(}$resp., non--accepting$\mathrm{)}$, if
  $r\in\Re_{PAR,F}$ $\mathrm{(}$resp.,
  $r\notin\Re_{PAR,F}$$\mathrm{)}$.
\end{Lemma}

\begin{proof}
  Let $r=t$\Rule{c}$t'\in\overline{\Re}_{PAR}\setminus\Re$, then
  $c\in\{\#,\$\}$ and $t,t'\in{}Var$. Moreover, $c=\#$ (resp., $c=\$$)
  if, and only if, $r\notin\Re_{PAR,F}$ (resp., $r\notin\Re_{PAR,F}$).
  Let us consider the Algorithm {\sc Build--parallel--BRS} (see
  Lemma~\ref{lemma:algo}). 
  Suppose that $r$ is the \emph{n}-th rule added to $\overline{\Re}$
  during the execution of the algorithm. Rule $r$ is added to
  $\overline{\Re}$ during an execution of the {\bf repeat} loop, where
  a rule $r'$ of the form $X$\Rule{a}$Y.(Z)\in\Re$ is considered.
  The proof is by induction on $n$.

\begin{description}
\item[\textbf{Base Step}] $n=1$. At this step of the algorithm the
  following holds:
\begin{enumerate}
\item $\Re_{PAR,AUX}=\{r\in\Re \mid \textrm{ $r$ is a PAR rule}\}$ and
  $\Re_{PAR,AUX,F}=\{r\in\Re_{F} \mid \textrm{ $r$ is}$ a PAR
  rule\}\label{cond1}
\end{enumerate}
First assume that $r\notin\Re_{PAR,F}$. Then, $c=\#$, and there are
two cases:
\begin{itemize}
\item $r$ is added to $\overline{\Re}$ in Line 19.  Then
  $r=X$\Rule{\#}$Y$, and the condition in the {\bf if} statement of
  Line 16 is satisfied. Therefore, $r'\notin\Re_{F}$ and $Z$
  \multiderparaux{\rho} $\varepsilon$, with $\rho$ devoid of
  (accepting) rules in $\Re_{PAR,AUX,F}$. From Property~\ref{cond1}
  above, $\rho$ must be a sequence of non--accepting rules in $\Re$.
  Therefore, $X$ \prsoneder{r'} $Y.(Z)$ \multider{\rho} $Y$ is a
  non--accepting derivation in $\Re$.
\item $r$ is added to $\overline{\Re}$ by the inner \textbf{for} loop
  in Lines 21--34, when a rule $r''$ of the form
  $Y.(W)$\Rule{b}$W'\in\Re$ is considered.  Then, $r=X$\Rule{\#}$W'$,
  and $r$ is added to $\overline{\Re}$ in Line 32. Hence, the
  condition of the \textbf{if} statement in Line 29 is satisfied.
  Therefore, $r',r''\notin\Re_{F}$ and $Z$ \multiderparaux{\rho} $W$,
  with $\rho$ devoid of (accepting) rules in $\Re_{PAR,AUX,F}$.
  Again, from Property~\ref{cond1}, $\rho$ must be a sequence of
  non--accepting rules in $\Re$.  It follows that $X$ \prsoneder{r'}
  $Y.(Z)$ \multider{\rho} $Y.(W)$ \prsoneder{r''} $W'$ is a
  non--accepting derivation in $\Re$.
\end{itemize}
Assume now that $r\in\Re_{PAR,F}$. Then, $c=\$$, and there are two
cases:
 \begin{itemize}
 \item $r$ is added to $\overline{\Re}$ in Line 12.  Thus,
   $r=X$\Rule{\$}$Y$, and the condition of the \textbf{if} statement
   in Line 9 must be satisfied. In particular, we have that
   $\prsderparaux{Z}{}{\varepsilon}$.  By Property~\ref{cond1}, it
   holds that $\prsder{Z}{}{\varepsilon}$.  Therefore, if
   $r'\in\Re_{F}$, the thesis is proved.  
   
   Otherwise, $Z$ \multiderparaux{\rho} $\varepsilon$, with $\rho$
   containing occurrences of (accepting) rules in $\Re_{PAR,AUX,F}$.
   In this case, the thesis immediately follows from
   Property~\ref{cond1}.
 \item $r$ is added to $\overline{\Re}$ by the inner \textbf{for} loop
   in Lines 22--34, when a rule $r''$ of the form
   $Y.(W)$\Rule{b}$W'\in\Re$ is considered.  Then, $r=X$\Rule{\$}$W'$,
   and $r$ is added to $\overline{\Re}$ in Line 25. Hence, the
   condition of the \textbf{if} statement in Line 22 must be
   satisfied. In particular, it holds that $\prsderparaux{Z}{}{W}$.
   By Property~\ref{cond1}, we have $\prsder{Z}{}{W}$.  Therefore, if
   $r'\in\Re_{F}$ or $r''\in\Re_{F}$, we obtain the thesis.
   
   Otherwise, it holds that $Z$ \multiderparaux{\rho} $W$, with $\rho$
   containing occurrences of (accepting) rules in $\Re_{PAR,AUX,F}$.
   In this case, $\rho$ is a sequence of accepting rules in $\Re$ by
   Property~\ref{cond1}, and the thesis immediately follows.
\end{itemize}

\item[\textbf{Induction Step}] $n>1$.  Let $\overline{\Re}'$ be the
  set of the rules in $\overline{\Re}$ after $n-1$ rules have been
  added.
  Then the following condition holds:

\begin{enumerate}\setcounter{enumi}{1}
\item $\Re_{PAR,AUX}$ $=\{r\in\Re \mid \textrm{ $r$ is a PAR
    rule}\}\cup\overline{\Re}'$, and\\ $\Re_{PAR,AUX,F}$
  $=\{r\in\Re_{F} \mid \textrm{ $r$ is a PAR
    rule}\}\cup\{X$\Rule{\$}$Y\in\overline{\Re}' \}$.\label{cond2}
\end{enumerate}
By inductive hypothesis, the thesis holds for every rule in
$\overline{\Re}'$.  Let us consider the case where
$r\notin\Re_{PAR,F}$ (the proof is similar in the case where
$r\in\Re_{PAR,F}$). Then, $c=\#$, and there are two cases:
\begin{itemize}
\item $r$ is added to $\overline{\Re}$ in Line 19.  Thus,
  $r=X$\Rule{\#}$Y$, and the condition of the \textbf{if} statement in
  Line 16 must be satisfied. Therefore, $r'\notin\Re_{F}$ and $Z$
  \multiderparaux{\rho} $\varepsilon$ with $\rho$ devoid of
  (accepting) rules in $\Re_{PAR,AUX,F}$. From Property~\ref{cond2},
  either $\rho$  contains occurrences of non--accepting rules in
  $\Re$, or it contains occurrences of rules in
  $\overline{\Re}'\setminus\Re_{PAR,F}$.  By inductive hypothesis, for
  every rule in $\overline{\Re}'\setminus\Re_{PAR,F}$ of the form
$t_1$\Rule{c}$t_2$ there exists a non accepting derivation in
$\Re$ of the form $t_1$ \multider{} $t_2$. As a consequence, 
 there exists a non
  accepting derivation in $\Re$ of the form 
  $Z$ \multider{\rho'} $\varepsilon$, with $\rho'$ non--accepting.
  Therefore, $X$ \prsoneder{r'} $Y.(Z)$ \multider{\rho'} $Y$ is a non
  accepting derivation in $\Re$. 
\item $r$ is added to $\overline{\Re}$ by the inner \textbf{for} loop,
  when a rule $r''$ of the form $Y.(W)$\Rule{b}$W'\in\Re$ is
  considered.  Then, $r=X$\Rule{\#}$W'$, and $r$ is added to
  $\overline{\Re}$ in Line 32. Therefore, the condition of the
  \textbf{if} statement in Line 29 must be satisfied. Hence,
  $r',r''\notin\Re_{F}$ and $Z$ \multiderparaux{\rho} $W$, with $\rho$
  devoid of (accepting) rules in $\Re_{PAR,AUX,F}$.  Again, from
  Property~\ref{cond2}, either $\rho$  contains occurrences of non
  accepting rules in $\Re$, or it  contains occurrences of rules
  in $\overline{\Re}'\setminus\Re_{PAR,F}$. By inductive hypothesis, it follows that
  $Z$ \multider{\rho'} $W$, with $\rho'$ non--accepting.  Therefore,
  the derivation $X$ \prsoneder{r'} $Y.(Z)$ \multider{\rho} $Y.(W)$
  \prsoneder{r'} $W'$ is a non--accepting derivation in $\Re$.\qedhere
\end{itemize}
\end{description}
\end{proof}



\begin{Lemma}\label{Lemma:From-RPAR-To-R}
  Let $p,p',p''\in{}T_{PAR}$, where $p'$ does not contain occurrences
  of $Z_{ACC}$ and $Z_{NOT\_ACC}$, and $p''$ does not contain
  occurrences of variables in $Var$.
  
  If $p$ \multiderpar{\sigma} $\parcomp{p'}{p''}$, then there exists a
  term $t\in{}T$ and a derivation $p$ \multider{\rho}
  $\parcomp{p'}{t}$ in $\Re$, with $|\rho|>0$, if $|\sigma|>0$.
  Moreover, if $\sigma$ is accepting $\mathrm{(}$resp., non
  accepting$\mathrm{)}$ then, $\rho$ can be chosen accepting
  $\mathrm{(}$resp., non--accepting$\mathrm{)}$.
\end{Lemma}
\begin{proof}
  The proof is by induction on the length of the rule sequence
  $\sigma$.
\begin{description}  
\item[\textbf{Base Step}] $|\sigma|=0$. In this case the conclusion 
  immediately follows.
\item[\textbf{Induction Step}] $|\sigma|>0$. In this case the
  derivation $p$ \multiderpar{\sigma} $\parcomp{p'}{p''}$ can be
  written in the following form:
\begin{displaymath}
\textrm{$p$ \multiderpar{\sigma'}
$\parcomp{\overline{p}'}{\overline{p}''}$ \onederpar{r}
$\parcomp{p'}{p''}$ }
\end{displaymath}
with $|\sigma'|<|\sigma|$, $r\in\Re_{PAR}$ and
$\overline{p}',\overline{p}''\in{}T_{PAR}$.  Moreover, $\overline{p}'$
does not contain occurrences of $Z_{ACC}$ and $Z_{NOT\_ACC}$, and
$\overline{p}''$ does not contain occurrences of variables in $Var$.

By inductive hypothesis, there exists a term $\overline{t}\in{}T$, and
a derivation $p$ \multider{\rho'}
$\parcomp{\overline{p}'}{\overline{t}}$, with $|\rho'|>0$, if
$|\sigma'|>0$, and $\rho'$ accepting $\mathrm{(}$resp., non
accepting$\mathrm{)}$, if $\sigma'$ is accepting $\mathrm{(}$resp., non
accepting$\mathrm{)}$.
Then, there are three possible cases:
\begin{enumerate}
\item $r$ is a PAR rule in $\Re$. From the definition of $\Re_{PAR}$,
  $r\in\Re_{PAR}$, and $r\in\Re_{F}$ if, and only if,
  $r\in\Re_{PAR,F}$. Moreover, $\overline{p}''= p''$ and
  $\overline{p}'$ \onederpar{r} $p'$. Therefore, $p$ \multider{\rho'}
  $\parcomp{\overline{p}'}{\overline{t}}$ \prsoneder{r}
  $\parcomp{p'}{\overline{t}}$, with $\rho'r$ accepting (resp., non
  accepting), if $\sigma'r$ is accepting (resp., non--accepting).
\item $r\in\overline{\Re}_{PAR}\setminus\Re$. Therefore,
  $r=X$\Rule{a}$Y$, with $X,Y\in{}Var$, $a\in\{\#,\$\}$ and $r$
  accepting (resp. non accepting), if $a=\$$
  (resp., $a=\#$).  From
  Lemma~\ref{Lemma:From-Signed-RPAR-To-R}, we have that $X$
  \multider{\rho''} $Y$, with $\rho''$ accepting (resp., non
  accepting), if $r\in\Re_{PAR,F}$ (resp., $r\notin\Re_{PAR,F}$), and
  $|\rho''|>0$.  Moreover, $\overline{p}''=p''$ and $\overline{p}'$
  \onederpar{r} $p'$.  Hence, there exists in $\Re$ the derivation $p$
  \multider{\rho'} $\parcomp{\overline{p}'}{\overline{t}}$
  \multider{\rho''} $\parcomp{p'}{\overline{t}}$, where the rule
  sequence $\rho'\rho''$ is accepting (resp., non--accepting), if
  $\sigma'r$ is accepting (resp., non--accepting).
\item $r\in\Re_{PAR}\setminus\overline{\Re}_{PAR}$. Therefore,
  $r=X$\Rule{}$\hat{Y}$, with $X\in{}Var$,
  $\hat{Y}\in\{Z_{ACC},Z_{NOT\_ACC}\}$ and $r$ accepting (resp. non
  accepting), if $\hat{Y}=Z_{ACC}$ (resp., $\hat{Y}=Z_{NOT\_ACC}$).
  From the definition of $\Re_{PAR}$, it follows that $X$
  \multider{\rho''} $t$, with $\rho''$ accepting (resp., not
  accepting), if $r\in\Re_{PAR,F}$ (resp., $r\notin\Re_{PAR,F}$), and
  $|\rho''|>0$.  Clearly, $p''= \parcomp{\overline{p}''}{\hat{Y}}$,
  and $\overline{p}'= \parcomp{p'}{X}$. Hence, $p$ \multider{\rho'}
  $\parcomp{\overline{p}'}{\overline{t}} =
  \parcomp{p'}{\parcomp{X}{\overline{t}}}$ \multider{\rho''}
  $\parcomp{p'}{\parcomp{t}{\overline{t}}}$, and the rule sequence
  $\rho'\rho''$ is accepting (resp., non--accepting), if $\sigma'r$ is
  accepting (resp., non--accepting).\qedhere
\end{enumerate}
\end{description}  
\end{proof}



\begin{Lemma}\label{Lemma:From-RPAR-To-R-Inf}
  For every $p\in{}T_{PAR}$, if $p$ \multiderpar{\sigma} is an
  infinite accepting derivation $\mathrm{(}$resp., an infinite
  derivation devoid of accepting rules$\mathrm{)}$ in $\Re_{PAR}$, %
  then there exists in $\Re$ an infinite accepting derivation
  $\mathrm{(}$resp., an infinite derivation devoid of accepting
  rules$\mathrm{)}$ from $p$.
\end{Lemma}
\begin{proof}
  To prove the lemma, we use the following property:
\begin{description}
\item[A] If $\parcomp{p'}{p''}$\multiderpar{\sigma} (with
  $p',p''\in{}T_{PAR}$), and $p''$ does not contain variables in
  $Var$, then $\prsderpar{p'}{\sigma}{}$.
\end{description}
Property~\textbf{A} easily follows from the observation that the
left-hand side of each rule in $\Re_{PAR}$ does not contain occurrences
of $Z_{ACC}$ and $Z_{NOT\_ACC}.$
  
Let now $p\in{}T_{PAR}$, and $p$ \multiderpar{\sigma} be an infinite
accepting derivation (resp., an infinite derivation devoid of
accepting rules) in $\Re_{PAR}$. We prove that there exists a sequence
of terms $(p_{n})_{n\in{N}}$ in $T_{PAR}$, and a sequence of terms
$(t_{n})_{n\in{}N\setminus\{0\}}$ satisfying the following properties:
\begin{description}
\item[i.] $p_{0}=p$.
\item[ii.] for all $n\in{}N$, there exists in $\Re_{PAR}$ an infinite
  accepting derivation (resp., an infinite derivation devoid of
  accepting rules) from $p_{n}$.
\item[iii.] for all $n\in{}N$, $p_{n}$ \multider{\rho_{n}}
  $\parcomp{p_{n+1}}{t_{n+1}}$, with $\rho_{n}$ non--null and
  accepting (resp., non--accepting).
\end{description}

Since, by setting $p_{0}=p$, Property~\textbf{ii} is satisfied for
$n=0$, it suffices to prove that the following property holds for any
$p\in{}T_{PAR}$:
\begin{description}
\item[B] If $p$ \multiderpar{\sigma} is an infinite accepting
  derivation (resp., an infinite derivation devoid of accepting rules)
  in $\Re_{PAR}$,  then, the following hold:
\begin{enumerate}
\item there exists a term $p'\in{}T_{PAR}$, and a term $t$, such that
  $p$ \multider{\rho} $\parcomp{p'}{t}$, with $\rho$ non--null and
  accepting (resp., non--accepting), and
\item there exists in $\Re_{PAR}$ an infinite accepting derivation
  (resp., an infinite derivation devoid of accepting rules) from $p'$.
\end{enumerate}
\end{description}
Let us prove Property~\textbf{B}. The infinite derivation
$\prsderpar{p}{\sigma}{}$ can be written in the form:
\begin{displaymath}
\textrm{$p$ \multiderpar{\lambda} $\overline{p}$
\multiderpar{\omega}}
\end{displaymath}
where $\prsderpar{\overline{p}}{\omega}{}$ is an infinite accepting
derivation (resp., an infinite derivation devoid of accepting rules)
in $\Re_{PAR}$ from $\overline{p}\in{}T_{PAR}$, and
$\prsderpar{p}{\lambda}{\overline{p}}$ is a non--null finite accepting
(resp., non--accepting) derivation in $\Re_{PAR}$.  Now,
$\overline{p}$ can be written in the form $\parcomp{p'}{p''}$, where
$p'$ does not contain occurrences of $Z_{ACC}$ and $Z_{NOT\_ACC}$, and
$p''$ does not contain occurrences of variables in $Var$. 
>From Property~\textbf{A}, $\prsderpar{p'}{\omega}{}$ is an infinite
accepting derivation (resp., an infinite derivation devoid of
accepting rules) in $\Re_{PAR}$, hence Property~\textbf{B}.2 holds.

>From Lemma~\ref{Lemma:From-RPAR-To-R}, applied to the accepting
derivation (resp., non--accepting derivation)
$\prsderpar{p}{\lambda}{\overline{p}=\parcomp{p'}{p''}}$, there exists
a term $t$ and a derivation $\prsder{p}{\rho}{\parcomp{p'}{t}}$, with
$\rho$ non--null and accepting (resp., non--accepting), hence
Property~\textbf{B}.1 holds.

Now, let $(p_{n})_{n\in{N}}$ and $(t_{n})_{n\in{}N\setminus{0}}$ be
the sequence satisfying Properties~\textbf{i--iii}. By Property
\textbf{iii}, the derivation
\begin{displaymath}
\textrm{%
$p_{0}$ 
\multider{\rho_{o}} 
$\parcomp{p_{1}}{t_{1}}$
\multider{\rho_{1}}
$\parcomp{p_{2}}{\parcomp{t_{1}}{t_{2}}}$
\multider{\rho_{2}}
$\ldots$ 
\multider{\rho_{n-1}}
$\parcomp{p_{n}}{\parcomp{t_{1}}{\parcomp{\ldots}{t_{n}}}}$
\multider{\rho_{n}}
$\parcomp{p_{n+1}}{\parcomp{t_{1}}{\parcomp{\ldots}{\parcomp{t_{n}}{t_{n+1}}}}}$
\multider{\rho_{n+1}}
$\ldots$
}
\end{displaymath}
is an infinite accepting derivation (resp. an infinite derivation
devoid of accepting rules) in $\Re$ from $p$. Hence the thesis.
\end{proof}



\begin{Lemma}\label{Lemma:From-RPAR-To-R-Inf2}
  If $p$ \multiderpar{\sigma} is an infinite derivation in $\Re_{PAR}$
  from $p\in{}T_{PAR}$ containing a finite non--null number of
  accepting rule occurrences, then there exists an infinite derivation
  in $\Re$ from $p$ containing a finite non--null number of accepting
  rule occurrences.
\end{Lemma}

\begin{proof}
  The infinite derivation $p$ \multiderpar{\sigma} can be written in
  the form:
\begin{displaymath}
\textrm{$p$ \multiderpar{\lambda} $\overline{p}$
\multiderpar{\omega}}
\end{displaymath}
where $\overline{p}$ \multiderpar{\omega} is an infinite derivation in
$\Re_{PAR}$ from $\overline{p}\in{}T_{PAR}$ devoid of accepting rule occurrences,
and $p$ \multiderpar{\lambda} $\overline{p}$ is an accepting finite
derivation in $\Re_{PAR}$. Now $\overline{p}$ can be written in the
form $\parcomp{p'}{p''}$, where $p'$ does not contain occurrences of
$Z_{ACC}$ and $Z_{NOT\_ACC}$, and $p''$ does not contain occurrences
of variables in $Var$. By Property~\textbf{A} in the proof of
Lemma~\ref{Lemma:From-RPAR-To-R-Inf}, we have that $p'$
\multiderpar{\omega} is an infinite derivation in $\Re_{PAR}$ devoid
of accepting rule occurrences. From Lemma~B.3, there exists an infinite derivation
$p'$ \multider{\mu} in $\Re$ from $p'$ devoid of accepting rule occurrences.
Finally, from Lemma~\ref{Lemma:From-RPAR-To-R}, applied to the
accepting derivation $p$ \multiderpar{\lambda}
$\overline{p}=\parcomp{p'}{p''}$, there exists a term $t$ and a
derivation $p$ \multider{\rho} $\parcomp{p'}{t}$, with $\rho$
accepting.  Hence, the derivation
$\prsder{p}{\rho}{\prsder{\parcomp{p'}{t}}{\mu}{}}$ is an infinite
derivation in $\Re$ containing a finite non-null number of accepting
rule occurrences.
\end{proof}

\begin{Lemma}\label{Lemma:From-RSEQ-To-R} 
  Let $t,t'\in{}T_{SEQ}$ and $s$ be any term in $T$ such that
  $t\in{}SEQ(s)$. The following results hold:
\begin{enumerate}
\item If $t$ \onederseq{r} $t'$, then there exists a term $s'\in{}T$,
  with $t'\in{}SEQ(s')$, such that $s$ \multider{\sigma} $s'$, and
  $|\sigma|>0$.  
  Moreover, if $r\in\Re_{SEQ,F}$ $\mathrm{(}$resp.,
  $r\notin\Re_{SEQ,F}\mathrm{)}$, then $\sigma$ can be chosen
  accepting $\mathrm{(}$resp., non--accepting$\mathrm{)}$.
\item If $\prsderseq{t}{\sigma}{t'}$ and $t\neq\varepsilon$, then
  there exists a $s'\in{}T$, with $t'\in{}SEQ(s')$, such that
  $\prsder{s}{\sigma'}{s'}$, and $|\sigma'|>0$, if $|\sigma|>0$.
  Moreover, if $\sigma$ is accepting $\mathrm{(}$resp.,
  non--accepting$\mathrm{)}$, then $\sigma'$ is accepting
  $\mathrm{(}$resp., non--accepting$\mathrm{)}$.
\item If $t$ \multiderseq{\sigma} is an accepting infinite derivation
  $\mathrm{(}$resp., an infinite derivation devoid of accepting
  rules$\mathrm{)}$ in $\Re_{SEQ}$ from $t\in{}T_{SEQ}$, then there
  exists in $\Re$ an accepting infinite derivation $\mathrm{(}$resp.,
  an infinite derivation devoid of accepting rules$\mathrm{)}$ from
  $s$.
\item If $\prsderseq{t}{\sigma}{}$ is an infinite derivation in
  $\Re_{SEQ}$ from $t\in{}T_{SEQ}$ containing a finite non--null
  number of accepting rule occurrences, then there exists an infinite
  derivation in $\Re$ from $s$ containing a finite non--null number of
  accepting rule occurrences.
\end{enumerate}
\end{Lemma}
\begin{proof}
  Let us first prove Property~\emph{1}. We use the following two
  properties, whose proofs are immediate. Let $t\in{}SEQ(s)$,
  $s\in{}T$ and $t=X_{1}.(X_{2}.(\ldots{}X_{n}.(Y)\ldots))$, with
  $n\geq0$. Then:
\begin{description}
\item[A.] if $st\in{}T_{SEQ}\setminus\{\varepsilon\}$ and
  $t'=X_{1}.(X_{2}.(\ldots{}X_{n}.(st)\ldots))$, then there exists a
  $s'\in{}s[Y\rightarrow{}st]$ (notice that $Y$ is a subterm of $s$) such
  that $t'\in{}SEQ(s')$.
\item[B.] if $Z\in{}Var$, $st'\in{}T$, and $st = \parcomp{st'}{Z}$,
  then there exists a $s'\in{}s[Y\rightarrow{}st]$  such that
  $X_{1}.(X_{2}.(\ldots{}X_{n}.(Z)\ldots))\in{}SEQ(s')$.
\end{description}

\noindent We can now distinguish the following two cases:
\begin{itemize}
\item $r=\prsrule{Y}{a}{Z_{1}.(Z_{2})}\in{}\Re$.  From the definition
  of $\Re_{SEQ}$, it follows that $r\in\Re_{SEQ}$, and $r\in\Re_{F}$
  if, and only if, $r\in\Re_{SEQ,F}$. Moreover,
  $t=X_{1}.(X_{2}.(\ldots{}X_{n}.(Y)\ldots))$ and
  $t'=X_{1}.(X_{2}.(\ldots{}X_{n}.(Z_{1}.(Z_{2}))\ldots))$.  Let
  $s\in{}T$ be such that $t\in{}SEQ(s)$. From Property~\textbf{A}
  above, there exists a $s'\in{}s[Y\rightarrow{}Z_{1}.(Z_{2})]$ such
  that $t'\in{}SEQ(s')$.  Since $Y$ \prsoneder{r} $Z_{1}.(Z_{2})$, by
  Proposition~\ref{Prop:Subterms1} it follows that $s$ \prsoneder{r}
  $s'$, with $r\in\Re_{F}$ if, and only if, $r\in\Re_{SEQ,F}$, and the
  thesis is proved.
%
\item $r=\prsrule{Y}{a}{Z}$ with $Y,Z\in{}Var$ and $a\in\{\#,\$\}$. 
  Moreover $t=X_{1}.(X_{2}.(\ldots{}X_{n}.(Y)\ldots))$ and
  $t'=X_{1}.(X_{2}.(\ldots{}X_{n}.(Z)\ldots))$. From the definition
of $\Re_{SEQ}$  there exists a
  derivation in $\Re_{PAR}$ of the form $Y$ \multiderpar{\sigma}
  $\parcomp{p}{Z}$ for some $p\in{}T_{PAR}$, with $|\sigma|>0$.
  Moreover, if $r\in\Re_{SEQ,F}$ (resp., $r\notin\Re_{SEQ,F}$), then
  $Y$ \multiderpar{\sigma} $\parcomp{p}{Z}$ can be chosen accepting
  (resp., non--accepting). From Lemma~\ref{Lemma:From-RPAR-To-R},
  there exists a term $st$ such that $Y$ \multider{\rho}
  $\parcomp{st}{Z}$, with $|\rho|>0$ and $\rho$ accepting (resp.,
  non--accepting) if $\sigma$ is accepting (resp., non--accepting).
  Let $s\in{}T$ be such that $t\in{}SEQ(s)$. From Property~\textbf{B}
  above, there exists a term $s'\in{}s[Y\rightarrow{}\parcomp{st}{Z}]$
  such that $t'\in{}SEQ(s')$.  Now, $Y$ \multider{\rho}
  $\parcomp{st}{Z}$.  From Proposition~\ref{Prop:Subterms1}, we
  conclude that $\prsder{s}{\rho}{s'}$, with $|\rho|>0$, and $\rho$
  accepting (resp., non--accepting), if $r\in\Re_{SEQ,F}$ (resp.,
  $r\notin\Re_{SEQ,F}$). Hence the thesis.
\end{itemize}

Property~\emph{2} can easily be proved by induction on the length of
$\sigma$, and using Property\emph{1} above.

Let us now consider Property~\emph{3}. The infinite accepting
derivation (resp., the infinite derivation devoid of accepting rules)
$t$ \multiderseq{\sigma} can be written in the form:
\begin{displaymath}
\prsderseq{t}{\rho}{\prsderseq{\overline{t}}{\omega}{}}
\end{displaymath}
with $\prsderseq{t}{\rho}{\overline{t}}$ a non--null finite accepting
derivation (resp., finite non--accepting derivation), and
$\prsderseq{\overline{t}}{\omega}{}$ an infinite accepting derivation
(resp., infinite derivation devoid of accepting rules) from
$\overline{t}\in{}T_{SEQ}$. Let $s\in{}T$ be such that $t\in{}SEQ(s)$.
>From Property~\emph{2} of the lemma, there exists a term
$\overline{s}\in{}T$, with $\overline{t}\in{}SEQ(\overline{s})$ and
such that $\prsder{s}{\lambda}{\overline{s}}$, $|\lambda|>0$, and
$\lambda$ accepting (resp., non--accepting). By repeating the
reasoning above, it follows that there exists a sequence
of terms, $(s_{n})_{n\in{}N}$, such that for all $n\in{}N$:
\begin{itemize}
\item $\prsder{s_{n}}{\lambda_{n}}{s_{n+1}}$, with $\lambda_{n}$
  accepting (resp., non--accepting), $|\lambda_{n}|>0$ and $s_{0}=s$.
\end{itemize}
Therefore, the following derivation
\begin{displaymath}
\textrm{$s=s_{0}$  \multider{\lambda_{0}} $s_{1}$
\multider{\lambda_{1}}$\ldots$ $s_{n}$ \multider{\lambda_{n}}
$s_{n+1}\ldots$}
\end{displaymath}
is an accepting infinite derivation (resp., an infinite derivation
devoid of accepting rules) in $\Re$ from $s$. This proves the thesis.

We now prove Property~\emph{4}. The infinite derivation
$\prsderseq{t}{\sigma}{}$, containing a finite non--null number of
accepting rule occurrences, can be written in the form:
\begin{displaymath}
\prsderseq{t}{\rho}{\prsderseq{\overline{t}}{\omega}{}}
\end{displaymath}
where $\prsderseq{t}{\rho}{\overline{t}}$ is a finite accepting
derivation, and $\prsderseq{\overline{t}}{\omega}{}$ is an infinite
derivation from $\overline{t}\in{}T_{SEQ}$ devoid of accepting rules.
Let $s\in{}T$ be such that $t\in{}SEQ(s)$. From Property~\emph{2} of
the lemma, there exists a term $\overline{s}\in{}T$, with
$\overline{t}\in{}SEQ(\overline{s})$, such that
$\prsder{s}{\lambda}{\overline{s}}$, with $\lambda$ accepting.  From
Property~\emph{3} of the lemma, there exists an infinite derivation in
$\Re$ from $\overline{s}$ devoid of accepting rules. From this
observation the thesis immediately follows.
\end{proof}



We are now ready to prove the \emph{if} direction 
of Theorem~\ref{theo:sound2}. Let $X\in{}Var$ and assume that one of
the following conditions holds:
\begin{description}
\item[C1] there exists a variable $Y$ reachable $\mathrm{(}$resp.,
  reachable through a non--accepting derivation, reachable$\mathrm{)}$
  from $X$ in $\Re_{SEQ}$, and there exists in $\Re_{PAR}$ an infinite
  accepting derivation $\mathrm{(}$resp,. an infinite derivation
  devoid of accepting rules, an infinite derivation containing a
  finite non--null number of accepting rule occurrences$\mathrm{)}$
  from $Y$.
\item[C2] there exists in $\Re_{SEQ}$ an infinite accepting derivation
  $\mathrm{(}$resp., an infinite derivation devoid of accepting rules,
  an infinite derivation containing a finite non--null number of
  accepting rule occurrences$\mathrm{)}$ from $X$.
\end{description}
We have to prove that there exists in $\Re$ an infinite accepting
derivation $\mathrm{(}$resp., an infinite derivation devoid of
accepting rules, an infinite derivation with a finite non--null number
of accepting rules$\mathrm{)}$ from $X$.

First, assume that Condition~\textbf{C2} holds. In this case the
thesis follows from Property~\emph{3} of
Lemma~\ref{Lemma:From-RSEQ-To-R} (resp., Property~\emph{3} of
Lemma~\ref{Lemma:From-RSEQ-To-R}, Property~\emph{4} of
Lemma~\ref{Lemma:From-RSEQ-To-R}), since $X\in{}SEQ(X)$.

Assume that Condition~\textbf{C1} holds instead. Then, by
Lemma~\ref{Lemma:From-RPAR-To-R-Inf} (resp.,
Lemma~\ref{Lemma:From-RPAR-To-R-Inf},
Lemma~\ref{Lemma:From-RPAR-To-R-Inf2}), there exists a term
$t\in{}T_{SEQ}$ of the form $X_{1}.(X_{2}.(\ldots{}X_{n}.(Y)$
$\ldots))$ (with $n\geq0$), and a variable $Y$ such that:
\begin{itemize}
\item $\prsderseq{X}{\rho}{t}$, for some rule sequence $\rho$ (resp.,
  with $\rho$ non--accepting, for some rule sequence $\rho$)
\item $\prsder{Y}{\sigma}{}$, with $\sigma$ infinite and accepting
  (resp., devoid of accepting rules, containing a finite non--null
  number of accepting rule occurrences).
\end{itemize}
>From Property~\emph{2} of Lemma~\ref{Lemma:From-RSEQ-To-R}, and the
fact that $X\in{}SEQ(X)$, there exists a term $s\in{}T$, with
$t\in{}SEQ(s)$ and $\prsder{X}{\lambda}{s}$, for some rule sequence
$\lambda$ (resp., with $\lambda$ non--accepting, for some rule
sequence $\lambda$). Since $Y\in{}SubTerms(s)$, from
Proposition~\ref{Prop:Subterms1}, it follows that
$\prsder{s}{\sigma}{}$ is an infinite derivation in $\Re$, with
$\sigma$ accepting (resp., devoid of accepting rules, containing a
finite non--null number of accepting rule occurrences), hence the
thesis.

\section{Proof of the necessary condition of  Theorem \ref{theo:sound2}}

In order to prove \emph{only if} direction of Theorem~\ref{theo:sound2} we
need the following Lemmata~C.1--C.6.

\begin{Lemma}\label{Lemma:Subderivations1}
 Let $t$$\parallel$$X.(s)$ \multider{\sigma} be a
derivation in $\Re$, and let $s$ \multider{\sigma{}'} be the
subderivation of  $t$$\parallel$$X.(s)$ \multider{\sigma} from
$s$. Then, the following properties are satisfied:
\begin{enumerate}
    \item If $s$ \multider{\sigma{}'} is infinite, then it holds that
    $t$ \multider{\sigma\setminus\sigma{}'}.
     Moreover, if $t$$\parallel$$X.(s)$ \multider{\sigma}
    is in $\Pi_{PAR}$ $\mathrm{(}$resp., in $\Xi_{PAR}$$\mathrm{)}$, 
then also  $t$ \multider{\sigma\setminus\sigma{}'} is in
    $\Pi_{PAR}$ $\mathrm{(}$resp., in $\Xi_{PAR}$$\mathrm{)}$.
    \item If $s$ \multider{\sigma{}'} leads to
    $\varepsilon$ then, the derivation
    $t$$\parallel$$X.(s)$ \multider{\sigma} can be written in the form
    \begin{displaymath}
     \textrm{$t$$\parallel$$X.(s)$ \multider{\sigma_{1}} $t'$$\parallel$$X$ \multider{\sigma_{2}}}
    \end{displaymath}
    with  $t$ \multider{\lambda} $t'$ and $\sigma_{1}\in{}Interleaving(\lambda,\sigma{}')$.
    \item If $s$ \multider{\sigma{}'} leads to
    a term $s'\neq\varepsilon$, then one of the following conditions is satisfied:
    \begin{itemize}
        \item There is a derivation $t$ \multider{\sigma\setminus\sigma{}'}. Moreover,
            if  $t$$\parallel$$X.(s)$ \multider{\sigma}
            is in $\Pi_{PAR}$ $\mathrm{(}$resp., in $\Xi_{PAR}$$\mathrm{)}$, then also $t$ \multider{\sigma\setminus\sigma{}'}
            is in
            $\Pi_{PAR}$ $\mathrm{(}$resp., in $\Xi_{PAR}$$\mathrm{)}$. If
            $t$$\parallel$$X.(s)$ \multider{\sigma} is finite and leads to
            $\overline{t}$, then
            $\overline{t}=X.(s')$$\parallel$$t'$ with $t$
            \multider{\sigma\setminus\sigma{}'} $t'$.
        \item $s'=W\in{}Var$ and the derivation
        $t$$\parallel$$X.(s)$ \multider{\sigma} can be written in the
        form
        \begin{displaymath}
        \textrm{$t$$\parallel$$X.(s)$ \multider{\sigma_{1}}
        $t'$$\parallel$$X.(W)$ \prsoneder{r}
        $t'$$\parallel$$W'$ \multider{\sigma_{2}}}
        \end{displaymath}
        where  $r=X.(W)$\Rule{a}$W'\in{}\Re$. Moreover,
         $t$ \multider{\lambda} $t'$
          with $\sigma_{1}\in{}Interleaving(\lambda,\sigma{}')$.
    \end{itemize}
\end{enumerate}
\end{Lemma}

\begin{proof}
 The assertion  follows directly from the definition of subderivation.
\end{proof}

\begin{Lemma}\label{Lemma:Subderivations2}
 Let $t$ \multider{\sigma} be a derivation in $\Pi_{PAR}$ $\mathrm{(}$resp., in $\Xi_{PAR}$$)$.
 The following properties hold:
 \begin{enumerate}
    \item If $t$ \multider{\sigma} can be written in the form
    $t$ \multider{\sigma_{1}} $t'$ \multider{\sigma_{2}}, then
    $t'$ \multider{\sigma_{2}} is in $\Pi_{PAR}$ $\mathrm{(}$resp., in $\Xi_{PAR}$$)$.
    \item For every finite derivation of the form
    $t'$ \multider{\omega} $t$, the derivation
    $t'$ \multider{\omega} $t$ \multider{\sigma} is in $\Pi_{PAR}$ $\mathrm{(}$resp., in $\Xi_{PAR}$$)$.
    \item For every term $p\in{}T_{PAR}\quad$$t$$\parallel$$p$ \multider{\sigma}
    is in $\Pi_{PAR}$ $\mathrm{(}$resp., in $\Xi_{PAR}$$)$.
 \end{enumerate}
\end{Lemma}

\begin{proof}
 The assertion  follows directly from the definition of subderivation.
\end{proof}


\begin{Lemma}\label{Lemma:Base}
\setcounter{equation}{0}
 Let $p$ \multider{\sigma}
$t$$\parallel$$Y.(s)$ \multider{\omega}  be a derivation with $s\neq\varepsilon$
and $p\in{}T_{PAR}$. Then,  $p$ \multider{\sigma}
$t$$\parallel$$Y.(s)$ can be written in the form
\begin{equation}
\textrm{$p$ \multider{\sigma_{1}} $t'$$\parallel$$Z$ \prsoneder{r}
$t'$$\parallel$$Y.(Z')$ \multider{\sigma_{2}}
$t$$\parallel$$Y.(s)$}
\end{equation}
with $r=Z$\Rule{a}$Y.(Z')$, and
\begin{equation}
\textrm{$Z'$ \multider{\sigma_{2}'} $s\quad$ and $t'$
\multider{\sigma_{2}''} $t$}
\end{equation}
with $\sigma_{2}\in{}Interleaving(\sigma_{2}',\sigma_{2}'')$.
Moreover, the following property is satisfied:
\begin{description}
    \item[A] The subderivation of $t'$$\parallel$$Y.(Z')$ \multider{\sigma_{2}}
    $t$$\parallel$$Y.(s)$ \multider{\omega}  from $Z'$ can be written
    in the form
    \begin{displaymath}
        \textrm{$Z'$ \multider{\sigma_{2}'} $s$
        \multider{\omega'}}
    \end{displaymath}
    where $s$ \multider{\omega'} is the subderivation of $t$$\parallel$$Y.(s)$
    \multider{\omega}  from $s$.
\end{description}
\end{Lemma}
\begin{proof}
  The proof is by induction on the length of
  $\sigma$.
\begin{description}
\item[\textbf{Base Step}] $|\sigma|=1$.  In this case, there exists a
  rule $r=Z$\Rule{a}$Y.(Z')\in\Re$ with $p=t$$\parallel$$Z$ and
  $Z'=s$. So, the first part of the assertion holds, with $\sigma_{1}$
  and $\sigma_{2}$ the empty sequences. As far as Property~\textbf{A}
  is concerned, it suffices to observe that in this case $\sigma_{2}$
  is the empty sequence.
\item[\textbf{Induction Step}] $|\sigma|>1$. The derivation $p$
  \multider{\sigma} $t$$\parallel$$Y.(s)$ can be written in the form
\begin{displaymath}
\textrm{$p$ \multider{\sigma'} $\overline{t}$ \prsoneder{r'}
$t$$\parallel$$Y.(s),\quad\quad\quad$ with $r'\in\Re$ and
$|\sigma'|=|\sigma|-1$.}
\end{displaymath}
There are three cases:
\begin{itemize}
\item $\overline{t}=t$$\parallel$$Y.(\overline{\overline{t}})$, with
  $\overline{\overline{t}}$ \prsoneder{r'} $s$. It immediately follows
  that $\overline{\overline{t}}\neq\varepsilon$.  By inductive
  hypothesis, $p$ \multider{\sigma'}
  $t$$\parallel$$Y.(\overline{\overline{t}})$ can be written in the
  form $p$ \multider{\sigma_{1}} $t'$$\parallel$$Z$ \prsoneder{r}
  $t'$$\parallel$$Y.(Z')$ \multider{\rho_{2}}
  $t$$\parallel$$Y.(\overline{\overline{t}})$, with
  $r=Z$\Rule{a}$Y.(Z')$.  Moreover, he have that $Z'$
  \multider{\rho_{2}'} $\overline{\overline{t}}$ and $t'$
  \multider{\rho_{2}''} $t$, with
  $\rho_{2}\in{}Interleaving(\rho_{2}',\rho_{2}'')$.  As a
  consequence, we have that $p$ \multider{\sigma}
  $t$$\parallel$$Y.(s)$ can be written in the form $p$
  \multider{\sigma_{1}} $t'$$\parallel$$Z$ \prsoneder{r}
  $t'$$\parallel$$Y.(Z')$ \multider{\sigma_{2}} $t$$\parallel$$Y.(s)$,
  with $\sigma_{2}=\rho_{2}r'$.  Moreover, taking
  $\sigma_{2}'=\rho_{2}'r'$ and $\sigma_{2}''=\rho_{2}''$ we have that
  $Z'$ \multider{\sigma_{2}'} $s$, $t'$ \multider{\sigma_{2}''} $t$
  and $\sigma_{2}\in{}Interleaving(\sigma_{2}',\sigma_{2}'')$.  The
  first part of the assertion is proved.  \newline We consider now
  Property~\textbf{A}.  By inductive hypothesis, the subderivation of
  $t'$$\parallel$$Y.(Z')$ \multider{\rho_{2}}
  $t$$\parallel$$Y.(\overline{\overline{t}})$ \multider{r'\omega} from
  $Z'$ can be written in the form $Z'$ \multider{\rho_{2}'}
  $\overline{\overline{t}}$ \multider{\eta}, where
  $\overline{\overline{t}}$ \multider{\eta} is the subderivation of
  $t$$\parallel$$Y.(\overline{\overline{t}})$ \multider{r'\omega} from
  $\overline{\overline{t}}$. Notice that $\overline{\overline{t}}$
  \multider{\eta} can be written in the form $\overline{\overline{t}}$
  \prsoneder{r'} $s$ \multider{\omega'}, where $s$ \multider{\omega'}
  is the subderivation of $t$$\parallel$$Y.(s)$ \multider{\omega} from
  $s$.  Considering that $\sigma_{2}'=\rho_{2}'r'$, the thesis holds.
\item $\overline{t}=t$$\parallel$$Z$, $r'=Z$\Rule{a}$Y.(Z')$ and
  $s=Z'$.  In this case the first part of the assertion holds taking
  $\sigma_{1}=\sigma'$, $\sigma_{2}= \varepsilon$ and $r=r'$. As far
  as Property~\textbf{A} is concerned, it suffices to observe that
  $\sigma_{2}$ is the empty sequence.
    \item $\overline{t}=\overline{\overline{t}}$$\parallel$$Y.(s)$,
    with
    $\overline{\overline{t}}$ \prsoneder{r'} $t$. By inductive hypothesis, $p$
     \multider{\sigma'}
    $\overline{\overline{t}}$$\parallel$$Y.(s)$ can be written in the form
    $p$ \multider{\sigma_{1}} $t'$$\parallel$$Z$ \prsoneder{r} $t'$$\parallel$$Y.(Z')$
    \multider{\rho_{2}}
    $\overline{\overline{t}}$$\parallel$$Y.(s)$, with
    $r=Z$\Rule{a}$Y.(Z')$.
    Moreover, it holds that
     $Z'$ \multider{\rho_{2}'} $s$  and  $t'$ \multider{\rho_{2}''} $\overline{\overline{t}}$,
    with
    $\rho_{2}\in{}Interleaving(\rho_{2}',\rho_{2}'')$.
    As a consequence, it holds that  $p$ \multider{\sigma}
    $t$$\parallel$$Y.(s)$ can be written in the form
    $p$ \multider{\sigma_{1}} $t'$$\parallel$$Z$
    \prsoneder{r} $t'$$\parallel$$Y.(Z')$ \multider{\sigma_{2}} $t$$\parallel$$Y.(s)$,
    with
    $\sigma_{2}=\rho_{2}r'$. Moreover, taking  $\sigma_{2}'=\rho_{2}'$
    and
    $\sigma_{2}''=\rho_{2}''r'$, it holds that $Z'$ \multider{\sigma_{2}'} $s$,
    $t'$ \multider{\sigma_{2}''}
     $t$  and $\sigma_{2}\in{}Interleaving(\sigma_{2}',\sigma_{2}'')$.
    This proves the first part of the assertion. We consider now the
     property \textbf{A}. By inductive hypothesis,
    the subderivation of
    $t'$$\parallel$$Y.(Z')$ \multider{\rho_{2}} $\overline{\overline{t}}$$\parallel$$Y.(s)$
     \multider{r'\omega} from $Z'$ can be written in the form
      $Z'$ \multider{\rho_{2}'} $s$ \multider{\eta}, where  $s$
     \multider{\eta} is the subderivation
    of $\overline{\overline{t}}$$\parallel$$Y.(s)$ \multider{r'\omega}
    from $s$. Now,  $s$ \multider{\eta} is also the subderivation of
    $t$$\parallel$$Y.(s)$ \multider{\omega}  from $s$.
    Considering that $\sigma_{2}'=\rho_{2}'$, the thesis holds.\qedhere
\end{itemize}
\end{description}
\end{proof}


\setcounter{equation}{0}
\begin{Lemma}\label{Lemma:From-R-To-RPAR}
 If $p$ \multider{\sigma} $t$$\parallel$$p'$,
  with $p,p'\in{}T_{PAR}$, then the following properties hold:
\begin{description}
\item[A.] There exists a term $s{}\in{}T_{PAR}$ such that $p$
  \multiderpar{\rho} $s$$\parallel$$p'$ where $s$ \multider{\mu} $t$,
  $\mu$ is non--accepting if $\sigma$ is not accepting, and
  $s=\varepsilon$ if $t=\varepsilon$.
\item[B.] There exists a term $s{}\in{}T_{PAR}$ such that $p$
  \multiderpar{\rho} $s$$\parallel$$p'$ where $|\rho|>0$ if
  $|\sigma|>0$, $s=\varepsilon$ if $t=\varepsilon$, and $\rho$ is
  accepting $\mathrm{(}$resp., non--accepting$\mathrm{)}$ if $\sigma$
  is accepting $\mathrm{(}$resp., non--accepting$\mathrm{)}$.
\end{description}
\end{Lemma}
\begin{proof}
  The proof is by induction on the length of finite derivations $p$
  \multider{\sigma} in $\Re$ from terms in $T_{PAR}$.
\begin{description}
\item[\textbf{Base Step}] $|\sigma|=0$. In this case the assertion is
  obvious.
\item[\textbf{Induction Step}] $|\sigma|>0$. The derivation $p$
  \multider{\sigma} can be written in the form
\begin{equation}
\textrm{$p$ \prsoneder{r} $\overline{t}$ \multider{\sigma'}
$t$$\parallel$$p'$ $\quad$ }
\end{equation}
with $r\in\Re$, and $|\sigma'|<|\sigma|$. There are two cases:
\begin{enumerate}
\item r is a PAR rule. Then, we have that $\overline{t}\in{}T_{PAR}$
  and $r\in\Re_{PAR}$.  Moreover, it holds that $r\in{}\Re_{F}$ iff
  $r\in{}\Re_{PAR,F}$.  Let us consider property \textbf{A}.  By
  inductive hypothesis, there exists a $s\in{}T_{PAR}$ such that
  $\overline{t}$ \multiderpar{\rho} $s$$\parallel$$p'$ where $s$
  \multider{\mu} $t$, $\mu$ is non--accepting if $\sigma'$ is not
  accepting, and $s=\varepsilon$ if $t=\varepsilon$. Therefore, it
  holds that $p$ \onederpar{r} $\overline{t}$ \multiderpar{\rho}
  $s$$\parallel$$p'$, where $s$ \multider{\mu} $t$, $\mu$ is non
  accepting if $\sigma$ is not accepting, and $s=\varepsilon$ if
  $t=\varepsilon$.  Let us consider \textbf{B}. By inductive
  hypothesis, there exists a $s\in{}T_{PAR}$ such that $\overline{t}$
  \multiderpar{\rho'} $s$$\parallel$$p'$ where $\rho'$ is accepting
  (resp., non--accepting) if $\sigma'$ is accepting (resp., non
  accepting), and $s=\varepsilon$ if $t=\varepsilon$. Therefore, it
  holds that $p$ \onederpar{r} $\overline{t}$ \multiderpar{\rho'}
  $s$$\parallel$$p'$, where $r\rho'$ is accepting (resp., non
  accepting) if $\sigma=r\sigma'$ is accepting (resp.,
  non--accepting), and $s=\varepsilon$ if $t=\varepsilon$. Thus, the
  assertion is proved.
\item $r=Z$\Rule{a}$Y.(Z')$. In this case, we have
  $p=p''$$\parallel$$Z$ and $\overline{t}=p''$$\parallel$$Y.(Z')$,
  with $p''\in{}T_{PAR}$.  From Equation (1), let $Z'$
  \multider{\lambda} $t_{1}$ be the subderivation of
  $\overline{t}=p''$$\parallel$$Y.(Z')$ \multider{\sigma'} from $Z'$.
  By Lemma \ref{Lemma:Subderivations1}, we can distinguish three subcases:
    \begin{itemize}
    \item $t_{1}\neq\varepsilon$ and $p''$ \multider{\sigma'\setminus\lambda}
      $t'$. Moreover, we have that
      $t$$\parallel$$p'=t'$$\parallel$$Y.(t_{1})$,
      $t'=p'$$\parallel$$t''$, for some term $t''$, and
      $t=t''$$\parallel$$Y.(t_{1})$ (in particular,
      $t\neq\varepsilon$). Let us consider Property \textbf{A}. Since
      $|\sigma'\setminus\lambda|<|\sigma|$, by inductive hypothesis, there
      exists a term $s\in{}T_{PAR}$ such that $p''$
      \multiderpar{\rho'} $s$$\parallel$$p'$, where $s$ \multider{\mu}
      $t''$, and $\mu$ is non--accepting if $\sigma'\setminus\lambda$ is not
      accepting.  Therefore, we have $p''$$\parallel$$Z$
      \multiderpar{\rho'} $s$$\parallel$$Z$$\parallel$$p'$, where
      $s$$\parallel$$Z$ \multider{\mu} $t''$$\parallel$$Z$
      \prsoneder{r} $t''$$\parallel$$Y.(Z')$ \multider{\lambda}
      $t''$$\parallel$$Y.(t_{1})=t$, and $\mu{}r\lambda$ is non
      accepting if $\sigma$ is not accepting, thus proving the
      assertion.  \newline Let us consider now Property \textbf{B}.
      By inductive hypothesis there exists a term $s\in{}T_{PAR}$ such
      that $p''$ \multiderpar{\rho'} $s$$\parallel$$p'$ where $\rho'$
      is accepting (resp., non--accepting) if $\sigma'\setminus\lambda$ is
      accepting (resp., non--accepting). Now, by definition of
      $\Re_{PAR}$, it holds that $r'=Z\rightarrow\hat{Z}\in\Re_{PAR}$
      with $\hat{Z}\in\{Z_{ACC},Z_{NOT\_ACC}\}$, and $\hat{Z}=Z_{ACC}$
      (resp., $\hat{Z}=Z_{NOT\_ACC}$) if $r\lambda$ is accepting
      (resp., non--accepting).  Then, we have that
      $p=p''$$\parallel$$Z$ \onederpar{r'} $p''$$\parallel$$\hat{Z}$
      \multiderpar{\rho'} $s$$\parallel$$p'$$\parallel$$\hat{Z}$ where
      $r'\rho'$ is accepting (resp., non--accepting) if
      $r\lambda(\sigma'\setminus\lambda)$ is accepting (resp., non--accepting).
      Since $\sigma$ is a reordering of $r\lambda(\sigma'\setminus\lambda)$,
      we obtain the assertion.
    \item $t_{1}=\varepsilon$ and the derivation
      $p''$$\parallel$$Y.(Z')$ \multider{\sigma'} $t$$\parallel$$p'$
      can be written in the following form:
    \begin{equation}
     \textrm{$p''$$\parallel$$Y.(Z')$
     \multider{\sigma_{1}} $t'$$\parallel$$Y$ \multider{\sigma_{2}} $t$$\parallel$$p',\quad$}
     \end{equation}
     with $p''$ \multider{\sigma'_{1}} $t'$, and
     $\sigma_{1}\in{}Interleaving(\lambda,\sigma'_{1})$.  Now, it
     holds that $Z'$ \multider{\lambda} $\varepsilon$, with
     $|\lambda|<|\sigma|$. By inductive hypothesis, we have $Z'$
     \multiderpar{\rho} $\varepsilon$ where $\rho$ is accepting
     (resp., non--accepting) if $\lambda$ is accepting (resp., non
     accepting).  By Remark \ref{remark:2}, it follows that
     $r'=Z$\Rule{c}$Y\in\Re_{PAR}$, with $c=\$$
     (resp., $c=\#$) if
     $r\lambda$ is accepting (resp., non--accepting). So, it holds that
     $r'\in\Re_{PAR,F}$ (resp., $r'\notin\Re_{PAR,F}$) if $r\lambda$
     is accepting (resp., non--accepting).  Now, $p''$$\parallel$$Y$
     \multider{\sigma'_{1}} $t'$$\parallel$$Y$ \multider{\sigma_{2}}
     $t$$\parallel$$p'$ with $|\sigma'_{1}\sigma_{2}|<|\sigma|$.  Let
     us consider Property \textbf{A}.  By inductive hypothesis, there
     exists a $s\in{}T_{PAR}$ such that $p''$$\parallel$$Y$
     \multiderpar{\rho'} $s$$\parallel$$p'$, where $s$ \multider{\mu}
     $t$, $\mu$ is non--accepting if $\sigma'_{1}\sigma_{2}$ is not
     accepting, and $s=\varepsilon$ if $t=\varepsilon$.  So, we have
     $p=p''$$\parallel$$Z$ \onederpar{r'} $p''$$\parallel$$Y$
     \multiderpar{\rho'} $s$$\parallel$$p'$, where $s$ \multider{\mu}
     $t$, $\mu$ is non--accepting if $\sigma$ is not accepting, and
     $s=\varepsilon$ if $t=\varepsilon$.  Thus, the
     assertion is proved.  Let us consider Property \textbf{B}. By
     inductive hypothesis, there exists a $s\in{}T_{PAR}$ such that
     $p''$$\parallel$$Y$ \multiderpar{\rho'} $s$$\parallel$$p'$, where
     $\rho'$ is accepting (resp., non--accepting) if
     $\sigma'_{1}\sigma_{2}$ is accepting (resp., non--accepting), and
     $s=\varepsilon$ if $t=\varepsilon$.  Therefore, we have
     $p=p''$$\parallel$$Z$ \onederpar{r'} $p''$$\parallel$$Y$
     \multiderpar{\rho'} $s$$\parallel$$p'$ with $r'\rho'$ accepting
     (resp., non--accepting) if $r\lambda\sigma'_{1}\sigma_{2}$ is
     accepting (resp., non--accepting), and $s=\varepsilon$ if
     $t=\varepsilon$. Since $\sigma$ is a reordering of
     $r\lambda\sigma'_{1}\sigma_{2}$, the assertion is proved.
   \item $t_{1}=W\in{}Var$ and the derivation $p''$$\parallel$$Y.(Z')$
     \multider{\sigma'} $t$$\parallel$$p'$ can be written in the form
      \begin{eqnarray}
       \textrm{$p''$$\parallel$$Y.(Z')$
                \multider{\sigma_{1}} $t'$$\parallel$$Y.(W)$
                \prsoneder{r'} $t'$$\parallel$$W'$ \multider{\sigma_{2}}
                $t$$\parallel$$p'$,}
     \end{eqnarray}
     with $p''$ \multider{\sigma'_{1}} $t'$, $r'=Y.(W)$\Rule{b}$W'$
     and $\sigma_{1}\in{}Interleaving(\lambda,\sigma'_{1})$.  Now, we
     have that $Z'$ \multider{\lambda} $W$, with $|\lambda|<|\sigma|$.
     By inductive hypothesis, it holds that $Z'$ \multiderpar{\rho}
     $W$ with $\rho$ accepting (resp., non--accepting) if $\lambda$ is
     accepting (resp., non--accepting).  Now,
     $r=Z$\Rule{a}$Y.(Z')\in\Re$ and $r'=Y.(W)$\Rule{b}$W'\in\Re$.  By
     remark \ref{remark:2}, it follows that
     $r''=Z$\Rule{c}$W'\in\Re_{PAR}$, with $c=\$$
     (resp., $c=\#$) if
     $rr'\lambda$ is accepting (resp., non--accepting).  So, it follows
     that $r''\in\Re_{PAR,F}$ (resp., $r'\notin\Re_{PAR,F}$) if
     $rr'\lambda$ is accepting (resp., non--accepting).  Now, we have a
     derivation $p''$$\parallel$$W'$ \multider{\sigma'_{1}}
     $t'$$\parallel$$W'$ \multider{\sigma_{2}} $t$$\parallel$$p'$,
     with $|\sigma'_{1}\sigma_{2}|<|\sigma|$.  Let us consider
     Property \textbf{A}.  By inductive hypothesis, there exists a
     term $s\in{}T_{PAR}$ such that $p''$$\parallel$$W'$
     \multiderpar{\rho'} $s$$\parallel$$p'$, where $s$ \multider{\mu}
     $t$, $\mu$ is non--accepting if $\sigma'_{1}\sigma_{2}$ is not
     accepting, and $s=\varepsilon$ if $t=\varepsilon$.  So, we have
     $p=p''$$\parallel$$Z$ \onederpar{r''} $p''$$\parallel$$W'$
     \multiderpar{\rho'} $s$$\parallel$$p'$, where $s$ \multider{\mu}
     $t$, $\mu$ is non--accepting if $\sigma$ is not accepting, and
     $s=\varepsilon$ if $t=\varepsilon$, thus proving the assertion.
     Let us consider Property \textbf{B}.  By inductive hypothesis,
     there exists a term $s\in{}T_{PAR}$ such that
     $p''$$\parallel$$W'$ \multiderpar{\rho'} $s$$\parallel$$p'$, with
     $\rho'$ accepting (resp., non--accepting) if
     $\sigma'_{1}\sigma_{2}$ is accepting (resp., non--accepting), and
     $s=\varepsilon$ if $t=\varepsilon$. As a consequence, we have
     $p=p''$$\parallel$$Z$ \onederpar{r''} $p''$$\parallel$$W'$
     \multiderpar{\rho'} $s$$\parallel$$p'$ with $r''\rho'$ accepting
     (resp., non--accepting) if $rr'\lambda\sigma'_{1}\sigma_{2}$ is
     accepting (resp., non--accepting), and $s=\varepsilon$ if
     $t=\varepsilon$. Since $\sigma$ is a reordering of
     $rr'\lambda\sigma'_{1}\sigma_{2}$, the assertion is proved.\qedhere
    \end{itemize}
\end{enumerate}
\end{description}
\end{proof}

\begin{Lemma}\label{Lemma:From-R-To-RPAR-Inf1}
 For $p\in{}T_{PAR}$, let $p$ \multider{\sigma} be an accepting infinite
 derivation $\mathrm{(}$resp., an infinite derivation devoid of accepting rules,
 an  infinite derivation containing a finite non--null number of
 accepting rules$\mathrm{)}$ in $\Re$ from $p$ belonging to
  $\Pi_{PAR}$ $\mathrm{(}$resp., $\Xi_{PAR}$$\mathrm{)}$. Then,
   there exists an  accepting infinite
 derivation $\mathrm{(}$resp., an infinite derivation devoid of accepting rules,
 an  infinite derivation containing a finite non--null number of
 accepting rules$\mathrm{)}$ in $\Re_{PAR}$ from $p$.
 \end{Lemma}
\setcounter{equation}{0}

\begin{proof}
We give the proof in the case in which $p$
\multider{\sigma} is an accepting infinite derivation belonging to
 $\Pi_{PAR}$. The proof for the other two cases is similar. We have to
prove that there exists an accepting infinite derivation in
$\Re_{PAR}$ from $p$. First of all, we show that there
  exists a term
 $p'\in{}T_{PAR}$ satisfying the following conditions:
 \begin{enumerate}
    \item $p$ \multiderpar{\rho}
    $p',\quad$ with $\rho$ accepting;
    \item There exists an accepting  infinite derivation in $\Re$
      from $p'$ belonging to $\Pi_{PAR}$.
 \end{enumerate}
 The derivation
    $p$ \multider{\sigma} can be written in the form
\begin{equation}
\label{eqC5.1}
\textrm{$p$ \multider{\xi} $t_{1}$ \prsoneder{r}
 $t_{2}$ \multider{\lambda}}
\end{equation}
 where $r$ is an accepting rule in $\Re$  and $t_{2}$ \multider{\lambda}
 is an accepting infinite derivation. By Property 1 of Lemma
 \ref{Lemma:Subderivations2}, the derivation $t_{2}$ \multider{\lambda} belongs to $\Pi_{PAR}$.\\
Let us consider the case in which
  $r$ is a PAR rule applied at  level zero in the
    one--step derivation $t_{1}$ \prsoneder{r} $t_{2}$. In this case,
we have that
    $t_{1}=t$$\parallel$$s$, $t_{2}=t$$\parallel$$s'$, with
    $s,s'\in{}T_{PAR}$ and $r=s\rightarrow{}s'$.
    By Property \textbf{A} of Lemma \ref{Lemma:From-R-To-RPAR} applied to the
     derivation $p$ \multider\xi{}
    $t_{1}=t$$\parallel$$s$, there exists a term $\overline{t}\in{}T_{PAR}$
    such that $p$ \multiderpar{} $\overline{t}$$\parallel$$s$ and
    $\overline{t}$ \multider{} $t$.  Then, we have a derivation $p$
    \multiderpar{} $\overline{t}$$\parallel$$s$ \onederpar{r}
    $\overline{t}$$\parallel$$s'$, where $r$ is an accepting rule
    in $\Re_{PAR}$. By taking $p'=\overline{t}$$\parallel$$s'\in{}T_{PAR}$,
     we obtain $p$ \multiderpar{\rho} $p'$, with $\rho$
    accepting. Moreover, the following derivation from $p'$
    \begin{equation}
\label{eqC5.2}
    \textrm{$p'=\overline{t}$$\parallel$$s'$ \multider{}
    $t$$\parallel$$s'=t_{2}$ \multider{\lambda}}
    \end{equation}
    is an  infinite accepting derivation. Considering that  $t_{2}$
    \multider{\lambda} is in $\Pi_{PAR}$, from Property 2 of Lemma
    \ref{Lemma:Subderivations2} it follows that the derivation
of Eq. \ref{eqC5.2} is in $\Pi_{PAR}$.
    As a consequence, we have that $p'$  satisfies the desired properties.
\newline
    Let us consider now the case in which
  $r$ is not a PAR rule applied at level zero in the
    derivation $t_{1}$ \prsoneder{r} $t_{2}$. Then, we deduce that
    $t_{1}=t$$\parallel$$w$, $t_{2}=t$$\parallel$$X.(s)$, with $w$
    \prsoneder{r} $X.(s)$ (with $s$ possibly equal to $\varepsilon$).
    Moreover, either $w=X.(s')$ with $s'$ \prsoneder{r} $s$,
        or
        $r=w$\Rule{a}$X.(s)$ and $r$ is a $SEQ$ rule. 
Let us consider the first case.  (The second case can be
dealt with analogously.) From Lemma \ref{Lemma:Base},
        applied to the  derivation
        $p$ \multider{\xi}
        $t_{1}=t$$\parallel$$X.(s')$, it follows that
        there are two variables $Z,Z'$ such that
    \begin{equation}
    \textrm{$Z$\Rule{a}$X.(Z')\in\Re$,
        $p$ \multider{} $t$$\parallel$$Z$, and
        $Z'$ \multider{} $s'$.}
    \end{equation}
        By Property \textbf{A} of Lemma \ref{Lemma:From-R-To-RPAR}, applied to the
        derivation
        $p$ \multider{} $t$$\parallel$$Z$,  there exists
        a term $\overline{t}\in{}T_{PAR}$ such that
    \begin{equation}
    \textrm{$p$ \multiderpar{} $\overline{t}$$\parallel$$Z$ and
        $\overline{t}$ \multider{} $t$.}
    \end{equation}
   With reference to Eq. \ref{eqC5.1}, let $s$ \multider{\lambda'} be the subderivation
    of $t_{2}=t$$\parallel$$X.(s)$
    \multider{\lambda} from $s$. Notice  that $s$ \multider{\lambda'}
    is not an  accepting infinite derivation. By  
Lemma \ref{Lemma:Subderivations1},
    we distinguish the following cases:
    \begin{itemize}
        \item $t$
        \multider{\lambda\setminus \lambda'}, and this derivation is in $\Pi_{PAR}$.
        Since   $s$ \multider{\lambda'}
         is not an  accepting infinite derivation, it follows  that
         $t$
        \multider{\lambda\setminus\lambda'} is an accepting infinite derivation
        (belonging to $\Pi_{PAR}$). From Eq.(3), we have that
         $Z'$ \multider{} $s'$ \prsoneder{r} $s$. From the definition
of $\Re_{PAR}$,  we
         have  that
        $r'=Z\rightarrow{}Z_{ACC}\in\Re_{PAR}$ (namely, an
        accepting rule in $\Re_{PAR}$).
        By Eq. (4), we have a derivation $p$
        \multiderpar{} $\overline{t}$$\parallel$$Z$ \onederpar{r'}
        $\overline{t}$$\parallel$$Z_{ACC}$. Taking 
        $p'=\overline{t}$$\parallel$$Z_{ACC}$, we obtain that
        $p$ \multiderpar{\rho} $p'$, with $\rho$
        accepting. Moreover, from Eq. (4) we have a derivation
     \begin{equation}
      \textrm{$p'=\overline{t}$$\parallel$$Z_{ACC}$ \multider{}
        $t$$\parallel$$Z_{ACC}$ \multider{\lambda\setminus\lambda'}}
     \end{equation}
        which is an  infinite accepting derivation. 
Considering that the derivation  $t$
        \multider{\lambda\setminus\lambda'} belongs to
        $\Pi_{PAR}$, by properties 2 and 3 of Lemma \ref{Lemma:Subderivations2}
        it follows that the derivation of Eq. (5) belongs to $\Pi_{PAR}$.\\
         This shows that  $p'$  satisfies the required properties.
        \item $s$ \multider{\lambda'}
        leads to term $st\neq\varepsilon$, and the second condition of Property 3
        of Lemma \ref{Lemma:Subderivations1} holds.
        Therefore, the derivation $t$$\parallel$$X.(s)$ \multider{\lambda}
        can be written in the form
        \begin{equation}
        \textrm{$t$$\parallel$$X.(s)$ \multider{\lambda_{1}}
            $\overline{\overline{t}}$$\parallel$$Y$ \multider{\lambda_{2}} $\quad$
            with $Y\in{}Var$ and}
        \end{equation}
        \begin{equation}
        \textrm{$t$ \multider{\lambda_{1}'} $\overline{\overline{t}}$,
           $X.(s)$ \multider{\lambda''} $Y$ with $\lambda'_{1}$ subsequence of $\lambda_{1}$}
        \end{equation}
        By Property 1 of Lemma \ref{Lemma:Subderivations2},
the derivation  $\overline{\overline{t}}$$\parallel$$Y$ \multider{\lambda_{2}} belongs to
        $\Pi_{PAR}$.\\
        By Eq. (3) and Eq. (7), we have an accepting derivation
        $Z$ \multider{} $X.(s')$ \prsoneder{r} $X.(s)$
        \multider{\lambda''} $Y$. By Property \textbf{B} of 
Lemma \ref{Lemma:From-R-To-RPAR},
         we obtain
        \begin{equation}
        \textrm{$Z$ \multiderpar{\eta} $Y$ with $\eta$
        accepting}
        \end{equation}
        From Eq. (4) and Eq. (8), we have a derivation $p$ \multiderpar{}
        $\overline{t}$$\parallel$$Z$ \multiderpar{\eta}
        $\overline{t}$$\parallel$$Y$. Taking
        $p'=\overline{t}$$\parallel$$Y$, we obtain
        $p$ \multiderpar{\rho} $p'$, with $\rho$
        accepting. Moreover, from Eq. (4), (6) and (7) we have an 
infinite accepting derivation
        \begin{equation}
        \textrm{
            $p'=\overline{t}$$\parallel$$Y$ \multider{}
            $t$$\parallel$$Y$ \multider{\lambda_{1}'} $\overline{\overline{t}}$$\parallel$$Y$
            \multider{\lambda_{2}}}
        \end{equation}
        Considering that the derivation $\overline{\overline{t}}$$\parallel$$Y$
        \multider{\lambda_{2}} belongs to $\Pi_{PAR}$, from Property 2 of Lemma
        \ref{Lemma:Subderivations2} it follows that the derivation of Eq. (9) 
belongs to $\Pi_{PAR}$.\\
         This shows that  $p'$  satisfies the required properties.
        \item $s$ \multider{\lambda'}
        leads to $\varepsilon$, and the derivation $t$$\parallel$$X.(s)$ \multider{\lambda}
        can be written in the form
        \begin{equation}
        \textrm{$t$$\parallel$$X.(s)$ \multider{\lambda_{1}}
            $\overline{\overline{t}}$$\parallel$$X$ \multider{\lambda_{2}}}
\end{equation}
where 
$t$ \multider{\lambda_{1}'} $\overline{\overline{t}}$, with  $\lambda_{1}'$
            subsequence
            of $\lambda_{1}$.
        
        Moreover, from Property 1 of Lemma \ref{Lemma:Subderivations2}, the
derivation
        $\overline{\overline{t}}$$\parallel$$X$ \multider{\lambda_{2}} belongs to
        $\Pi_{PAR}$.\\
        From Eq. (3) we have an  accepting derivation
        $Z$ \multider{} $X.(s')$ \prsoneder{r} $X.(s)$
        \multider{\lambda'} $X$.
        From  Property \textbf{B} of Lemma \ref{Lemma:From-R-To-RPAR}
         we obtain a derivation
        \begin{equation}
        \textrm{$Z$ \multiderpar{\eta} $X$ with $\eta$
        accepting}
        \end{equation}
        From Eq. (4) and Eq. (11), we have a derivation $p$ \multiderpar{}
        $\overline{t}$$\parallel$$Z$ \multiderpar{\eta}
        $\overline{t}$$\parallel$$X$. Taking
        $p'=\overline{t}$$\parallel$$X$, we obtain a derivation
        $p$ \multiderpar{\rho} $p'$, with $\rho$
        accepting. Moreover, from Eq. (4) and Eq. (10)
we have the following  infinite accepting derivation
        \begin{equation}
        \textrm{
            $p'=\overline{t}$$\parallel$$Y$ \multider{}
            $t$$\parallel$$X$ \multider{\lambda_{1}'} $\overline{\overline{t}}$$\parallel$$X$
            \multider{\lambda_{2}}}.
        \end{equation}
        
Considering that the derivation $\overline{\overline{t}}$$\parallel$$X$
        \multider{\lambda_{2}}  belongs to $\Pi_{PAR}$,  from Property 2 of Lemma
        \ref{Lemma:Subderivations2} it follows that the derivation of Eq. (12) 
belongs to $\Pi_{PAR}$.\\
         This proves that  $p'$  satisfies the required properties.
    \end{itemize}
    
    Now, by exploiting Properties 1 and 2, we can prove the thesis of
    the lemma.  By Properties 1 and 2, it follows that there exists a
    sequence of terms in $T_{PAR}$, $(p_{n})_{n\in{}N}$, satisfying
    the following properties:
 \begin{description}
    \item[i.] $p_{0}=p$;
    \item[ii.]  $p_{n}$ \multiderpar{\rho_{n}}
    $p_{n+1}\quad$, with $\rho_{n}$ accepting, for all $n\in{}N$;
    \item[iii.] there exists an accepting infinite derivation  in $\Re$
    from $p_{n}$ belonging to  $\Pi_{PAR}$, for all $n\in{}N$ .
 \end{description}
The existence of such a sequence $(p_{n})_{n\in{}N}$ immediately
implies the thesis.
\end{proof}


\begin{Lemma}\label{Lemma:From-R-To-RPAR-Inf2}
  Let $X\in{}Var$ and $X$ \multider{\sigma} be an infinite accepting
  derivation $\mathrm{(}$resp., an infinite derivation devoid of
  accepting rules, an infinite derivation containing a finite
  non--null number $n$ of accepting rule occurrences$\mathrm{)}$ in
  $\Re$ from $X$.  Then, one of the following conditions is satisfied:
\begin{enumerate}
\item there exists a variable $Y$ reachable $\mathrm{(}$resp.,
  reachable through a non--accepting derivation, reachable$\mathrm{)}$
  from $X$ in $\Re_{SEQ}$, and there exists an accepting infinite
  derivation $\mathrm{(}$resp., an infinite derivation devoid of
  accepting rules, an infinite derivation containing a finite
  non--null number of accepting rule occurrences$\mathrm{)}$ in
  $\Re_{PAR}$ from $Y$.
\item there exists a term $t\in{}T_{SEQ}\setminus\{\varepsilon\}$ with
  $t=X_{1}.(X_{2}.(\ldots{}X_{k}.(Y)\ldots{}))$ $\mathrm{(}$with
  $k\geq0$$\mathrm{)}$ such that $X$ \multiderseq{\rho} $t$, with
  $\rho$ accepting $\mathrm{(}$resp., non--accepting,
  accepting$\mathrm{)}$ in $\Re_{SEQ}$, and there exists an accepting
  infinite derivation $\mathrm{(}$resp., an infinite derivation devoid
  of accepting rules, an infinite derivation containing a finite
  number $m$, with $0\leq{}m<n$, of accepting rule
  occurrences$\mathrm{)}$ in $\Re$ from $Y$.
\end{enumerate}
\end{Lemma}

\begin{proof}
  We give the proof for the case where $X$ \multider{\sigma} is an
  infinite accepting derivation (the proof for the other two cases is
  similar). We have to prove that one of the following properties is
  satisfied:
\begin{description}
\item[A] there exists a variable $Y$ reachable from $X$ in
  $\Re_{SEQ}$, and there exists an accepting infinite derivation in
  $\Re_{PAR}$ from $Y$.
\item[B] there exists a term $t\in{}T_{SEQ}\setminus\{\varepsilon\}$ with
  $t=X_{1}.(X_{2}.(\ldots{}X_{k}.(Y)\ldots{}))$ $\mathrm{(}$with
  $k\geq0$$\mathrm{)}$ such that $X$ \multiderseq{\rho} $t$, with
  $\rho$ accepting in $\Re_{SEQ}$, and there exists an accepting
  infinite derivation in $\Re$ from $Y$.
\end{description}
The proof is by induction on the level $k$ of application of the first
occurrence of an accepting rule $r$, in an infinite accepting
derivation in $\Re$ from a variable.

\begin{description}
\item[Base Step] $k=0$. If $X$ \multider{\sigma} belongs to the class
  $\Pi_{PAR}$, from Lemma~\ref{Lemma:From-R-To-RPAR-Inf1},
  Property~\textbf{A} follows, setting $Y=X$. Otherwise, from
  Lemma~\ref{Lemma:Base}, it follows that the derivation
  $\prsder{X}{\sigma}{}$ can be written in the form
  $$\prsder{X}{\sigma_{1}}{\prssimpder{\parcomp{t}{Z}}{r'}
    {\prsder{\parcomp{t}{Y.(Z')}}{\sigma_{2}}{}}}$$
  where $r'=Z$\Rule{a}$Y.(Z')$, and the subderivation of
  $\prsder{\parcomp{t}{Y.(Z')}}{\sigma_{2}}{}$ from $Z'$, namely
  $\prsder{Z'}{\sigma_{2}'}{}$, is an infinite accepting derivation.
  By noticing that every rule occurrence in $\sigma_{2}'$ is applied
  to level greater then zero in $\prsder{X}{\sigma}{}$, and that we
  are considering the case where $k=0$, it follows that $r$ must
  occurr in the rule sequence $\sigma_{1}r'\xi$, where $\xi =
  \sigma_{2}\setminus\sigma_{2}'$.  From
  Lemma~\ref{Lemma:Subderivations1}, we have that $\prsder{t}{\xi}{}$.
  Therefore, there exists in $\Re$ a derivation of the form
  $\prsder{X}{\lambda}{\prssimpder{\parcomp{t'}{Z}}{r'}{\parcomp{t'}{Y.(Z')}}}$,
  with $\lambda$ accepting, if $r'$ is not accepting.
  From Property~\textbf{B} 
  of Lemma~\ref{Lemma:From-R-To-RPAR} applied to the derivation
  $\prsder{X}{\lambda}{\parcomp{t'}{Z}}$, there exists a term
  $p\in{}T_{PAR}$, and a derivation
  $\prsderpar{X}{\rho}{\parcomp{p}{Z}}$, with $\rho$ accepting, if
  $\lambda$ is accepting.
  From the definition of $\Re_{SEQ}$, we have that
  $\prsderseq{X}{\mu}{Z\,\onederseq{r'}Y.(Z')}$, with $\mu$ accepting
  if $\rho$ is accepting. 
  Now, either $r'$ is an accepting rule in $\Re$, and is an accepting
  rule in $\Re_{SEQ}$ as well, or $\lambda$ is accepting, and $\mu$ is
  accepting as well. Therefore, variable $Z'$ is reachable from $X$ in
  $\Re_{SEQ}$ through an accepting derivation, and there exists an
  accepting infinite derivation in $\Re$ from $Z'$. This is exactly
  what Property~\textbf{B} states.
\item[Induction Step] $k>0$. If $\prsder{X}{\sigma}{}$ belongs to the
  class $\Pi_{PAR}$, from Lemma~\ref{Lemma:From-R-To-RPAR-Inf1},
  Property~\textbf{A} follows, by setting $Y = X$. 
  
  Otherwise, by Lemma~\ref{Lemma:Base}, the derivation
  $\prsder{X}{\sigma}{}$ can be written in the form
$$
\prsder{X}{\sigma_{1}}{\prssimpder{\parcomp{t}{Z}}{r'}{\prsder{\parcomp{t}{Y.(Z')}}{\sigma_{2}}{}}}
$$
where $r'=Z$\Rule{a}$Y.(Z')$, and the subderivation of
$\prsder{\parcomp{t}{Y.(Z')}}{\sigma_{2}}{}$ from $Z'$, namely
$\prsder{Z'}{\sigma_{2}'}{}$ is an infinite accepting derivation. Let
$\xi$ be the rule sequence $\sigma_{2}\setminus\sigma_{2}'$. There can
be two cases:
\begin{itemize}
\item the rule sequence $\sigma_{1}r'\xi$ contains an occurrence of
  the accepting rule $r$. In this case, the thesis follows by
  reasoning as in the base step.
\item $\sigma_{2}'$ contains the first occurrence of $r$ in $\sigma$.
  Clearly, this occurrence is the first accepting rule occurrence in
  the infinite derivation $\prsder{Z'}{\sigma_{2}'}{}$, and it is
  applied to level $k-1$ in $\prsder{Z'}{\sigma_{2}'}{}$.  By
  inductive hypothesis, the thesis holds of the derivation
  $\prsder{Z'}{\sigma_{2}'}{}$. Therefore, it suffices to prove that
  $Z'$ is reachable from $X$ in $\Re_{SEQ}$. By Property~\textbf{A} of
  Lemma~\ref{Lemma:From-R-To-RPAR}, applied to derivation
  $\prsder{X}{\sigma_{1}}{\parcomp{t}{Z}}$, there exists a term
  $p\in{}T_{PAR}$, and a derivation of the form
  $\prsderpar{X}{}{\parcomp{p}{Z}}$ in $\Re_{PAR}$.  From the
  definition of $\Re_{SEQ}$, we finally have that
  $\prsderseq{X}{}{Z\,\onederseq{r'}Y.(Z')}$. Hence the thesis.
  \qedhere
\end{itemize}
\end{description}
\end{proof}


Now, we can prove the \emph{only if} direction of
Theorem~\ref{theo:sound2}. Let $X\in{}Var$ and $\prsder{X}{\sigma}{}$
be an infinite accepting derivation $\mathrm{(}$resp., an infinite
derivation devoid of accepting rules, an infinite derivation with a
finite non--null number of accepting rules$\mathrm{)}$ in $\Re$ from
$X$. We have to prove that one of the following conditions holds:
\begin{itemize}
\item there exists a variable $Y$ reachable $\mathrm{(}$resp.,
  reachable through a non--accepting derivation, reachable$\mathrm{)}$
  from $X$ in $\Re_{SEQ}$, and there exists in $\Re_{PAR}$ an infinite
  accepting derivation $\mathrm{(}$resp,. an infinite derivation
  devoid of accepting rules, an infinite derivation containing a
  finite non--null number of accepting rule occurrences$\mathrm{)}$
  from $Y$.
\item there exists in $\Re_{SEQ}$ an infinite accepting derivation
  $\mathrm{(}$resp., an infinite derivation devoid of accepting rules,
  an infinite derivation containing a finite non--null number of
  accepting rule occurrences$\mathrm{)}$ from $X$.
\end{itemize}

In the following, we give the proof for the case where $X$
\multider{\sigma} is an infinite accepting derivation (the proof for
the other two cases is similar). We have to prove that one of the
following conditions holds:
\begin{description}
\item[C1] there exists a variable $Y$ reachable from $X$ in $\Re_{SEQ}$,
  and there exists an accepting infinite derivation in $\Re_{PAR}$
  from $Y$.
\item[C2] there exists an accepting infinite derivation in $\Re_{SEQ}$
  from $X$.
\end{description}
It suffices to prove that, assuming that Condition~\textbf{C1} does not
hold, Condition~\textbf{C2} must hold.  Under this
hypothesis, we show that there exists a sequence of terms
$(t_{n})_{n\in{}N}$ in $T_{SEQ}\setminus\{\varepsilon\}$, satisfying the
following properties:
\begin{description}
\item[i.] $t_{0}=X$
\item[ii.] for all $n\in{}N$,
  $\prsderseq{last(t_{n})}{\rho_{n}}{t_{n+1}}$, with $\rho_{n}$
  accepting.
\item[iii.] for all $n\in{}N$, there exists an infinite accepting
  derivation in $\Re$ from $last(t_{n})$.
\item[iv.] for all $n\in{}N$, the term $last(t_{n})$ is reachable from
  $X$ in $\Re_{SEQ}$.
\end{description}
For $n=0$, Properties~\textbf{i}, ~\textbf{iii} and ~\textbf{iv} are
satisfied, by setting $t_{0}=X$.  

Assume now the existence of a finite sequence of terms
$t_{0},t_{1},\ldots,t_{n}$ in $T_{SEQ}\setminus\{\varepsilon\}$, satisfying
Properties~\textbf{i--iv}. It suffices to prove that there exists a
term $t_{n+1}$ in $T_{SEQ}\setminus\{\varepsilon\}$ satisfying the
\textbf{iii} and \textbf{iv}, and a derivation
$\prsderseq{last(t_{n})}{\rho_{n}}{t_{n+1}}$, with $\rho_{n}$
accepting.  

By inductive hypothesis, $last(t_{n})$ is reachable from $X$ in
$\Re_{SEQ}$, and there exists in $\Re$ an infinite accepting
derivation from $last(t_{n})$.  From
Lemma~\ref{Lemma:From-R-To-RPAR-Inf2} applied to the variable
$last(t_{n})$, and the fact that Condition~\textbf{C1} does not hold,
it follows that there exists a term $t\in{}T_{SEQ}\setminus\{\varepsilon\}$ such that
$\prsderseq{last(t_{n})}{\rho_{n}}{t}$, with $\rho_{n}$
accepting, and there exists an infinite accepting derivation in $\Re$ from $last(t)$.
The term $last(t)$ is reachable in $\Re_{SEQ}$ from $last(t_{n})$, and
$last(t_{n})$ is reachable from $X$ in $\Re_{SEQ}$. Therefore,
$last(t)$ is reachable in $\Re_{SEQ}$ from $X$. Thus, by setting
$t_{n+1} = t$, we obtain the result.
    
Let now $(t_{n})_{n\in{}N}$ be the sequence of terms in
$T_{SEQ}\setminus\{\varepsilon\}$ satisfying Properties
\textbf{i}--\textbf{iv}. Then, by Property~\emph{P1} of
Proposition~\ref{Prop:Subterms2}, we have that for every $n\in N$:
\begin{displaymath}
\prsderseq{t_{n}}{\rho_{n}}{t_{n}\circ{}t_{n+1}}
\end{displaymath}
that is an accepting derivation.
Moreover, by Property~\emph{P2} of Proposition~\ref{Prop:Subterms2},
we have that, for all $n\in{}N$:
\begin{displaymath}
\prsderseq{t_{0}\circ t_{1}\circ \ldots\circ t_{n}}{\rho_{n}}
     {t_{0}\circ t_{1}\circ\ldots\circ t_{n}\circ_{n+1}}
\end{displaymath}
that is an accepting derivation. Therefore, the following derivation
\begin{eqnarray}
\textrm{$X=t_{0}$ \multiderseq{\rho_{0}} $t_{0}$$\circ$$t_{1}$
\multiderseq{\rho_{1}} $t_{0}$$\circ$$t_{1}$$\circ$$t_{2}$
\multiderseq{\rho_{2}} $\ldots$ \multiderseq{\rho_{n-1}}
$t_{0}$$\circ$$t_{1}$$\circ$$\ldots$$\circ$$t_{n}$}\nonumber\\
\textrm{\multiderseq{\rho_{n}}
$t_{0}$$\circ$$t_{1}$$\circ$$\ldots$$\circ$$t_{n}$$\circ$$t_{n+1}$
\multiderseq{\rho_{n+1}}$\ldots$}\nonumber
\end{eqnarray}
is an infinite accepting derivation in $\Re_{SEQ}$ from $X$. Hence
Condition~\textbf{C2} holds.

\end{document}